\documentclass[a4paper,fleqn]{cas-dc}

\usepackage[numbers]{natbib}

\def\tsc#1{\csdef{#1}{\textsc{\lowercase{#1}}\xspace}}
\tsc{WGM}
\tsc{QE}
\tsc{EP}
\tsc{PMS}
\tsc{BEC}
\tsc{DE}
\begin{document}
	\let\WriteBookmarks\relax
	\def\floatpagepagefraction{1}
	\def\textpagefraction{.001}
	\shorttitle{Causality-informed dynamic mode decomposition for unsteady flow analysis}
	\shortauthors{Y. Pan and J. Kou}
	
	\title [mode = title]{Causality-informed dynamic mode decomposition for unsteady flow modeling and analysis}

	\author[1,2,3]{Yihua Pan}
	
	\credit{Methodology, Validation, Formal analysis, Data Curation, Writing - Original Draft, Writing - Review \& Editing, Visualization}
	
	\affiliation[1]{organization={School of Aeronautics, Northwestern Polytechnical University},
		city={Xi'an},
		postcode={710072},
		country={China}}
	
	\author[1,2,3]{Jiaqing Kou}[style=chinese,
	orcid=0000-0001-0000-0000]
	\cormark[1]
	\ead{jqkou@nwpu.edu.cn}
	\cortext[cor1]{Corresponding author}
    \credit{Conceptualization, Formal analysis, Writing - Review \& Editing, Supervision, Funding acquisition}
	
	\affiliation[2]{organization={International Joint Institute of Artificial Intelligence on Fluid Mechanics, Northwestern Polytechnical University},
		city={Xi'an},
		postcode={710072},
		country={China}}
	
	\affiliation[3]{organization={National Key Laboratory of Aircraft Configuration Design, Northwestern Polytechnical University},
		city={Xi'an},
		postcode={710072},
		country={China}}

\begin{abstract}
Causal analysis characterizes directional dependencies in dynamical systems, while such information is not typically used to inform reduced-order modeling of unsteady flows. This study proposes a Granger Causality-informed Dynamic Mode Decomposition (GC-DMD) framework that incorporates such information into data-driven modeling and modal analysis of high-dimensional flow data. The framework uses conditional Granger causality estimates to regularize the identification of the DMD operator, preferentially suppressing weakly supported interactions while preserving dynamically relevant couplings. This regularization promotes sparse operators informed by directional predictive dependencies, leading to improved predictive accuracy and identification of dynamically relevant modes. GC-DMD is evaluated on a linear system with a known coupling structure, lid-driven cavity flows with varying dynamical complexity, and a broadband transonic turbulent jet. Compared with standard DMD, GC-DMD improves operator identification accuracy in the linear benchmark and improves the robustness of the identified spectrum while reducing long-horizon prediction errors in the cavity flows. For the turbulent jet, it yields a sparser operator with improved prediction accuracy. These results demonstrate the potential of incorporating causality information to guide the identification of sparse, predictive reduced-order models of complex flows. The \href{https://github.com/JiaqingKou/GC-DMD-Granger-Causality-informed-DMD}{source codes} \footnote{https://github.com/JiaqingKou/GC-DMD-Granger-Causality-informed-DMD} from this research are publicly available in the GitHub repository.
\end{abstract}

% \includegraphics{figs/fig3.pdf}

% \begin{highlights}
% \item Research highlights item 1
% \item Research highlights item 2
% \item Research highlights item 3
% \end{highlights}


\maketitle

\section{Introduction}
Unsteady flows exhibit high-dimensional spatiotemporal dynamics involving interactions across multiple spatial and temporal scales \cite{taira2017modal}. Although high-fidelity simulations and time-resolved experiments provide detailed flow-field information, the resulting large-scale datasets pose substantial challenges for computational efficiency, long-term prediction, and physical interpretation \cite{rowley2017model}. Nevertheless, the dominant dynamics of many flows can be represented by a limited number of coherent structures, motivating the development of Reduced-Order Modeling (ROM) techniques \cite{schmid2022dynamic}. Among these, data-driven methods construct reduced representations directly from observations without explicitly projecting the governing equations. Proper Orthogonal Decomposition (POD), for instance, provides an energy-optimal basis for representing flow fields and has been widely employed for dimensionality reduction \cite{rowley2004model,braconnier2011towards}. However, its energy-based criterion does not explicitly distinguish temporal frequencies, potentially overlooking dynamically important but energetically weak structures \cite{noack2016snapshots}. Moreover, spatial dimensionality reduction alone does not determine the temporal evolution of reduced coordinates, motivating methods that identify dynamical models directly from sequential flow-field observations.

Dynamic Mode Decomposition (DMD) has become a widely used framework for analyzing and modeling unsteady flows. By extracting spatiotemporal modes associated with characteristic frequencies and growth or decay rates, DMD provides a data-driven description of flow dynamics \cite{schmid2022dynamic,schmid2010dynamic}. Beyond modal analysis, DMD identifies a finite-dimensional linear evolution operator from sequential snapshots, whose eigendecomposition connects the temporal dynamics with their spectral and modal characteristics \cite{tu2013dynamic,kutz2016dynamic}. This dual perspective enables DMD to support modal analysis, reduced-order modeling, and temporal prediction \cite{hemati2014dynamic,zhang2019online,bai2020dynamic}.

However, standard DMD may suffer from redundant dynamical representations and limited predictive generalization as system complexity increases. Several extensions have addressed these limitations. Sparsity-Promoting DMD (SPDMD) selects dynamically important modes through sparse amplitude regularization \cite{jovanovic2014sparsity}, while Nuclear-Norm Regularized DMD (NNR-DMD) and Optimal Mode Decomposition (OMD) seek compact low-dimensional representations \cite{wang2016nuclear,wynn2013optimal}. Other approaches focus on predictive reliability: consistent DMD improves forward–backward dynamical consistency \cite{azencot2019consistent}, whereas constrained optimized DMD imposes spectral stability constraints to mitigate spurious unstable dynamics \cite{rains2024constrained}. These developments primarily address dynamical compactness and predictive reliability.

From an operator perspective, DMD dynamics are governed by the identified linear evolution operator, whose entries characterize directional couplings among reduced coordinates and whose eigenspectrum determines the associated frequencies and growth or decay rates \cite{tu2014spectral}. Its structure therefore influences model complexity, spectral properties, and long-term prediction. Recent methods have incorporated constraints directly into operator identification. Regularized DMD (ReDMD) introduces operator regularization to balance prediction accuracy and stability \cite{xie2024regularized}, while physics-informed DMD (piDMD) constrains the operator according to prescribed physical properties, including conservation, symmetry, and locality \cite{baddoo2021physics}. Nevertheless, the underlying coupling structure is often unknown for complex unsteady flows, motivating the use of data-derived directional information to guide operator identification.

Causal analysis provides a means of inferring directional dependencies from data, offering potentially informative structural priors for operator identification \cite{ding2026understanding}. Among established methods, Granger Causality (GC) evaluates whether the past of one variable improves the prediction of another beyond the information provided by other relevant variables \cite{guo2010granger,shojaie2022granger}. Other approaches, including Transfer Entropy (TE) \cite{bossomaier2016transfer}, Peter and Clark Momentary Conditional Independence Plus (PCMCI+) \cite{miersch2025evaluating}, and Synergistic Unique Redundant Decomposition (SURD) \cite{martinez2024decomposing}, extend causal analysis to nonlinear dependencies and multivariate information interactions. In fluid mechanics, causal methods have been applied to investigate directional relationships among flow structures, spatial scales, and reduced-order variables \cite{martinez2023causality,massaro2026interpreting,wang2026causal}. Applications include the analysis of energy redistribution and bursting dynamics in wall-bounded turbulence \cite{tissot2014granger}, as well as self-sustaining processes and interactions among coherent structures \cite{towne2023database}.

Despite these advances, causal analysis has primarily been used to diagnose dynamical interactions rather than to guide evolution-operator identification. Recent studies have explored related connections, using GC for model-order selection in Hankel-DMD \cite{gunjal2023granger} or Koopman/DMD models to infer causal relationships \cite{nathaniel2025deep}. In linear dynamical systems, operator couplings are closely related to Granger-causal dependencies under suitable assumptions \cite{partington2004linear,srivastava2026operator}. However, conventional DMD identifies these couplings primarily through least-squares fitting without explicitly accounting for causal structure. Estimation errors, limited observations, and correlations among reduced coordinates may therefore introduce spurious couplings or obscure meaningful interactions, compromising the interpretability and predictive generalization of the identified operator. Incorporating data-derived causal information into operator regularization therefore offers a means of promoting structurally meaningful sparsity while preserving important directional couplings.

Motivated by this gap, the present study develops a Granger Causality-informed Dynamic Mode Decomposition (GC-DMD) framework that incorporates data-driven directional information directly into the reduced-order operator identification. GC-derived dependencies are translated into coupling-dependent regularization, preferentially suppressing weakly supported interactions while retaining dynamically relevant couplings. In this way, causal information serves as a data-driven structural prior for learning a compact and predictive evolution operator. The proposed framework is examined from three complementary perspectives. A linear system with known coupling structure is first used to investigate the relationship between GC and operator identification. Lid-driven cavity flows with increasing dynamical complexity are then considered to assess spectral stability and long-term prediction, followed by a broadband transonic jet flow to examine operator sparsification and compact dynamical representation. Together, these cases evaluate GC-DMD from operator identification to predictive dynamics and dynamical compactness. 

The remainder of this paper is organized as follows. Section 2 presents GC, DMD, and the proposed GC-DMD formulation. Sections 3-5 investigate operator identification, cavity-flow prediction, and broadband jet dynamics, respectively. Section 6 summarizes the main conclusions.

\section{Methodology}

\subsection{Dynamic mode decomposition}

DMD provides a data-driven  representation of dynamical systems by identifying a linear evolution operator from sequential snapshots \cite{schmid2010dynamic,tu2014spectral}. In the present study, DMD serves as the baseline framework for reduced-order operator learning. The flow-field snapshots are first projected onto a low-dimensional Singular Value Decomposition (SVD) subspace, after which the temporal evolution of the reduced coordinates is identified through a linear operator. This formulation enables the coupling structure, spectral characteristics, and predictive behavior of the reduced dynamics to be analyzed within a unified framework.

Consider a snapshot matrix $\boldsymbol{X}=[\boldsymbol{x}_1,\ldots,\boldsymbol{x}_m]\in\mathbb{R}^{n\times m}$, where $\boldsymbol{x}_k\in\mathbb{R}^{n}$ denotes the mean-subtracted flow-field snapshot at the $k$th time instant, $n$ is the number of spatial degrees of freedom, and $m$ is the number of snapshots. Following Hirsh et al.~\cite{hirsh2020centering}, the temporal mean is subtracted to focus on the fluctuation dynamics and improve the identification of the linear evolution operator. To identify the linear operator at a manageable computational cost, a low-dimensional representation is first obtained through SVD.
\begin{equation}
    \boldsymbol{X}
    =
    \boldsymbol{U}\boldsymbol{\Sigma}\boldsymbol{V}^{H},
    \label{eq:svd}
\end{equation}
where $\boldsymbol{U}$ and $\boldsymbol{V}$ contain the left and right
singular vectors, respectively, $\boldsymbol{\Sigma}$ contains the
singular values, and the superscript $H$ denotes the conjugate transpose.  The decomposition is truncated to retain the leading $r$ modes, where
$r$ denotes the dimension of the reduced space and is selected according to the cumulative energy and reconstruction accuracy for each flow case. The first $r$ left singular vectors, denoted by
$\boldsymbol{U}_r\in\mathbb{R}^{n\times r}$, constitute the SVD basis. Each flow-field state is then represented by the reduced coordinates
\begin{equation}
    \boldsymbol{q}_k
    =
    \boldsymbol{U}_r^{H}\boldsymbol{x}_k,
    \qquad
    \boldsymbol{q}_k\in\mathbb{R}^{r}.
    \label{eq:reduced_coordinates}
\end{equation}

DMD approximates the temporal evolution of the full-order flow-field
states through a linear mapping
$\boldsymbol{x}_{k+1}\approx
\boldsymbol{A}^{\mathrm{full}}\boldsymbol{x}_k$,
where $\boldsymbol{A}^{\mathrm{full}}\in\mathbb{R}^{n\times n}$
denotes the full-order evolution operator. Projecting this mapping
onto the SVD subspace yields
\begin{equation}
    \boldsymbol{q}_{k+1}
    \approx
    \boldsymbol{A}\boldsymbol{q}_k,
    \qquad
    \boldsymbol{A}
    =
    \boldsymbol{U}_r^{H}
    \boldsymbol{A}^{\mathrm{full}}
    \boldsymbol{U}_r,
    \label{eq:reduced_operator}
\end{equation}
where $\boldsymbol{A}\in\mathbb{R}^{r\times r}$ denotes the reduced-order evolution operator. Throughout this study, $\boldsymbol{A}$ refers to the operator in the reduced coordinate space, while
$\boldsymbol{A}^{\mathrm{full}}$ denotes its full-order counterpart.
The reduced DMD operator is then identified by minimizing the
discrepancy between consecutive snapshot pairs,
\begin{equation}
    \boldsymbol{A}^{\mathrm{DMD}}
    =
    \arg\min_{\boldsymbol{A}}
    \left\|
    \boldsymbol{Q}_2-\boldsymbol{A}\boldsymbol{Q}_1
    \right\|_F^2,
    \label{eq:dmd_operator}
\end{equation}
where
$\boldsymbol{Q}_1=[\boldsymbol{q}_1,\ldots,\boldsymbol{q}_{m-1}]$
and $\boldsymbol{Q}_2=[\boldsymbol{q}_2,\ldots,\boldsymbol{q}_m]$,
and $\|\cdot\|_F$ denotes the Frobenius norm. The element $A_{ij}$
characterizes the one-step predictive coupling from the $j$th to
the $i$th reduced coordinate, providing a natural connection between
the operator structure and directional dependencies among the
reduced variables.

The spectral and modal characteristics of the identified operator
are obtained from
\begin{equation}
    \boldsymbol{A}\boldsymbol{W}
    =
    \boldsymbol{W}\boldsymbol{\Lambda},
    \qquad
    \boldsymbol{\Phi}
    =
    \boldsymbol{U}_r\boldsymbol{W},
    \qquad
    \omega_j
    =
    \frac{\log(\lambda_j)}{\Delta t},
    \label{eq:dmd_spectrum}
\end{equation}
where $\boldsymbol{W}$ contains the eigenvectors of $\boldsymbol{A}$,
$\boldsymbol{\Lambda}=\mathrm{diag}(\lambda_1,\ldots,\lambda_r)$
contains the corresponding discrete-time eigenvalues,
$\boldsymbol{\Phi}$ contains the DMD modes in the physical space,
and $\Delta t$ is the sampling interval. The continuous-time
eigenvalue $\omega_j$ characterizes the temporal behavior of the
$j$th mode, with $\mathrm{Re}(\omega_j)$ describing its growth or
decay rate and $\mathrm{Im}(\omega_j)/(2\pi)$ giving its frequency.
These quantities are subsequently used to examine the spectral
stability and dynamical characteristics of DMD and GC-DMD.

For comparisons of dominant dynamical structures, the DMD modes
are ranked using the criterion proposed by Kou et al.
\cite{kou2017improved}, which accounts for the accumulated temporal
contribution of each mode rather than its initial amplitude alone.
The importance of the $j$th mode is evaluated as
\begin{equation}
    I_j
    =
    \sum_{k=1}^{m}
    \left|b_j\lambda_j^{\,k-1}\right|
    \left\|\boldsymbol{\phi}_j\right\|_F^2
    \Delta t,
    \qquad
    \boldsymbol{b}
    =
    \boldsymbol{W}^{-1}\boldsymbol{q}_1,
    \label{eq:kou_rank}
\end{equation}
where $I_j$ denotes the importance of the $j$th mode,
$\boldsymbol{\phi}_j$ is the corresponding normalized DMD mode,
$b_j$ is its initial modal amplitude, $\boldsymbol{b}$ is the
vector of modal amplitudes, and $m$ is the number of snapshots
used for modal ranking. The modes are ranked in descending order
of $I_j$. Since this criterion has been shown to be effective in many previous studies\cite{liu2021data} , this criterion is consistently adopted in the following flow cases for the identification and comparison of representative DMD and GC-DMD modes.

\subsection{Granger causality}

Granger causality provides a statistical framework for quantifying
directional predictive dependencies in multivariate dynamical systems
\cite{granger1969investigating,barnett2014mvgc}. In the present study,
GC is evaluated among the reduced coordinates introduced in
Section~2.1. Its basic principle is to determine whether the past
information of one coordinate provides additional predictive
information for another when the remaining coordinates are taken
into account.

For each directed pair $q_j\rightarrow q_i$, where $q_j$ and $q_i$
denote the candidate source and target coordinates, respectively,
GC compares a full autoregressive model with a restricted model in
which $q_j$ is excluded from the predictors of $q_i$. Since the DMD operator considered here describes the first-order mapping from $\boldsymbol{q}_k$ to $\boldsymbol{q}_{k+1}$, a first-order vector autoregressive model, VAR(1), is adopted to maintain a consistent one-step dynamical representation, 
\begin{equation}
\begin{aligned}
q_i(k+1)
&=
c_i^{F}
+\sum_{\ell=1}^{r}a_{i\ell}q_\ell(k)
+\varepsilon_i^{F}(k),\\
q_i(k+1)
&=
c_i^{R}
+\sum_{\substack{\ell=1\\\ell\neq j}}^{r}
a_{i\ell}^{R}q_\ell(k)
+\varepsilon_i^{R}(k),
\end{aligned}
\label{eq:gc_models}
\end{equation}
where $k$ denotes the time index, $r$ is the reduced-order dimension, and the superscripts $F$ and $R$ refer to the full and restricted autoregressive models, respectively. The intercepts $c_i^{F}$ and $c_i^{R}$ and the regression coefficients $a_{i\ell}$ and
$a_{i\ell}^{R}$ are estimated separately for the corresponding models
by ordinary least squares. The terms $\varepsilon_i^{F}(k)$ and
$\varepsilon_i^{R}(k)$ denote their prediction residuals, which are
computed from one-step-ahead predictions using the observed historical states, rather than from recursively generated model trajectories. The restricted model excludes the past information of the candidate source coordinate $q_j$ while retaining the intercept and all remaining predictors.

The conditional GC strength is quantified by the increase in prediction error when $q_j$ is excluded,
\begin{equation}
    \mathcal{G}_{j\rightarrow i}
    =
    \ln
    \left(
    \frac{\sigma^2_{i,R}}
         {\sigma^2_{i,F}}
    \right),
    \label{eq:gc_strength}
\end{equation}
where $\mathcal{G}_{j\rightarrow i}$ denotes the conditional
GC strength from $q_j$ to $q_i$, and $\sigma^2_{i,F}$ and $\sigma^2_{i,R}$ are the variances of the full and restricted models of residual $\varepsilon$, respectively. A larger $\mathcal{G}_{j\rightarrow i}$ indicates a stronger predictive contribution from $q_j$ to $q_i$.

To distinguish statistically supported interactions from variations arising from finite samples, each directed pair is tested under the null hypothesis $H_0:q_j\nrightarrow q_i$. Since the full and restricted models are nested, their residual sums of squares are compared using an $F$-test,
\begin{equation}
    \mathcal{F}_{j\rightarrow i}
    =
    \frac{
    \left(\mathrm{RSS}_{i,R}
    -\mathrm{RSS}_{i,F}\right)/d_1
    }{
    \mathrm{RSS}_{i,F}/d_2
    }, \qquad
    d_1=1,\quad d_2=T-r-1,
    \label{eq:gc_ftest}
\end{equation}
where $\mathrm{RSS}_{i,F}$ and $\mathrm{RSS}_{i,R}$ denote the residual sums of squares of the full and restricted models, respectively, defined as
\begin{equation}
    \mathrm{RSS}_{i,F}
    =
    \sum_{k=1}^{T}
    \left[\varepsilon_i^{F}(k)\right]^2,
    \qquad
    \mathrm{RSS}_{i,R}
    =
    \sum_{k=1}^{T}
    \left[\varepsilon_i^{R}(k)\right]^2.
    \label{eq:gc_rss}
\end{equation}
Here, $T$ is the number of regression samples, while $d_1$ and $d_2$ are the numerator and denominator degrees of freedom, respectively.

Because significance testing is performed for all $M=r(r-1)$ directed off-diagonal pairs, independently thresholding the resulting $p$-values $p_{ij}$ may lead to multiple false-positive interactions. Here, $p_{ij}$ denotes the upper-tail probability of the observed statistic $\mathcal{F}_{j\rightarrow i}$ under the null hypothesis, evaluated from the $F(d_1,d_2)$ distribution. The Benjamini-Hochberg False Discovery Rate (BH-FDR) procedure \cite{benjamini1995controlling} is therefore applied to control the expected proportion of false discoveries among the retained interactions. Let $p_{(1)}\leq\cdots\leq p_{(M)}$ denote the ordered $p$-values. The largest index satisfying
\begin{equation}
    k^{*}
    =
    \max
    \left\{
    k:
    p_{(k)}
    \leq
    \frac{k}{M}q_{\mathrm{FDR}}
    \right\}
    \label{eq:bh_fdr}
\end{equation}
defines the critical value $p_{\mathrm{crit}}=p_{(k^{*})}$, where $q_{\mathrm{FDR}}$ is the prescribed target false discovery rate. If no index satisfies the criterion, no interaction is retained.
Otherwise, the interaction $q_j\rightarrow q_i$ is considered statistically significant when $p_{ij}\leq p_{\mathrm{crit}}$.
In this study, $q_{\mathrm{FDR}}=0.05$ is used throughout. The resulting significance-filtered GC structure provides the directional information used for the GC-informed operator regularization introduced in Section~2.3.

\subsection{Granger Causality-informed DMD}

Based on the directional dependencies identified above, the GC-DMD framework is proposed to incorporate data-driven causal information directly into operator learning. Rather than prescribing a fixed operator structure, GC-DMD translates the identified directional dependencies into coupling-dependent regularization weights. Interactions strongly supported by GC are therefore preferentially retained, while weakly supported or statistically insignificant couplings are more strongly suppressed. In principle, this framework can be applied to either the original state space or a reduced coordinate space. In the present study, the latter is adopted to enable efficient modeling of high-dimensional flow-field dynamics.

Following the BH-FDR correction described in Section~2.2, the GC strengths are first filtered according to
\begin{equation}
    G_{ij}^{\mathrm{sig}}
    =
    \begin{cases}
    \mathcal{G}_{j\rightarrow i},
    & p_{ij}\leq p_{\mathrm{crit}},\quad i\neq j,\\
    0, & \mathrm{otherwise},
    \end{cases}
    \label{eq:gc_filter}
\end{equation}
where $G_{ij}^{\mathrm{sig}}$ denotes the significance-filtered
GC strength from the $j$th to the $i$th coordinate. To express the relative contributions of different source coordinates to each target coordinate, the retained GC strengths are normalized row-wise as
\begin{equation}
    C_{ij}
    =
    \begin{cases}
    \displaystyle
    \frac{G_{ij}^{\mathrm{sig}}}
    {\sum_{\ell=1}^{r}G_{i\ell}^{\mathrm{sig}}},
    & i\neq j,\\[6pt]
    0, & i=j,
    \end{cases}
    \label{eq:gc_normalization}
\end{equation}
where the matrix $\boldsymbol{C}\in\mathbb{R}^{r\times r}$ represents the normalized conditional GC structure, with $C_{ij}$ quantifying the relative statistical support for the directed interaction $q_j\rightarrow q_i$ among the significant sources of $q_i$. Since self-causality is overlooked in the standard GC, diagonal entries are set to $C_{ii}=0$. In addition, If the denominator is zero, the corresponding row of $\boldsymbol{C}$ is set to zero. 

Using $\boldsymbol{C}$ as a data-driven structural prior, the GC-DMD operator is identified by solving
\begin{equation}
\begin{aligned}
\boldsymbol{A}^{\mathrm{GC-DMD}}
=\arg\min_{\boldsymbol{A}}\Bigg[
&\frac{1}{2(m-1)}
\left\|
\boldsymbol{Q}_2-\boldsymbol{A}\boldsymbol{Q}_1
\right\|_F^2\\
&+\lambda
\sum_{i=1}^{r}
\sum_{\substack{j=1\\j\neq i}}^{r}
(1-C_{ij})|A_{ij}|
\Bigg],
\end{aligned}
\label{eq:gcdmd_operator}
\end{equation}
where $\lambda\geq0$ controls the overall regularization strength. The first term measures the one-step fitting error, while the second term imposes a GC-weighted penalty on the off-diagonal operator entries. Statistically insignificant interactions, for which $C_{ij}=0$, receive the full penalty, whereas interactions with stronger GC support receive progressively smaller penalties. The diagonal entries are left unpenalized to preserve the self-dynamics of the reduced coordinates. For $\lambda=0$, the formulation reduces to standard DMD. As $\lambda$ increases, weakly supported off-diagonal couplings are preferentially suppressed, while interactions with stronger GC support are more readily retained. The optimization problem is solved using the Fast Iterative Shrinkage-Thresholding Algorithm (FISTA)\cite{beck2009fast}.

To identify the optimal regularization parameter $\lambda$, the flow snapshots are divided into training and validation set. The GC structure, as well as the operator $\boldsymbol{A}(\lambda)$, are estimated using the training data, while the performance of the identified operator is evaluated on the validation set through one-step prediction. This procedure is repeated for a prescribed set of candidate regularization parameters $\boldsymbol{\Lambda}$. The regularization parameter is selected as
\begin{equation}
    \lambda^{*}
    =
    \arg\min_{\lambda\in\boldsymbol{\Lambda}}
    \mathrm{RMSE}_{\mathrm{val}}
    \left[\boldsymbol{A}(\lambda)\right],
    \label{eq:lambda_selection}
\end{equation}
where $\mathrm{RMSE}_{\mathrm{val}}$ denotes the root mean square error of one-step prediction on the validation set, defined as
\begin{equation}
    \mathrm{RMSE}_{\mathrm{val}}(\boldsymbol{A})
    =
    \sqrt{
    \frac{1}{rT_{\mathrm{val}}}
    \left\|
    \boldsymbol{Q}_{2,\mathrm{val}}
    -
    \boldsymbol{A}\boldsymbol{Q}_{1,\mathrm{val}}
    \right\|_F^2
    },
    \label{eq:validation_rmse}
\end{equation}
where $T_{\mathrm{val}}$ is the number of validation snapshot pairs, $\boldsymbol{Q}_{1,val}=[\boldsymbol{q}_{1,val},\ldots,\boldsymbol{q}_{T,val}]$
and $\boldsymbol{Q}_{2,val}=[\boldsymbol{q}_{2,val},\ldots,\boldsymbol{q}_{T+1,val}]$ contain the corresponding snapshots in the reduced coordinates.

To assess the predictive generalization of the identified operator, the one-step prediction RMSE on the independent test set is defined as
\begin{equation}
    \mathrm{RMSE}_{\mathrm{test}}(\boldsymbol{A})
    =
    \sqrt{
    \frac{1}{rT_{\mathrm{test}}}
    \left\|
    \boldsymbol{Q}_{2,\mathrm{test}}
    -
    \boldsymbol{A}\boldsymbol{Q}_{1,\mathrm{test}}
    \right\|_F^2
    },
    \label{eq:test_rmse}
\end{equation}
where $T_{\mathrm{test}}$ denotes the number of test snapshot pairs, and $\boldsymbol{Q}_{1,\mathrm{test}}$ and $\boldsymbol{Q}_{2,\mathrm{test}}$ contain the corresponding reduced-coordinate snapshots.

Thus, GC determines the relative regularization imposed on individual operator entries, while validation determines the overall regularization strength. Since the proposed GC-DMD adopts a regularized optimization formulation similar to the standard $L_1$ penalty, its performance is further compared with conventional $L_1$ regularization using a linear system, as presented in Appendix~\ref{app:L1}.

\section{Simple linear system with known coupling structure}

Before considering realistic flow problems, the proposed framework is first examined using an explicit linear system for which the underlying evolution operator is designed and known exactly. This allows the relationship between GC and the true operator couplings to be directly assessed and provides a basis for examining how GC-derived structural information affects operator identification.

\subsection{System definition and causal analysis}

A first-order vector autoregressive system, VAR(1), is considered to examine the relationship between GC and the underlying operator couplings,
\begin{equation}
    \boldsymbol{x}_{k+1}
    =
    \boldsymbol{A}^{\mathrm{true}}\boldsymbol{x}_k
    +
    \boldsymbol{\varepsilon}_k,
    \label{eq:toy_var}
\end{equation}
where $\boldsymbol{x}_k\in\mathbb{R}^{n_s}$ is the system state, $\boldsymbol{A}^{\mathrm{true}}\in \mathbb{R}^{n_s\times n_s}$ is the prescribed evolution operator, and $\boldsymbol{\varepsilon}_k$ denotes thee stochastic forcing, which is independent Gaussian white noise with zero mean and standard deviation $\sigma=0.02$. The system dimension is set to $n_s=20$, with a prescribed off-diagonal coupling density of 30\%. The nonzero off-diagonal coefficients have randomly assigned signs and log-uniformly distributed magnitudes. The exact coefficients in the linear operator is given in the source code of this research in the Github repository. A total of 5000 snapshots are retained, with a unit discrete time step $\Delta t=1$. The data are divided chronologically into 3500 training, 750 validation, and 750 test snapshots. Since the system variables directly constitute the state coordinates, no SVD projection is required.

In this VAR(1) system, the coefficient $A_{ij}^{\mathrm{true}}$ represents the direct one-step coupling from $x_j$ to $x_i$, while $\mathcal{G}_{j\rightarrow i}$ quantifies the corresponding conditional predictive dependence. Although these quantities are not equivalent, their directional correspondence allows the inferred GC structure to be compared with the known operator. Since the linear operator $\boldsymbol{A}$ does not directly quantify causal strength, its columns and rows are labeled as 'Source'  and 'Target', respectively.

\begin{figure}[width=\columnwidth,pos=!htbp]
    \centering

    % Row 1: (a)
    \begin{minipage}[t]{0.90\columnwidth}
        \centering
        \includegraphics[width=\linewidth]{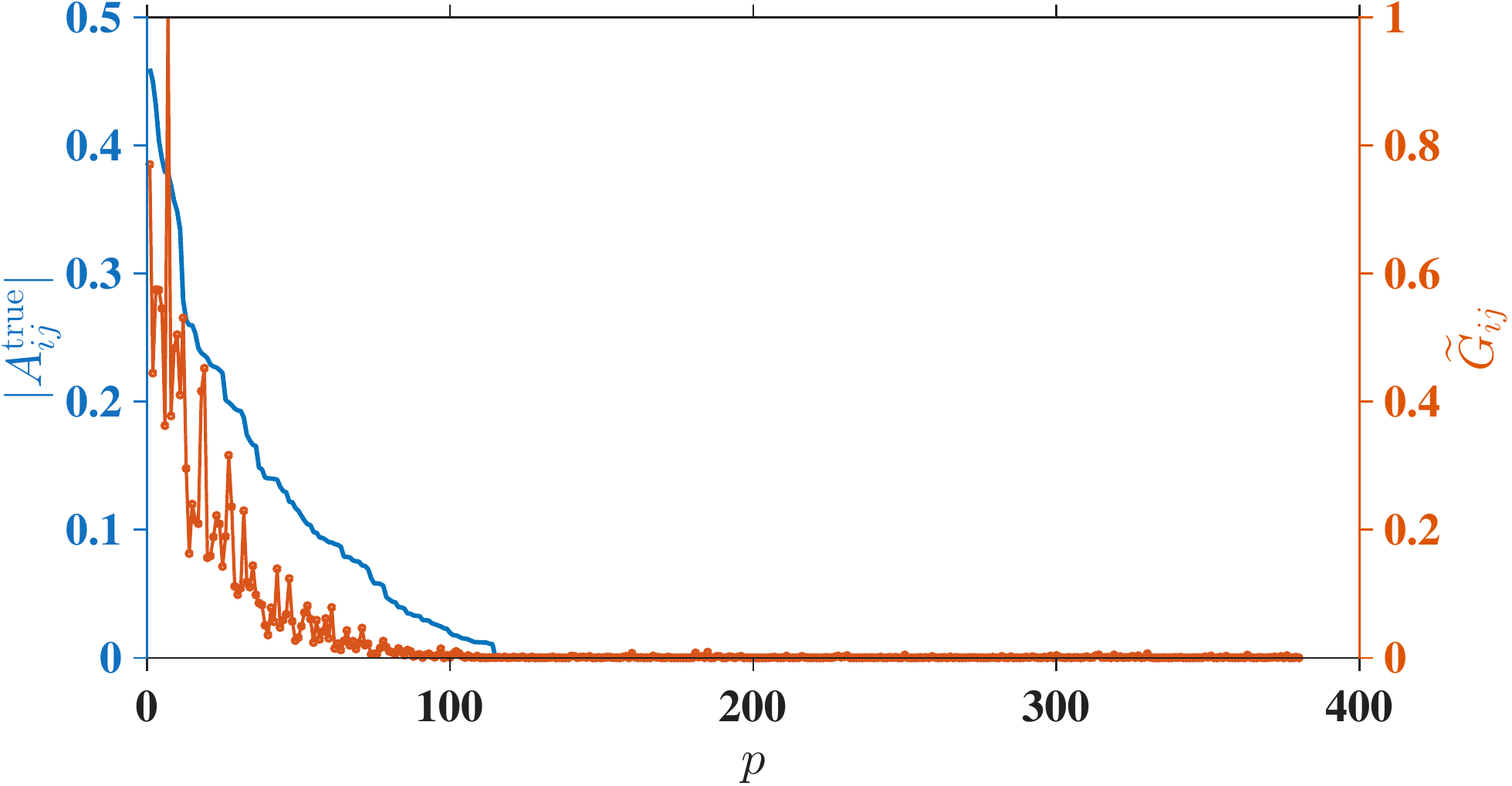}
        \par\vspace{-1mm}
        {\small (a)}
    \end{minipage}

    \par\vspace{1mm}

    % Row 2: (b) and (c)
    \begin{minipage}[t]{0.48\columnwidth}
        \centering
        \includegraphics[width=\linewidth]{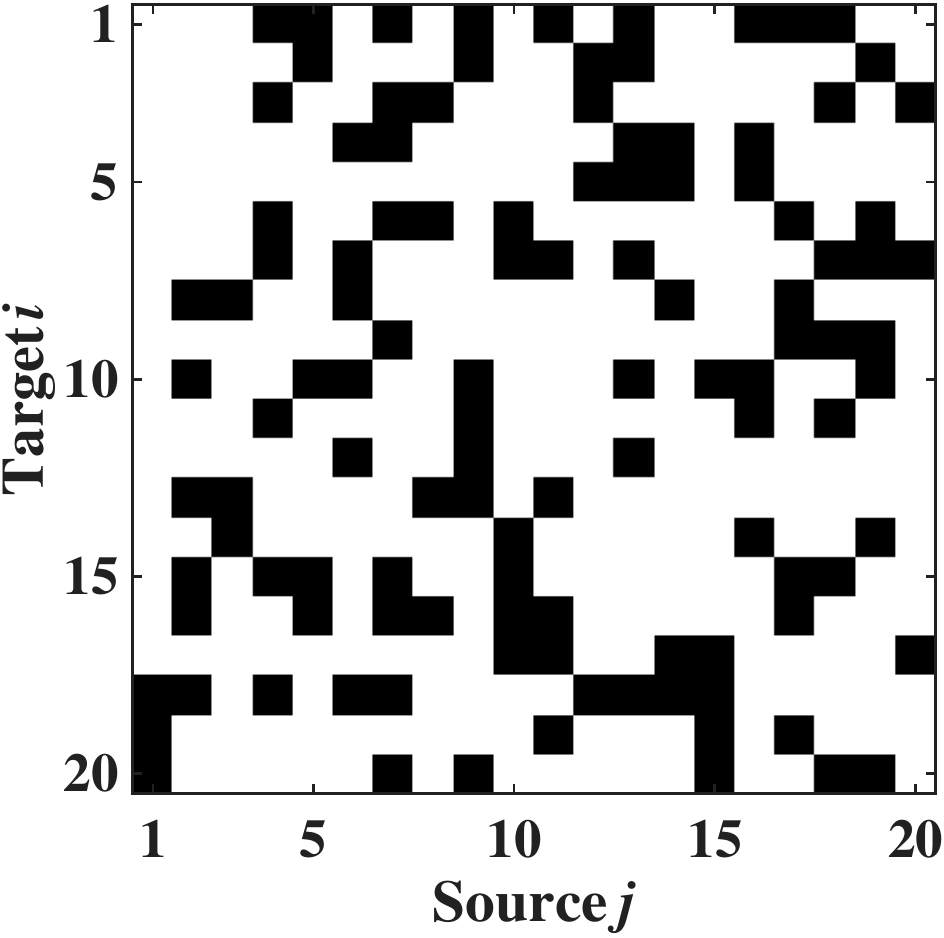}
        \par\vspace{-1mm}
        {\small (b)}
    \end{minipage}
    \hfill
    \begin{minipage}[t]{0.48\columnwidth}
        \centering
        \includegraphics[width=\linewidth]{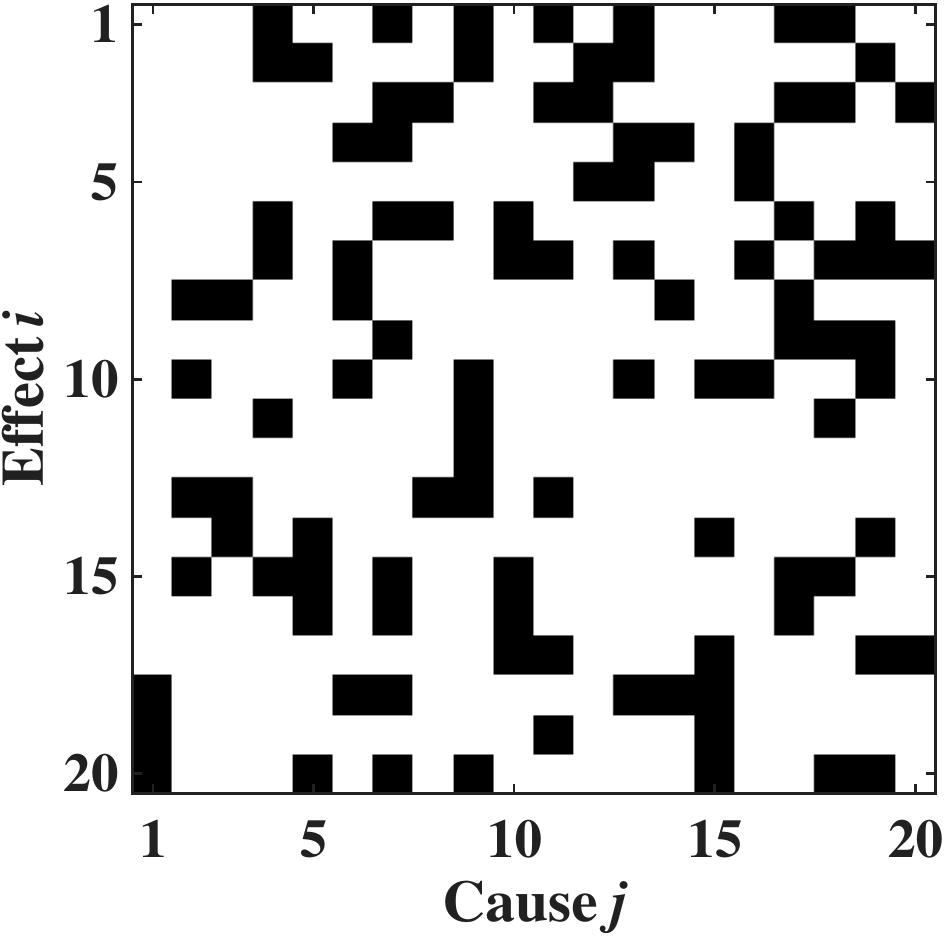}
        \par\vspace{-1mm}
        {\small (c)}
    \end{minipage}

    \caption{Relationship between the true operator couplings and the identified GC structure in the linear system. (a) True coupling magnitudes and globally normalized conditional GC strengths plotted against the sorted coefficient rank $p$, with the off-diagonal coefficients ranked in descending order of $|A_{ij}^{\mathrm{true}}|$. (b) Binary structure of the prescribed true operator couplings. (c) Binary structure of the statistically significant conditional GC interactions retained after BH-FDR correction.}
    \label{fig:toy_gc_structure}
\end{figure}

Figure~\ref{fig:toy_gc_structure} compares the prescribed operator couplings with the GC structure estimated from the training data. In Fig.~\ref{fig:toy_gc_structure}(a), the horizontal axis represents the sorted coefficient rank $p$, obtained by arranging the off-diagonal entries of the true operator in descending order of their absolute magnitudes. The left vertical axis shows the true coupling magnitude $|A_{ij}^{\mathrm{true}}|$, while the right vertical axis shows the normalized conditional GC strength $\widetilde{G}_{ij}=G_{ij}/\max_{u\neq v}G_{uv}$, where $G_{ij}$ denotes the conditional GC strength from the $j$th to the $i$th coordinate. As shown in Fig.~\ref{fig:toy_gc_structure}(a), stronger operator couplings generally correspond to higher conditional GC strengths, while weaker couplings exhibit diminished predictive contributions. Figures~\ref{fig:toy_gc_structure}(b) and \ref{fig:toy_gc_structure}(c) compare the binary structures of the true operator and the GC interactions, respectively. The former distinguishes zero (white) and nonzero (black) operator entries, while the latter distinguishes statistically insignificant (white) and significant (black) GC interactions according to the BH-FDR correction. These figures show that most statistically significant GC interactions coincide with nonzero entries of the true operator. However, some weak couplings are not identified as significant, since their predictive contributions may be limited in the presence of other correlated variables. Consequently, the inferred GC structure is generally more conservative than the original linear operator. Under ideal conditions of correct model specification, sufficient data, and consistent statistical identification, the binary structures of the true operator and the inferred GC interactions are expected to coincide. These observations support using GC as a directional soft prior for operator regularization rather than imposing a hard structural constraint.

\subsection{Linear operator identification}

Having established that the GC structure provides useful directional information about the underlying operator, its effect on operator identification is examined next. Standard DMD is used as the baseline, while GC-DMD incorporates GC-derived regularization weights according to Eq.~(\ref{eq:gcdmd_operator}). Since the true coupling structure is known in the present system, a reference result (named as Oracle) is additionally considered, in which the binary regularization weights are constructed directly from the known structure of the true operator. This results in the following binary GC matrix:

\begin{equation}
    C_{ij}^{\mathrm{oracle}}
    =
    \begin{cases}
    1, & A_{ij}^{\mathrm{true}}\neq 0,\quad i\neq j,\\
    0, & A_{ij}^{\mathrm{true}}=0,\quad i\neq j,\\
    0, & i=j,
    \end{cases}
    \label{eq:oracle_structure}
\end{equation}
where $A_{ij}^{\mathrm{true}}$ denotes the corresponding entry of the true evolution operator. The Oracle operator is then identified using
Eq.~(\ref{eq:gcdmd_operator}), with $\boldsymbol{C}$ replaced by
$\boldsymbol{C}^{\mathrm{oracle}}$. This idealized reference assumes perfect knowledge of the directional coupling structure and provides a benchmark for evaluating the effectiveness of GC-derived structural information in operator learning \cite{chang2026post}.

\begin{figure}[width=\columnwidth,pos=!htbp]
    \centering

    % Row 1
    \begin{minipage}[t]{0.48\columnwidth}
        \centering
        \includegraphics[width=\linewidth]{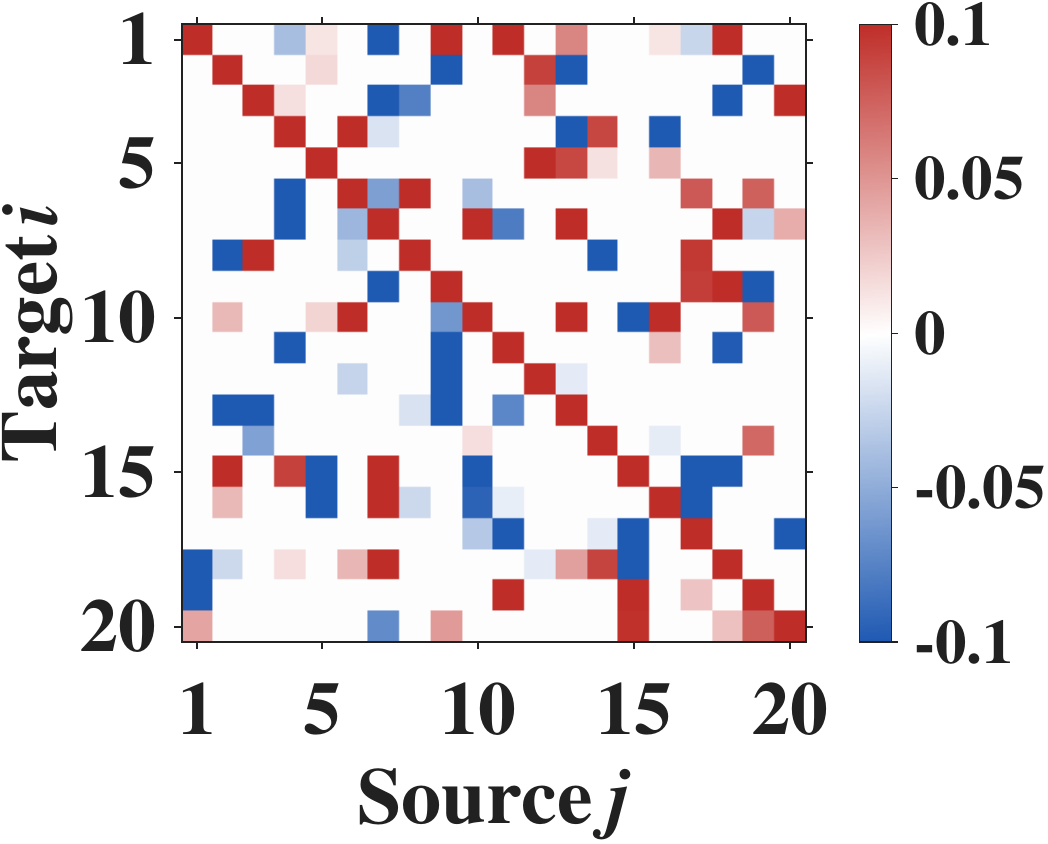}
        \par\vspace{-1mm}
        {\small (a)}
    \end{minipage}
    \hfill
    \begin{minipage}[t]{0.48\columnwidth}
        \centering
        \includegraphics[width=\linewidth]{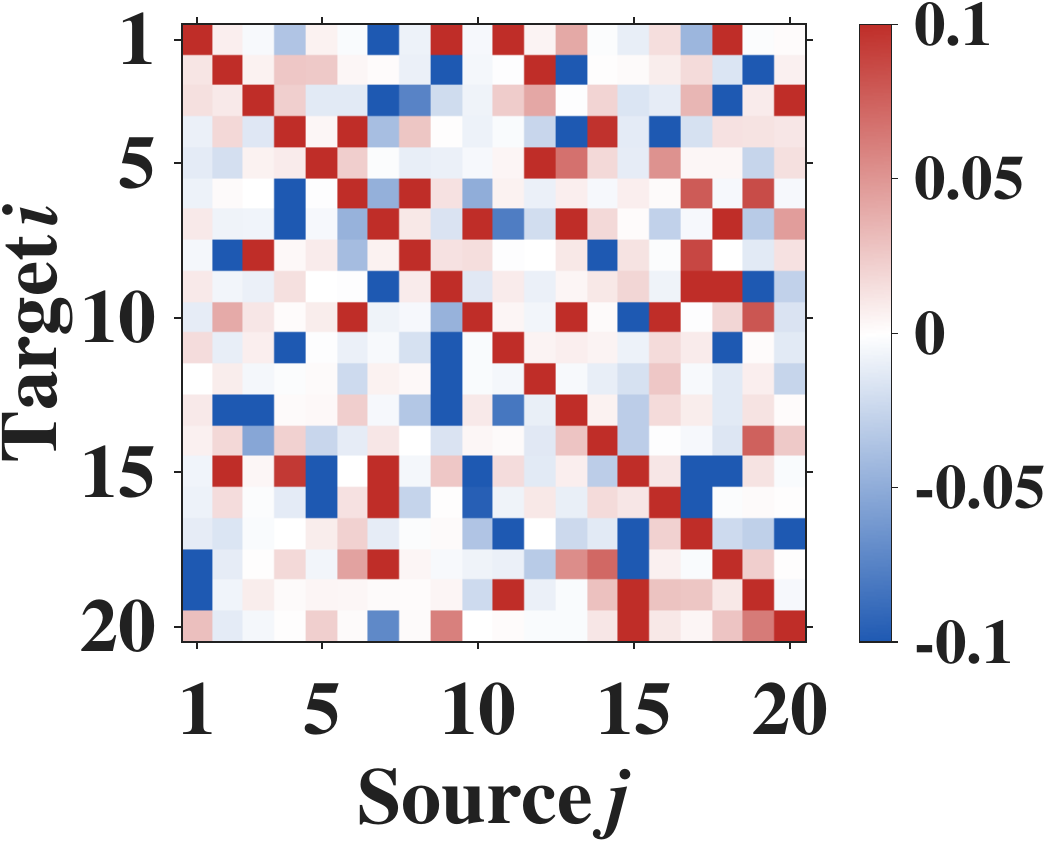}
        \par\vspace{-1mm}
        {\small (b)}
    \end{minipage}

    \par\vspace{1mm}

    % Row 2
    \begin{minipage}[t]{0.48\columnwidth}
        \centering
        \includegraphics[width=\linewidth]{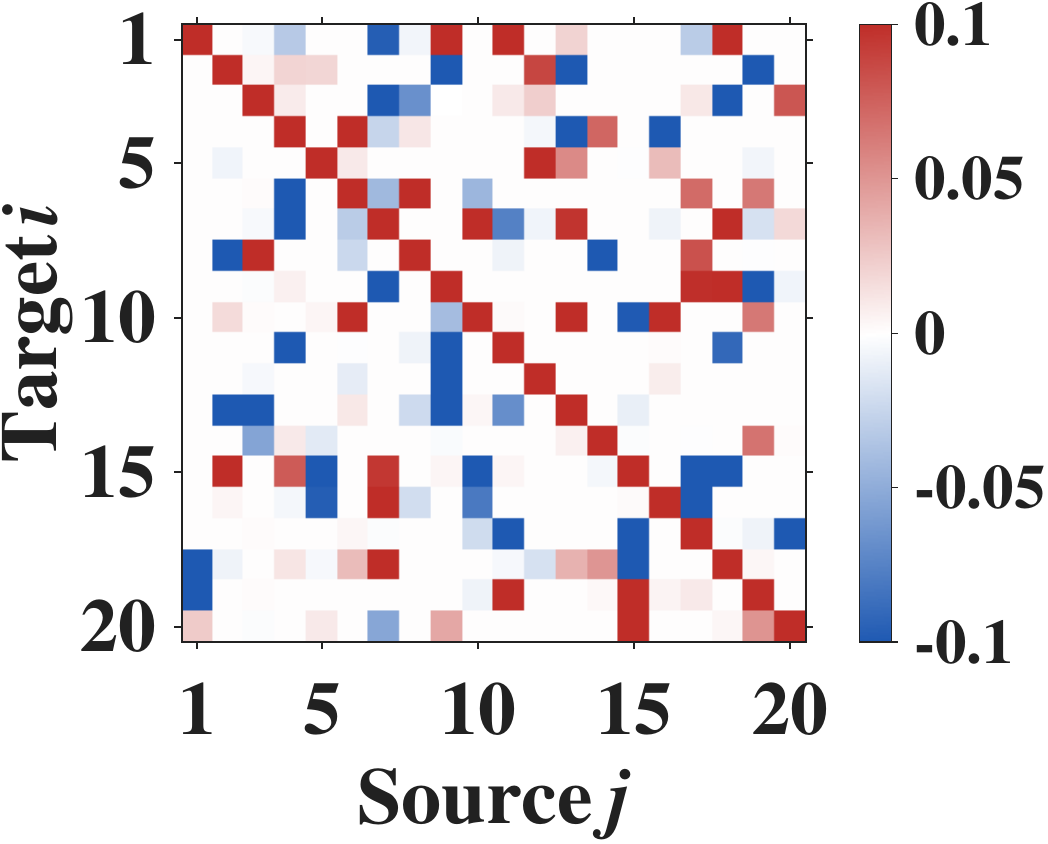}
        \par\vspace{-1mm}
        {\small (c)}
    \end{minipage}
    \hfill
    \begin{minipage}[t]{0.48\columnwidth}
        \centering
        \includegraphics[width=\linewidth]{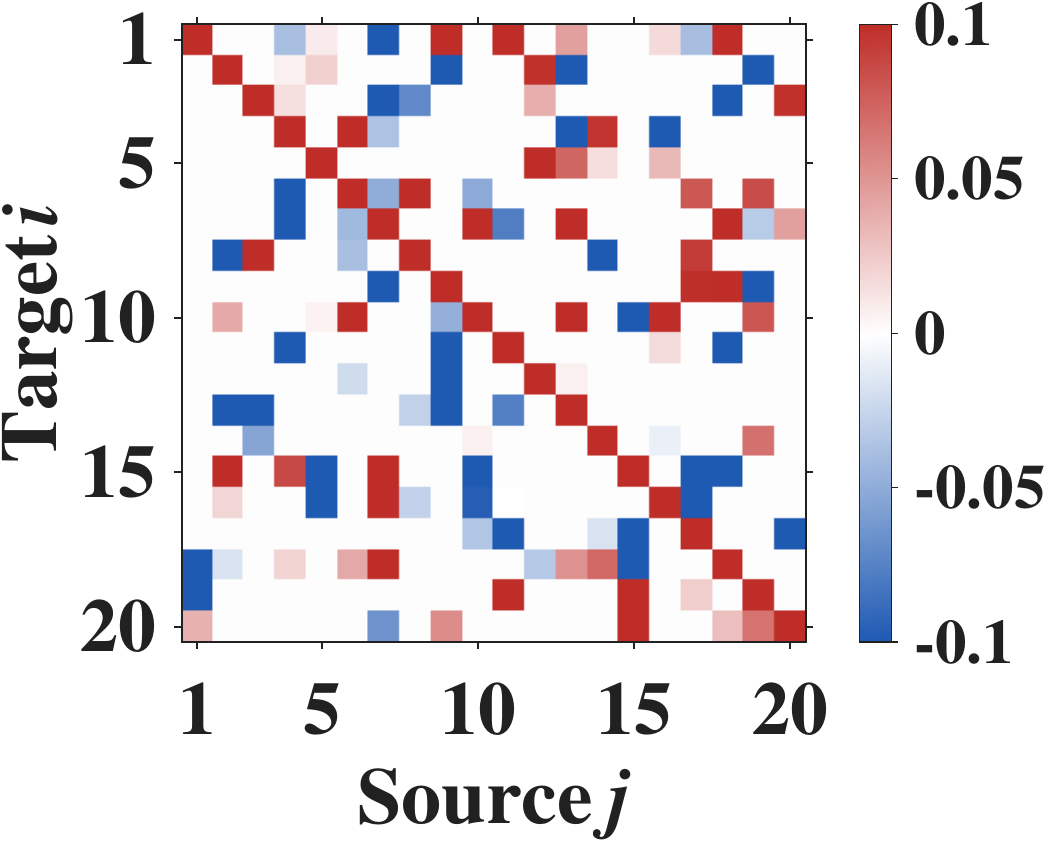}
        \par\vspace{-1mm}
        {\small (d)}
    \end{minipage}

    \caption{Comparison of the prescribed and identified evolution
    operators for the linear system.
    (a) Exact operator $\boldsymbol{A}^{\mathrm{true}}$;
    (b) standard DMD;
    (c) GC-DMD;
    (d) Oracle (GC-DMD guided by the prescribed coupling structure).
    All panels use the same color scale.}
    \label{fig:toy_operator_comparison}
\end{figure}

Figure~\ref{fig:toy_operator_comparison} compares the operators identified by DMD and GC-DMD with the prescribed operator and the Oracle reference. Standard DMD recovers the dominant entries of $\boldsymbol{A}^{\mathrm{true}}$, but also introduces numerous weak off-diagonal couplings that are absent or negligible in the true operator. In contrast, GC-DMD suppresses many of these weak couplings while retaining the dominant operator structure, resulting in a more compact representation that more closely resembles $\boldsymbol{A}^{\mathrm{true}}$. The Oracle reference exhibits a similar tendency, with the identified operator showing an even closer correspondence to the prescribed structure, which is expected.

To quantify the accuracy of operator identification, the relative Frobenius error is defined as
\begin{equation}
    E_A
    =
    \frac{
    \left\|
    \boldsymbol{A}-\boldsymbol{A}^{\mathrm{true}}
    \right\|_F
    }{
    \left\|
    \boldsymbol{A}^{\mathrm{true}}
    \right\|_F
    } \times 100\%.
    \label{eq:operator_error}
\end{equation}
This relative error decreases from $E_A=8.1\%$ for DMD to $6.4\%$ for GC-DMD, corresponding to a reduction of approximately $21\%$. The Oracle reference, representing the idealized case in which the true directional coupling pattern is known a priori, further reduces the error to $5.26\%$. These results show that incorporating directional coupling information into operator regularization improves the identification of the linear system. More importantly, GC-DMD achieves a substantial part of this improvement using directional information inferred entirely from the observed data, without prior knowledge of the true operator couplings. This demonstrates the potential of using data-driven directional information to guide the regularization of the DMD operator.

\subsection{Effect of regularization parameter}

The extent to which the directional information identified by GC affects operator identification is governed by the regularization level. In GC-DMD, the GC-derived structure determines the relative regularization applied to individual operator entries, whereas $\lambda$ controls the overall strength of this coupling-dependent regularization. Varying $\lambda$ therefore affects the GC-derived directional prior on the identified operator. 

\begin{figure*}[pos=!htbp]
    \centering
    \includegraphics[width=0.9\textwidth]{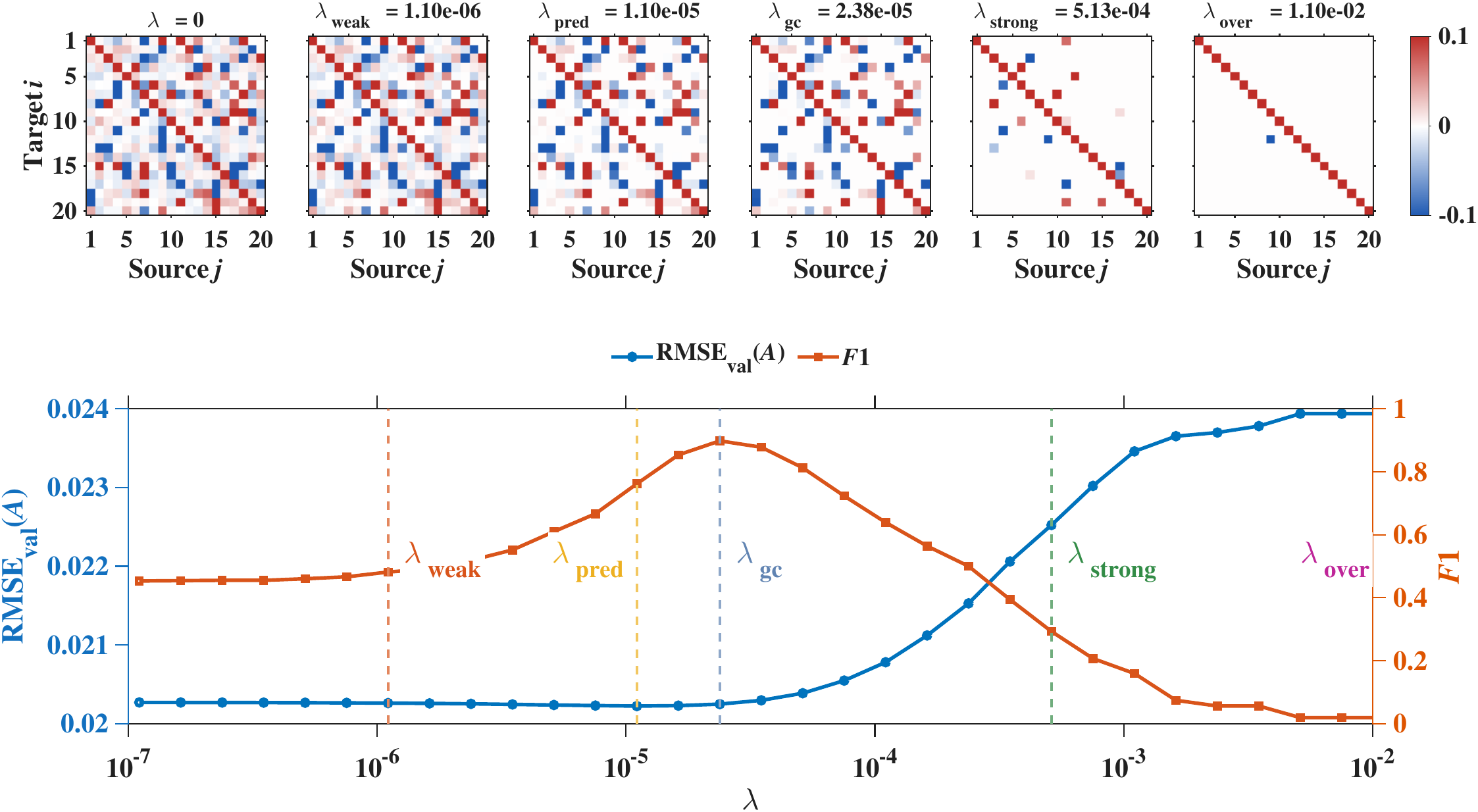}
    \caption{Effect of the regularization parameter on GC-DMD operator identification. The upper panels show the operators identified
    at representative regularization levels. The lower panel presents the variations in one-step validation error and structural consistency with the regularization parameter, where the structural consistency quantifies the agreement between the binary structures of the identified operator and the statistically significant GC interactions retained after BH-FDR correction.}
    \label{fig:toy_lambda_path}
\end{figure*}

Figure~\ref{fig:toy_lambda_path} illustrates how the one-step validation error and the structural consistency between the identified operator and the significant GC interactions vary with the regularization parameter $\lambda$, together with the operators identified by GC-DMD at representative regularization levels. To quantify this structural consistency, the F1 score, a metric commonly used in binary classification and network structure evaluation \cite{lozano2009grouped}, is adopted by treating the presence or absence of each off-diagonal interaction as a binary classification outcome. Specifically, two binary structures are constructed
according to the classification criteria introduced in Section~3.1: one representing the nonzero off-diagonal entries of the operator identified by GC-DMD, and the other representing the statistically significant GC interactions retained after BH-FDR correction. The $F1$ score is defined as
\begin{equation}
F1
=
\frac{2TP}{2TP+FP+FN},
\label{eq}
\end{equation}
where $TP$ denotes the number of off-diagonal interactions identified as nonzero in the operator and significant in the GC structure, $FP$ denotes those identified as nonzero in the operator but insignificant in the GC structure, and $FN$ denotes those identified as zero in the operator but significant in the GC structure. In this evaluation, off-diagonal coefficients of the identified operator with absolute values not exceeding $10^{-3}$ are treated as zero.

As shown in Fig.~\ref{fig:toy_lambda_path}, the unregularized DMD operator contains numerous weak off-diagonal coefficients. The dashed horizontal line indicates the support threshold of $10^{-3}$ adopted in this study. With the introduction of weak regularization, many of these coefficients are rapidly suppressed, while the dominant couplings are largely retained, bringing the identified operator closer to the prescribed true operator. As $\lambda$ increases further, the regularization begins to suppress these dominant couplings as well. Eventually, under sufficiently strong regularization, the off-diagonal entries are almost entirely eliminated, leaving an approximately diagonal operator. The validation error and structural consistency reach their optimal values at different regularization levels. The minimum validation error is obtained at $\lambda_{\mathrm{pred}}$, whereas the maximum structural consistency occurs at the larger $\lambda_{\mathrm{gc}}$. Two additional results at larger $\lambda$, including $\lambda_{\mathrm{strong}}$ and $\lambda_{\mathrm{over}}$ are shown, indicating strong regularization and over-regularized situations. As discussed in Section~3.1, the statistically significant GC structure is more conservative than the true coupling structure in the present benchmark. This difference arises because some genuine but weak couplings may not pass the GC significance tests, whose statistical power is influenced by finite-sample uncertainty and dependencies among variables. Consequently, the resulting GC structure is sparser than the true operator. Increasing $\lambda$ improves structural consistency by further suppressing couplings that lack significant GC support. However, this stronger regularization may also suppress dynamically relevant interactions, thereby increasing the validation error. Thus, maximizing structural consistency with the significant GC structure does not necessarily yield the best predictive generalization.

\begin{figure}[width=\columnwidth,pos=!htbp]
    \centering
    \includegraphics[width=\linewidth]{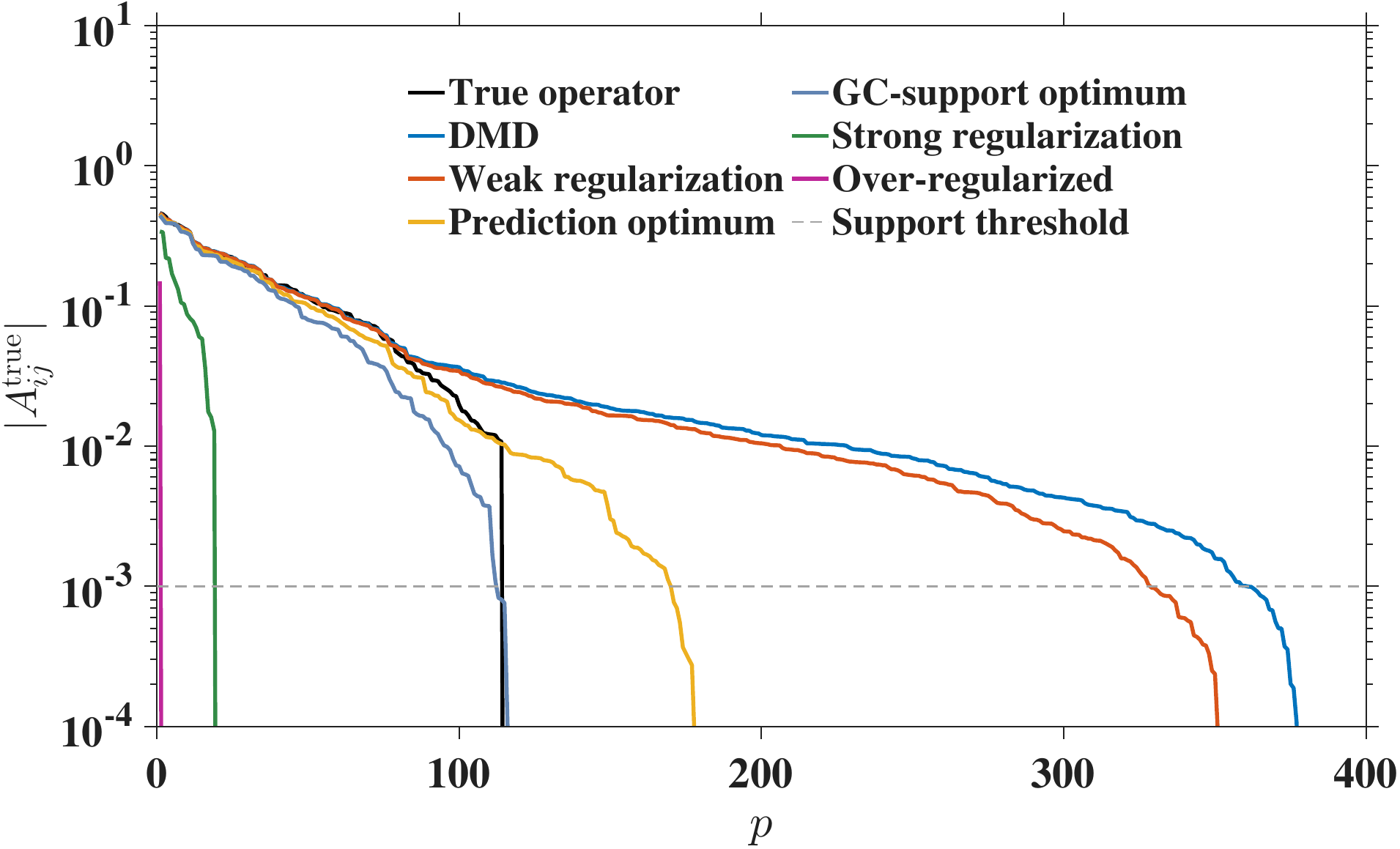}
    \caption{Off-diagonal coefficient magnitudes $|A_{ij}|$ versus sorted coefficient rank $p$ for the operators identified at representative regularization levels.} 
    \label{fig:toy_coefficient_distribution}
\end{figure}

To further examine how regularization modifies the identified operators, Fig.~\ref{fig:toy_coefficient_distribution} presents the off-diagonal coefficient magnitudes $|A_{ij}|$ against the sorted coefficient rank $p$, with the coefficients of each operator ranked independently in descending order. Standard DMD exhibits a long tail of small coefficients, whereas weak regularization primarily suppresses these small entries while preserving the dominant couplings. At $\lambda_{\mathrm{pred}}$, the coefficient distribution becomes more compact, with many weak entries removed and relatively large coefficients largely retained. Further increasing the regularization to $\lambda_{\mathrm{gc}}$ produces a shorter tail, indicating stronger suppression of weak couplings. However, the ranked magnitudes of some relatively large coefficients are also reduced compared with those at $\lambda_{\mathrm{pred}}$, revealing that the additional regularization affects more than the weakest interactions. Under stronger regularization, this suppression extends further into the dominant coefficients, eventually eliminating nearly all off-diagonal couplings. Together with the validation results in Fig.~\ref{fig:toy_lambda_path}, these observations suggest that the prediction-optimal regularization achieves a more favorable balance between removing weakly supported interactions and preserving dynamically relevant couplings.

Overall, the linear system demonstrates that GC can recover informative directional dependencies associated with the underlying operator structure and that embedding this information into DMD regularization suppresses weakly supported couplings and improves operator identification. These results establish GC as an effective data-driven structural prior for operator learning in the present linear system, providing the basis for evaluating GC-DMD in the realistic unsteady flow problems considered in the next sections.

\section{Cavity flows with increasing dynamical complexity}

The preceding linear system demonstrates the effectiveness of GC-derived directional information in guiding operator identification and improving predictive generalization. In realistic unsteady flows, however, the underlying evolution operators are unknown, and complex spatiotemporal dynamics present additional challenges for reduced-order modeling and long-horizon prediction. To further evaluate the proposed framework under more realistic conditions, lid-driven cavity flows exhibiting different levels of dynamical complexity are considered.

\subsection{Flow configuration, reduced-order representation, and evaluation metrics}

The dataset considered in this study is obtained from the two-dimensional lid-driven cavity flow simulated by Arbabi and Mezi\'c \cite{arbabi2017study}. The flow consists of an incompressible viscous fluid confined within a square cavity, $\Omega=[-1,1]^2$, with stationary walls except for the moving upper lid. A regularized velocity profile, $u_{\mathrm{lid}}(x)=(1-x^2)^2$, is prescribed along the upper boundary, where $x$ and $y$ denote the nondimensional spatial coordinates. The original flow solutions were obtained using a stream function formulation of the incompressible Navier--Stokes equations, with spatial discretization based on Chebyshev spectral collocation. The resulting post-transient snapshots of the stream function provide the high-dimensional flow-field data used for reduced-order operator identification. The snapshots are sampled at uniform intervals of $\Delta t=0.1$ in nondimensional time $t$, consistent with the temporal resolution used in the original spectral analysis \cite{arbabi2017study}.

The reduced-order models are constructed using the mean-subtracted stream function, with the discrete state vector defined as $\boldsymbol{q}=\boldsymbol{\psi}=\boldsymbol{\psi}^{\mathrm{raw}}-\overline{\boldsymbol{\psi}^{\mathrm{raw}}}$, where the overbar denotes the temporal mean calculated from the identification snapshots. Hereafter, $\psi$ denotes the mean-subtracted stream function field, while $\boldsymbol{\psi}$ represents its spatially discretized vector form. The corresponding vorticity field is used for spatial visualization and analysis. Following the vorticity-stream function formulation of Arbabi and Mezi\'c~\cite{arbabi2017study}, the instantaneous vorticity field is obtained from the original stream function as
\begin{equation}
\omega=-\nabla^2\psi^{\mathrm{raw}}=-\left(\frac{\partial^2\psi^{\mathrm{raw}}}{\partial x^2}+\frac{\partial^2\psi^{\mathrm{raw}}}{\partial y^2}\right),
\end{equation}
which is evaluated using the Chebyshev differentiation operators provided in Arbabi and Mezi\'c~\cite{arbabi2017study}. For visualization of the spatial modes, the same differentiation operators are applied directly to the corresponding stream-function modes.

\begin{figure*}[width=\textwidth,pos=!htbp]
    \centering
    \includegraphics[width=0.7\textwidth]{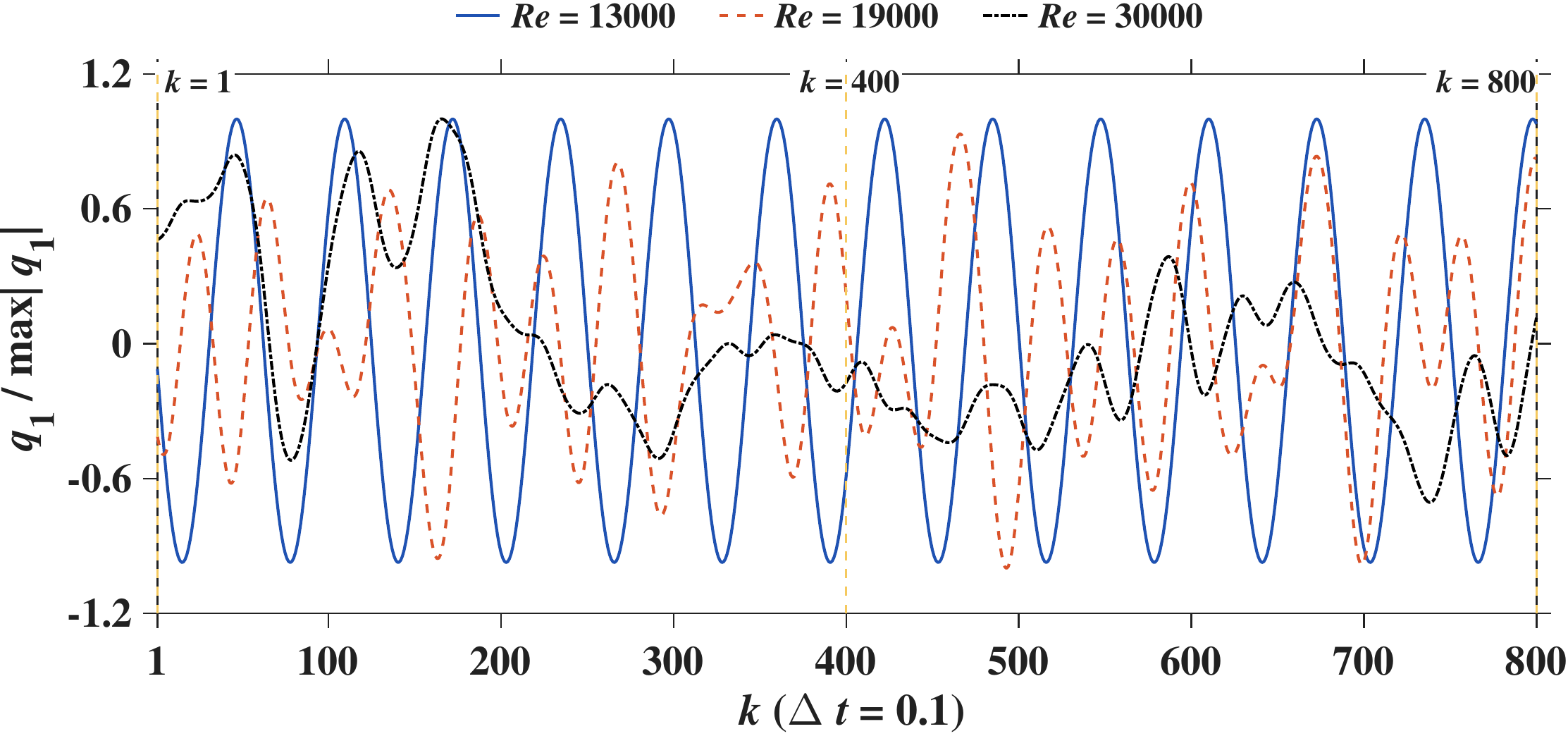}\\[-1mm]
    {\small (a)}
    \par\vspace{2mm}
    \includegraphics[width=0.7\textwidth]{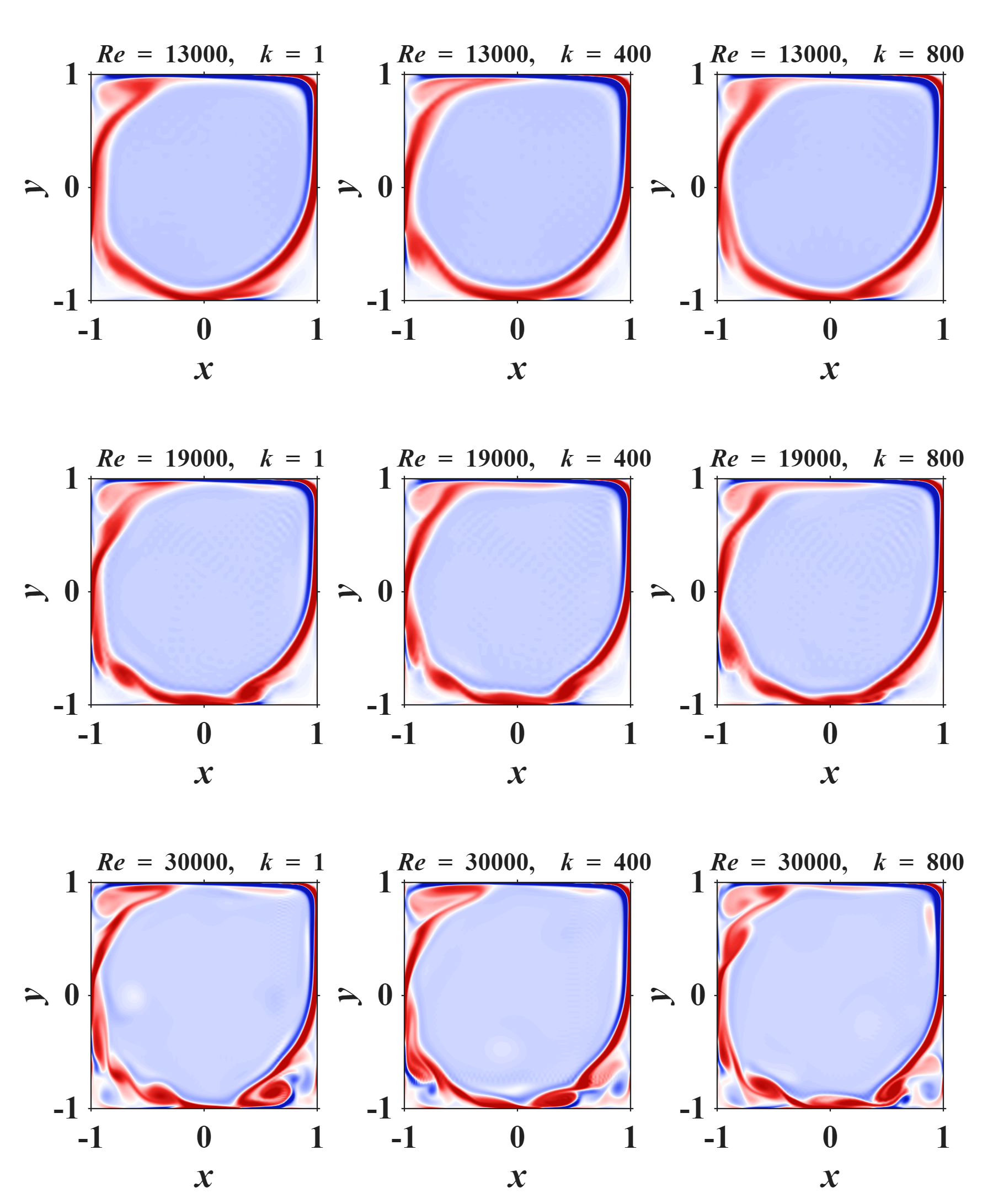}\\[-1mm]
    {\small (b)}
    \caption{Temporal evolution and representative vorticity structures of the lid-driven cavity flows at $Re=13{,}000$, $19{,}000$, and $30{,}000$. (a) Normalized first SVD coefficients, $q_1$, as functions of the snapshot index $k$, illustrating the periodic, mixed-spectrum, and aperiodic temporal behaviors, respectively. The coefficients are normalized by their respective maximum absolute values, and the uniform nondimensional sampling interval is $\Delta t=0.1$. The vertical dashed lines indicate the selected snapshot indices $k=1$, $400$, and $800$. (b) Corresponding instantaneous vorticity fields at the selected snapshots, arranged by Reynolds number (rows) and snapshot index (columns). The spatial coordinates $x$ and $y$ are nondimensional.}
    \label{fig:cavity_flow_configuration}
\end{figure*}

Three Reynolds numbers, $Re=13{,}000$, $19{,}000$, and $30{,}000$, are selected to represent different dynamical regimes. To illustrate their temporal and spatial characteristics, the first 800 snapshots of each available dataset of the post-transient flow are examined in Fig.~\ref{fig:cavity_flow_configuration}. Figure~\ref{fig:cavity_flow_configuration}(a) presents the temporal evolution of the normalized temporal coefficient associated with the leading SVD mode of the stream function, $q_1$, for the three cases, where $k$ is the snapshot index. The $Re=13{,}000$ flow exhibits a predominantly periodic response, whereas the $Re=19{,}000$ case displays more complex temporal dynamics with mixed spectral characteristics. At $Re=30{,}000$, the temporal evolution becomes increasingly irregular, corresponding to an aperiodic regime. Figure~\ref{fig:cavity_flow_configuration}(b) presents the corresponding instantaneous vorticity fields at three selected snapshot indices, $k=1$, $400$, and $800$, illustrating the spatial flow structures associated with these dynamical regimes. Together, these cases provide a progressive change of dynamical complexity for evaluating the identified reduced-order operators.

For each case, the temporal mean stream function field is subtracted from the original snapshots, and the resulting fluctuation fields are projected onto a reduced SVD basis before operator identification. The truncation rank is determined by considering both the cumulative SVD energy and the corresponding flow-field reconstruction error, ensuring adequate retention of the dominant spatial structures. A cumulative energy threshold of $99.999\%$ is adopted, resulting in reduced dimensions of $r=7$, $115$, and $162$ for $Re=13{,}000$, $19{,}000$, and $30{,}000$, respectively. The increasing number of retained SVD modes indicates that more spatial degrees of freedom are required to represent the higher-Reynolds-number flows at the prescribed energy threshold.

The SVD coefficient sequences are divided chronologically into
identification and prediction segments, with the identification data
further separated into training and validation sets. Consistent with
the procedure described in Section~2, the SVD basis, GC structure,
and evolution operator are estimated using the training data, while
the regularization parameter is selected according to the validation
prediction error. The prediction segment is reserved for evaluating
the generalization performance of the identified operators. Standard
DMD and GC-DMD are compared using the same reduced coordinates and data partitions for each case.

To evaluate the reconstruction and prediction performance in the following sections, three complementary error measures are introduced. The relative full-field error at snapshot $k$ is defined as
\begin{equation}
E_{\psi}(k)
=
\frac{
\left\|\widehat{\boldsymbol{\psi}}_k-\boldsymbol{\psi}_k\right\|_2
}{
\left\|\boldsymbol{\psi}_k\right\|_2
},
\label{eq:cavity_snapshot_error}
\end{equation}
where $\boldsymbol{\psi}_k$ and $\widehat{\boldsymbol{\psi}}_k$ denote the reference and reconstructed or predicted mean-subtracted stream function snapshots at the $k$th time instant, respectively. For a given reconstruction or prediction interval, the relative full-field error is further defined as
\begin{equation}
E_{\psi}^{\mathrm{int}}
=
\frac{
\left\|\widehat{\boldsymbol{\Psi}}-\boldsymbol{\Psi}\right\|_F
}{
\left\|\boldsymbol{\Psi}\right\|_F
}
\times 100\%,
\label{eq:cavity_interval_error}
\end{equation}
where $\boldsymbol{\Psi}$ and $\widehat{\boldsymbol{\Psi}}$ denote the reference and reconstructed or predicted mean-subtracted stream function snapshot matrices over the corresponding interval, respectively. For spatial error analysis, the signed vorticity error is defined as
\begin{equation}
\Delta\omega(x,y,k)
=
\widehat{\omega}(x,y,k)-\omega(x,y,k),
\label{eq:cavity_vorticity_error}
\end{equation}
where $\widehat{\omega}$ and $\omega$ denote the vorticity fields obtained from the reconstructed or predicted and reference full-flow stream function fields, respectively. The snapshot-wise relative error $E_{\psi}(k)$ quantifies the reconstruction or prediction accuracy at individual time instants, while the interval error $E_{\psi}^{\mathrm{int}}$ evaluates the overall accuracy across a sequence of snapshots. In contrast, the signed vorticity error $\Delta\omega$ reveals localized discrepancies in the reconstructed or predicted vortical structures.

\subsection{Periodic cavity flow at $Re=13{,}000$}

The cavity flow at $Re=13{,}000$ exhibits a predominantly periodic response with clearly identifiable dominant dynamical structures. A total of 3,000 consecutive snapshots are considered, covering the nondimensional time interval from approximately $t=3000$ to $t=3300$. The first 800 snapshots are used for operator identification, while the remaining 2,200 are reserved for recursive prediction. The identification segment is further divided chronologically into 640 training snapshots and 160 validation snapshots, corresponding to an 80\%/20\% split.

\begin{figure}[width=\columnwidth,pos=!htbp]
    \centering
    \includegraphics[width=\linewidth]{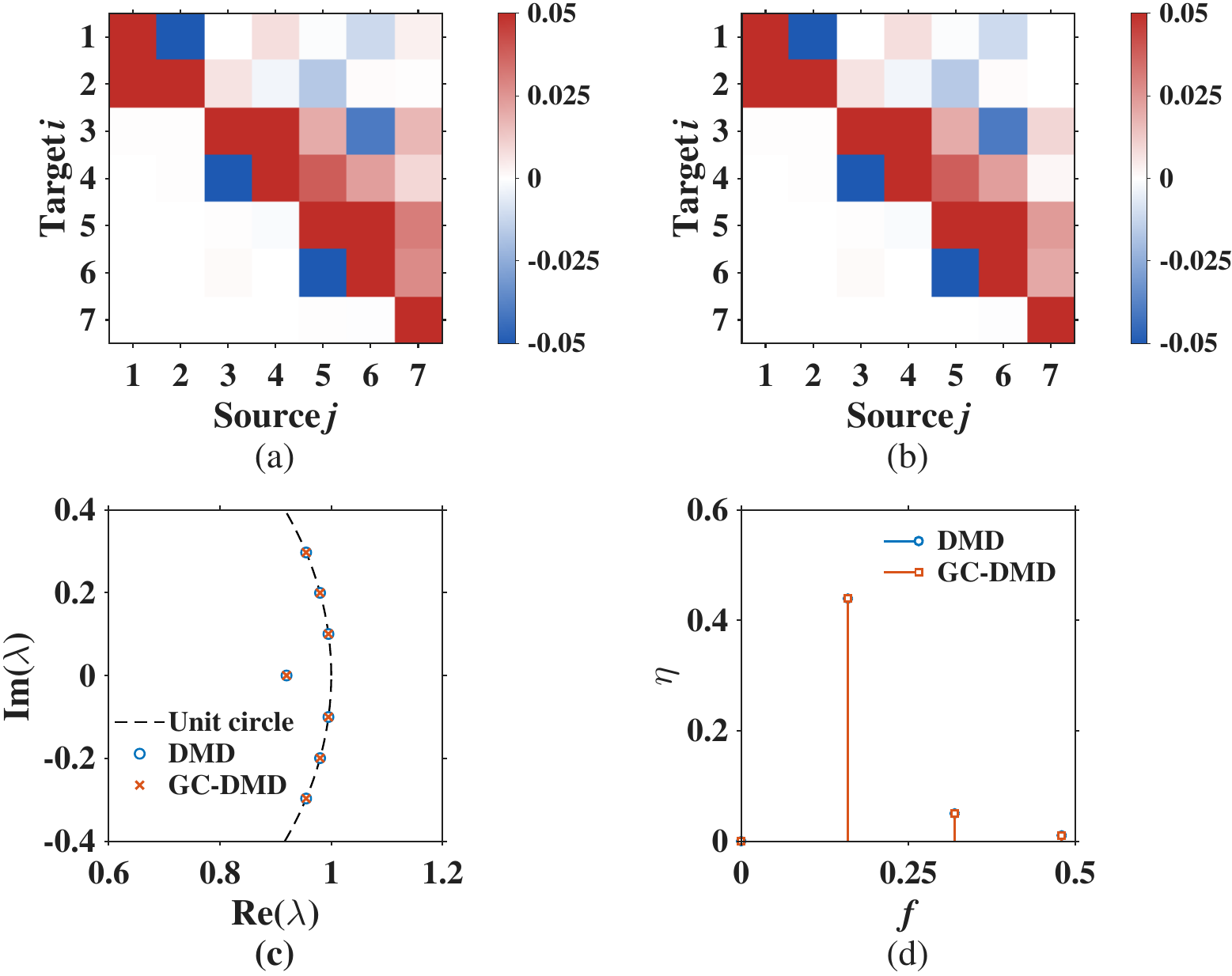}

    \caption{Operator and spectral characteristics of DMD and GC-DMD for the periodic cavity flow at $Re=13{,}000$. (a) DMD evolution operator. (b) GC-DMD evolution operator. (c) Ritz eigenvalue distributions. (d) Normalized modal contributions evaluated using the Kou criterion in Eq.~(6), plotted against the corresponding modal frequencies.}
    \label{fig:cavity13_operator}
\end{figure}

Figure~\ref{fig:cavity13_operator} compares the identified operators and their spectral characteristics. For this periodic flow with well-defined dominant dynamics, standard DMD already provides an effective representation of the underlying evolution. As shown in Figs.~\ref{fig:cavity13_operator}(a) and \ref{fig:cavity13_operator}(b), incorporating GC-derived directional information introduces moderate adjustments to selected operator coefficients while largely retaining the coupling structure identified by DMD. Figure~\ref{fig:cavity13_operator}(c) presents the Ritz eigenvalue distributions, while Fig.~\ref{fig:cavity13_operator}(d) shows the
normalized modal contributions as functions of their corresponding frequencies $f$. The contribution of the $j$th mode is defined as
$\eta_j=I_j/\sum_{\ell=1}^{r}I_\ell$, where $I_j$ is the cumulative modal importance evaluated using the criterion in Eq.~(6). Only modes with nonnegative frequencies are displayed, while the normalization includes all retained modes. The close agreement in both eigenvalue distributions and dominant modal contributions indicates that GC-informed regularization preserves the principal spectral characteristics of the periodic flow.

\begin{figure*}[width=\textwidth,pos=!htbp]
    \centering

    % (a) Temporal prediction error
    \includegraphics[width=0.95\textwidth]{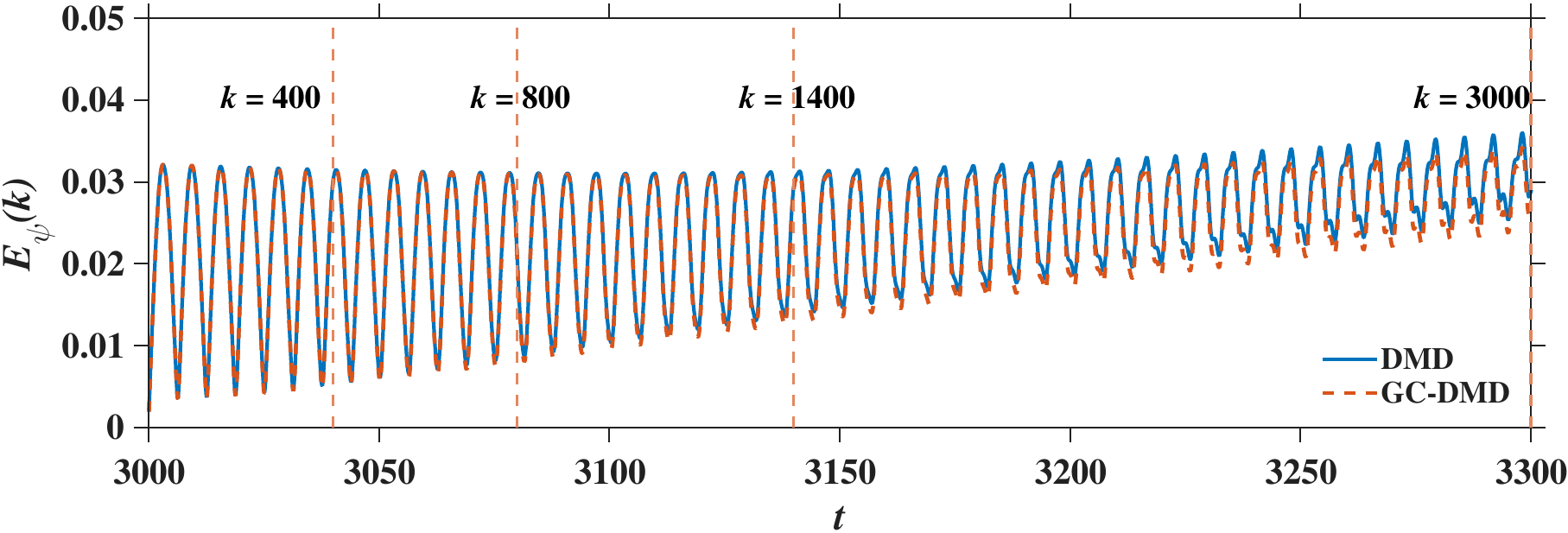}

    {\small (a)}

    \par\vspace{2mm}

    % (b) Spatial vorticity error distributions
    \includegraphics[width=0.95\textwidth]{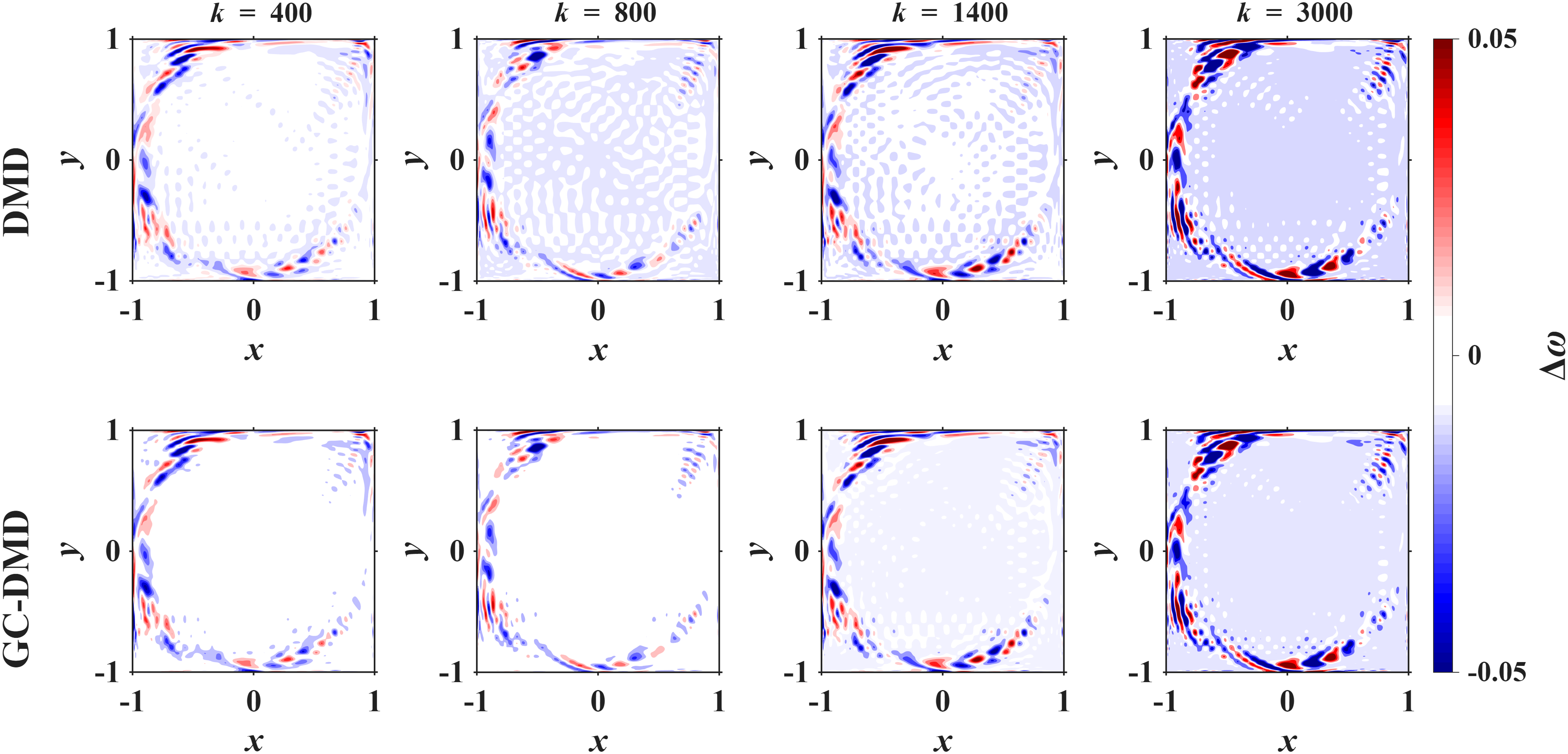}

    {\small (b)}

    \caption{Reconstruction and recursive prediction performance of DMD and GC-DMD for the periodic cavity flow at $Re=13{,}000$. (a) Evolution of the relative full-field error at snapshot $k$ $E_{\psi}(k)$ over nondimensional time. The reduced-order trajectories are recursively propagated from the first snapshot, and the identification and independent prediction evaluation intervals are separated at $k=800$ ($t\approx3080$). The vertical dashed lines indicate the selected snapshot indices. (b) The corresponding signed vorticity error distributions $\Delta\omega$, at $k=400$, $800$, $1400$, and $3000$, corresponding approximately to nondimensional times $t=3040$, $3080$, $3140$, and $3300$, respectively.}
    \label{fig:cavity13_prediction}
\end{figure*}

The predictive performance of the identified operators is further examined in Fig.~\ref{fig:cavity13_prediction}. The reduced-order trajectories are initialized from the first snapshot and propagated recursively over the entire observation interval, with the identification and prediction evaluation segments separated at $k=800$. Figure~\ref{fig:cavity13_prediction}(a) presents the relative full-field error at snapshot $k$ $E_{\psi}(k)$ defined in Section~4.1. Both methods exhibit comparable errors over the identification interval. During the independent prediction interval, the errors of both methods generally increase with the prediction horizon, reflecting the accumulation of deviations during recursive evolution. However, GC-DMD exhibits slower overall error growth than standard DMD, resulting in a more pronounced performance difference over longer prediction horizons. Despite the oscillatory variations in the error histories, GC-DMD generally maintains lower prediction errors. The spatial prediction results are further compared in Fig.~\ref{fig:cavity13_prediction}(b), which presents the signed errors $\Delta\omega$ at selected snapshots. Both methods reproduce the principal vortical structures of the periodic cavity flow, while GC-DMD generally yields smaller spatial errors. In particular, GC-DMD reduces the weak, spatially distributed vorticity errors within the cavity interior, leading to a more localized error distribution. Relatively large discrepancies remain near the cavity boundaries, particularly at later prediction times, but the GC-informed operator provides an overall improvement in the spatial accuracy of the predicted vorticity field.

Overall, the periodic cavity flow demonstrates that GC-informed regularization can improve the long-horizon predictive performance of an already effective DMD model without substantially altering its dominant dynamical characteristics. The moderate adjustments to the evolution operator preserve the principal spectral features while reducing error accumulation during recursive prediction and improving the spatial accuracy of the predicted vorticity fields.

\subsection{Multifrequency cavity flow at $Re=19{,}000$}

Unlike the predominantly periodic flow at $Re=13{,}000$, the cavity flows at $Re=19{,}000$ and $30{,}000$ exhibit more complex temporal dynamics involving multiple significant frequency components. The increased spatiotemporal complexity requires a substantially higher-dimensional POD representation and presents greater challenges for operator identification and long-horizon prediction. The analysis focuses on the $Re=19{,}000$ case. The more irregular flow at $Re=30{,}000$ exhibits similar overall trends in operator regularization and predictive performance, so its results are provided in Appendix~B as a complementary assessment.

\subsubsection{GC-informed operator identification}

A total of 1,200 consecutive snapshots are considered, covering the nondimensional time interval from approximately $t=5000$ to $t=5120$. The first 800 snapshots constitute the identification segment, while the remaining 400 snapshots form the independent prediction evaluation segment. The identification data are further divided chronologically into 640 training snapshots and 160 validation snapshots, corresponding to an 80\%/20\% split.

\begin{figure}[width=\columnwidth,pos=!htbp]
    \centering
    \includegraphics[width=\linewidth]{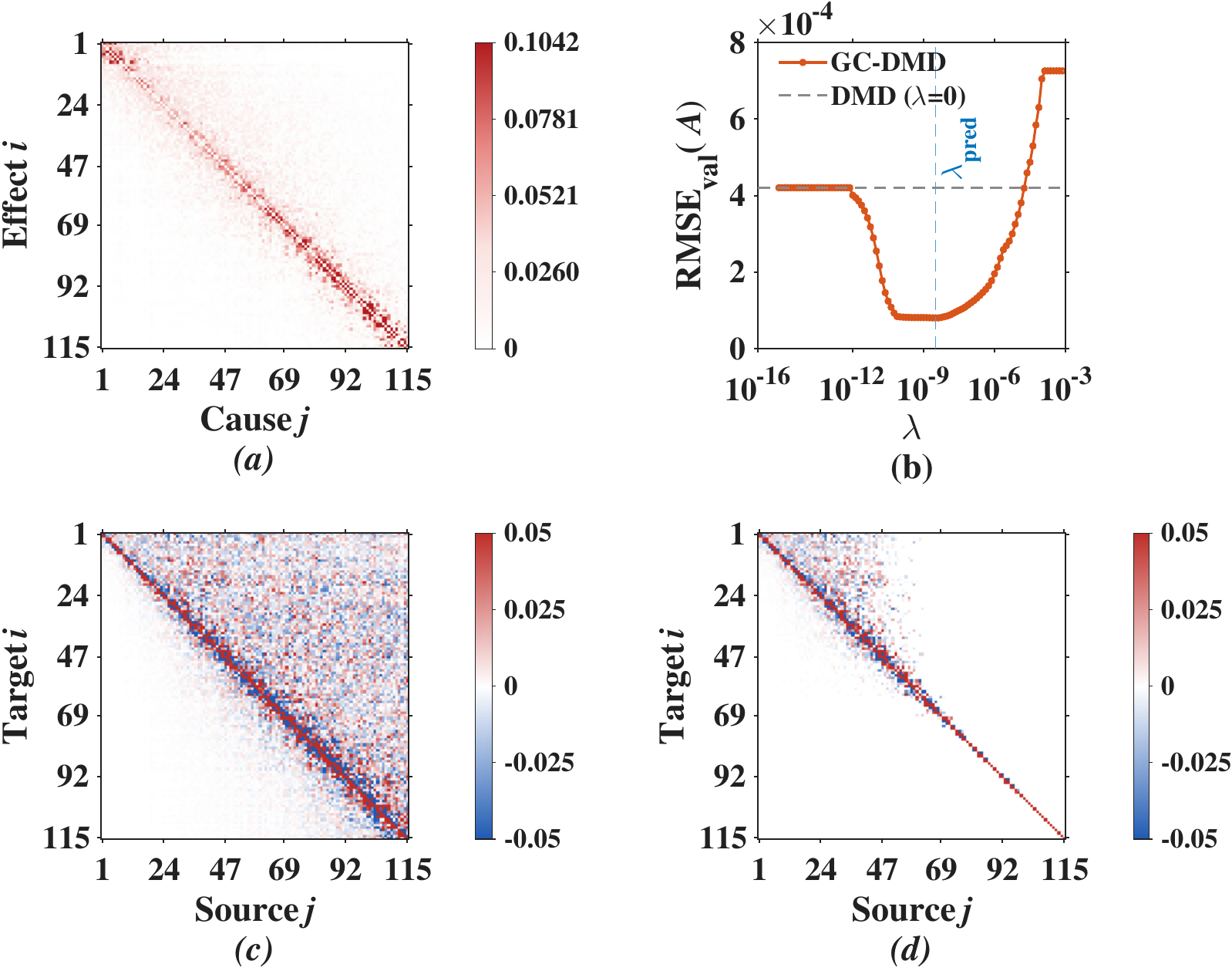}
    \caption{GC-informed operator identification for the multifrequency cavity flow at $Re=19{,}000$. (a) Normalized conditional GC strength. (b) One-step validation RMSE as a function of the regularization parameter $\lambda$. (c) DMD evolution operator. (d) GC-DMD evolution operator.}
    \label{fig:cavity19_operator}
\end{figure}

Figure~\ref{fig:cavity19_operator} presents the GC-derived directional dependencies, the influence of GC-informed regularization on validation performance, and the evolution operators identified by DMD and GC-DMD. As shown in Fig.~\ref{fig:cavity19_operator}(a), stronger GC interactions are predominantly concentrated near the main diagonal, indicating that the most pronounced directional predictive dependencies tend to occur between SVD coordinates with relatively close mode indices. Figure~\ref{fig:cavity19_operator}(b) illustrates the effect of the regularization parameter $\lambda$ on the one-step validation RMSE. As $\lambda$ increases, the validation error initially decreases, suggesting that an appropriate degree of GC-informed regularization can improve the predictive generalization of the identified operator. Excessive regularization, however, increases the validation error as dynamically relevant couplings are progressively suppressed. The validation-selected parameter therefore provides a balance between operator regularization and predictive accuracy. The corresponding evolution operators are compared in Figs.~\ref{fig:cavity19_operator}(c) and \ref{fig:cavity19_operator}(d). Standard DMD exhibits numerous weak off-diagonal coefficients distributed throughout the operator, whereas GC-DMD suppresses a substantial proportion of these interactions while retaining the more prominent coupling patterns. The resulting GC-DMD operator exhibits a markedly more concentrated, near-diagonal structure. Quantitatively, the off-diagonal sparsity increases from approximately 22.8\% for DMD to 85.2\% for GC-DMD, where sparsity is defined as the fraction of off-diagonal coefficients satisfying $|A_{ij}|<10^{-3}$. These results indicate that GC-derived directional information can guide the identification of a more compact evolution operator while maintaining favorable validation performance.

\subsubsection{Spectral and modal characteristics}

\begin{figure}[width=\columnwidth,pos=!htbp]
    \centering

    % Row 1: spectral characteristics
    \begin{minipage}[t]{0.48\linewidth}
        \centering
        \includegraphics[width=\linewidth]{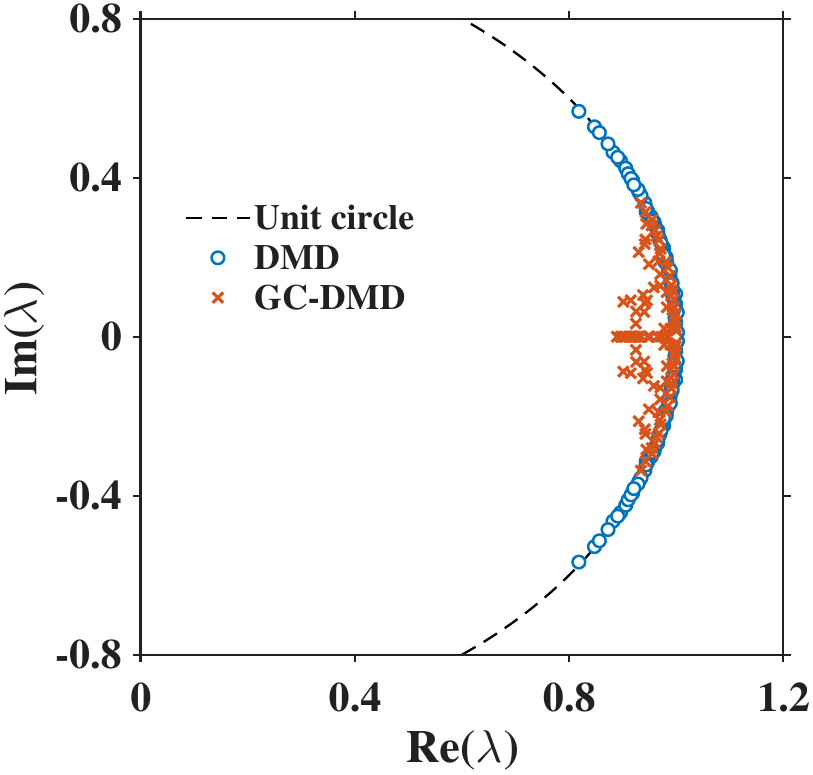}

        {\small (a)}
    \end{minipage}
    \hfill
    \begin{minipage}[t]{0.48\linewidth}
        \centering
        \includegraphics[width=\linewidth]{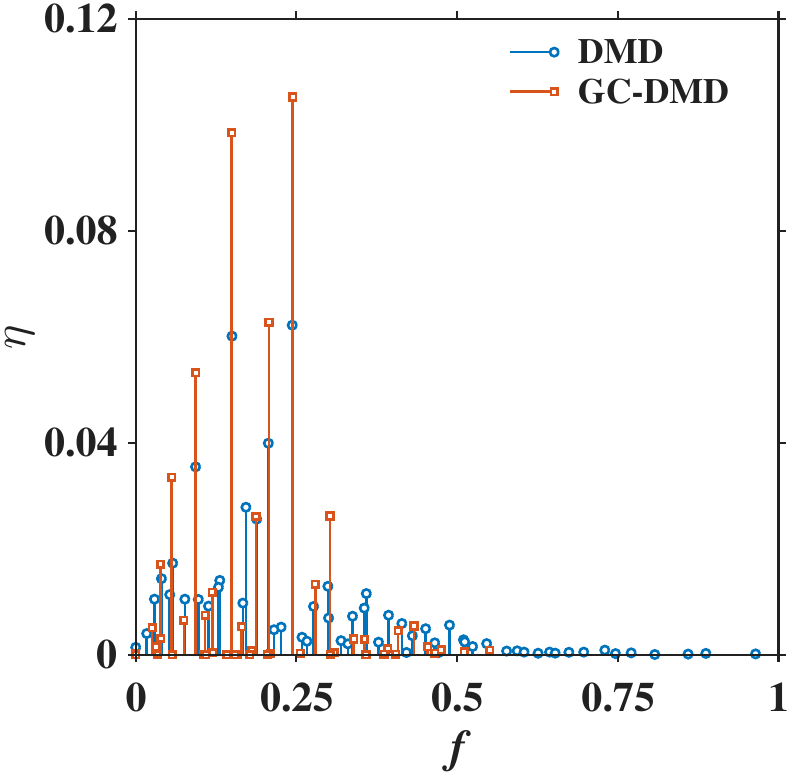}

        {\small (b)}
    \end{minipage}

    \par\vspace{1mm}

    % Row 2: representative spatial modes
    \includegraphics[width=\linewidth]{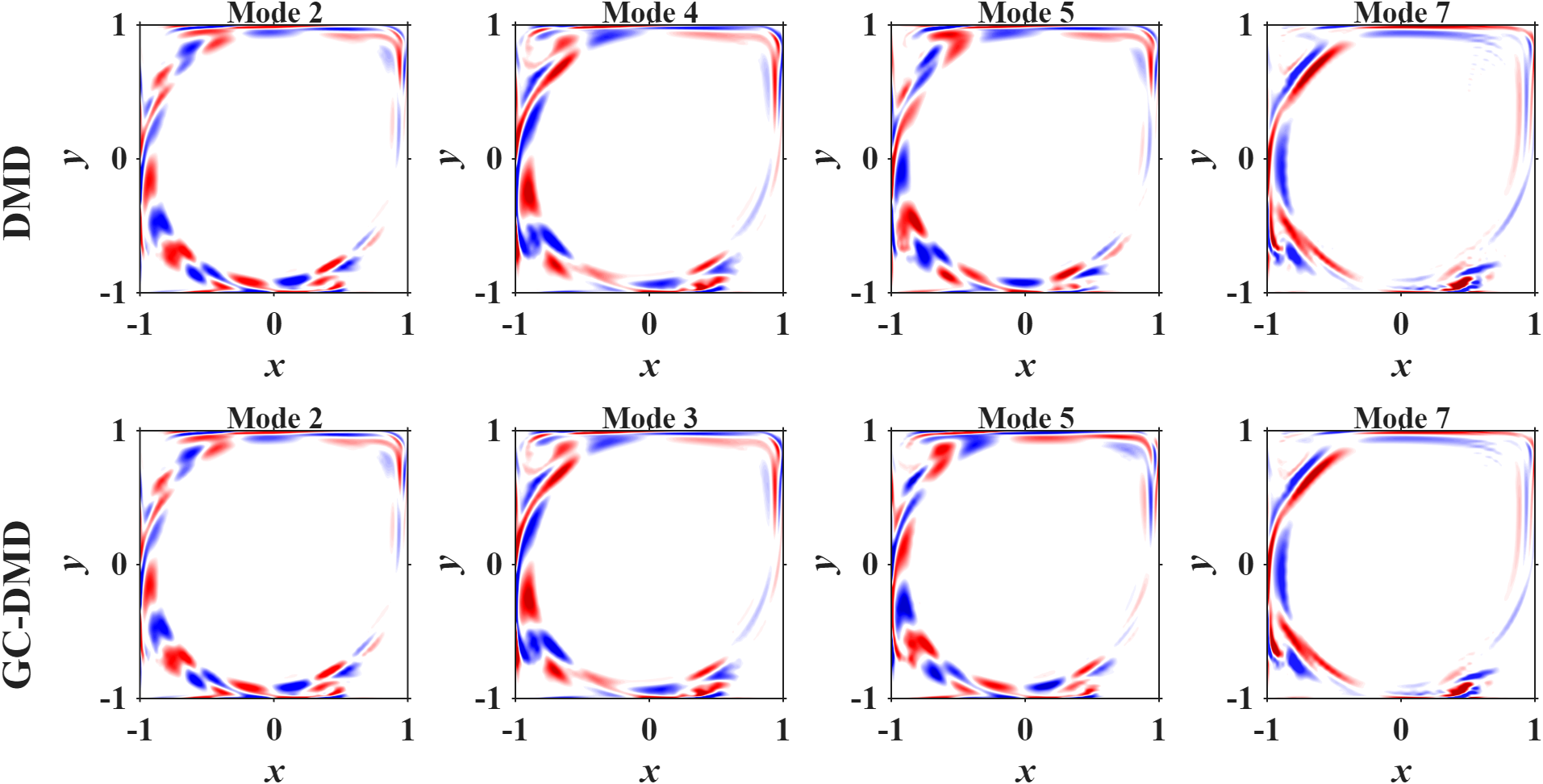}

    {\small (c)}

    \caption{Spectral and modal characteristics of DMD and GC-DMD for the multifrequency cavity flow at $Re=19{,}000$. (a) Ritz eigenvalue distributions. (b) Normalized modal contributions evaluated using the criterion in Eq.~(6), plotted against the corresponding modal frequencies. (c) Representative contribution-ranked vorticity modes identified by DMD and GC-DMD.}
    \label{fig:cavity19_modal}
\end{figure}

The spectral and modal characteristics of the identified operators are further examined to assess how GC-informed regularization affects the representation of the underlying flow dynamics. Figure~\ref{fig:cavity19_modal} compares the Ritz eigenvalue distributions, normalized modal contributions, and representative spatial modes obtained by DMD and GC-DMD.

As shown in Fig.~\ref{fig:cavity19_modal}(a), both methods retain similar dominant eigenvalue distributions near the unit circle, while the GC-DMD eigenvalues exhibit a more concentrated distribution, with some modes shifted further inside the unit circle, suggesting stronger decay of these components during recursive evolution. Figure~\ref{fig:cavity19_modal}(b) shows the normalized modal contributions as functions of their corresponding frequencies, evaluated using the criterion in Eq.~(6). As in Section~4.2, only modes with nonnegative frequencies are displayed, while the normalization includes all retained modes. Both methods identify prominent contributions within similar frequency ranges, whereas GC-DMD exhibits a greater concentration around several dominant frequencies, accompanied by reduced contributions from weaker components, particularly at higher frequencies. The representative spatial modes in Fig.~\ref{fig:cavity19_modal}(c), with only one mode displayed from each complex-conjugate pair, reveal broadly similar coherent vorticity structures concentrated primarily along the cavity boundaries. Overall, GC-informed regularization produces a more concentrated spectral and modal representation while largely preserving the dominant dynamical characteristics identified by standard DMD.

\subsubsection{Flow reconstruction and prediction}

\begin{figure*}[width=\textwidth,pos=!htbp]
    \centering

    % Row 1: error evolution and t = 800
    \begin{minipage}[t]{0.8\textwidth}
        \centering
        \includegraphics[width=\linewidth]{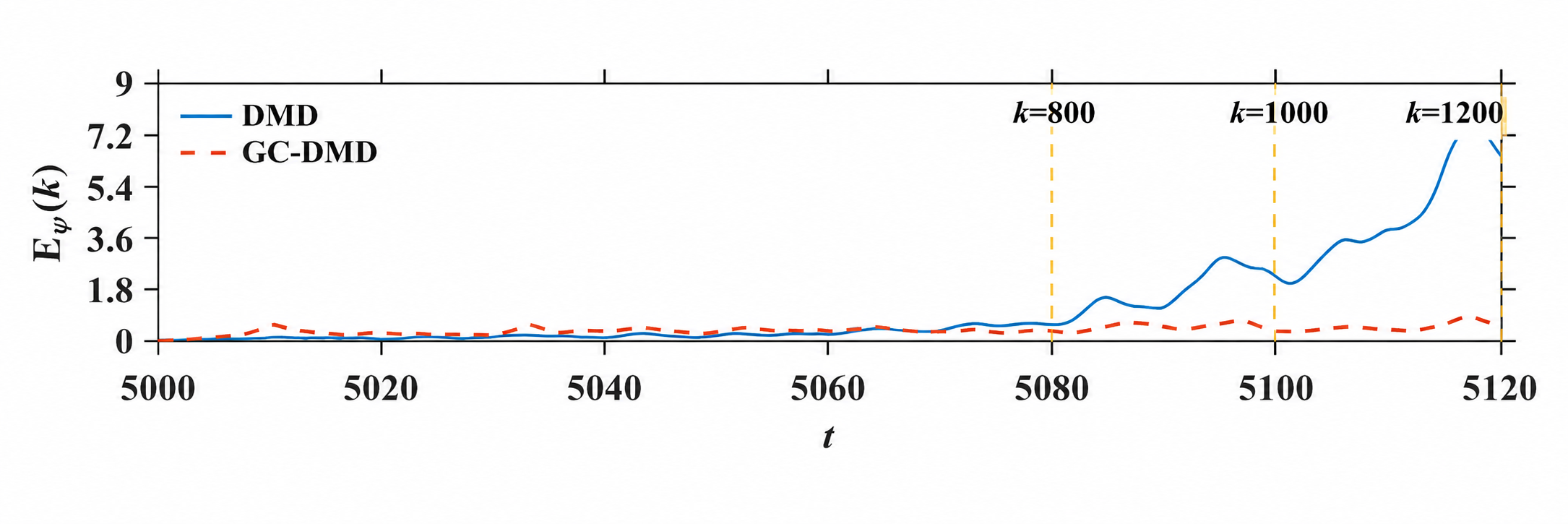}

        {\small (a)}
    \end{minipage}
    \hfill
    \begin{minipage}[t]{0.8\textwidth}
        \centering
        \includegraphics[width=\linewidth]{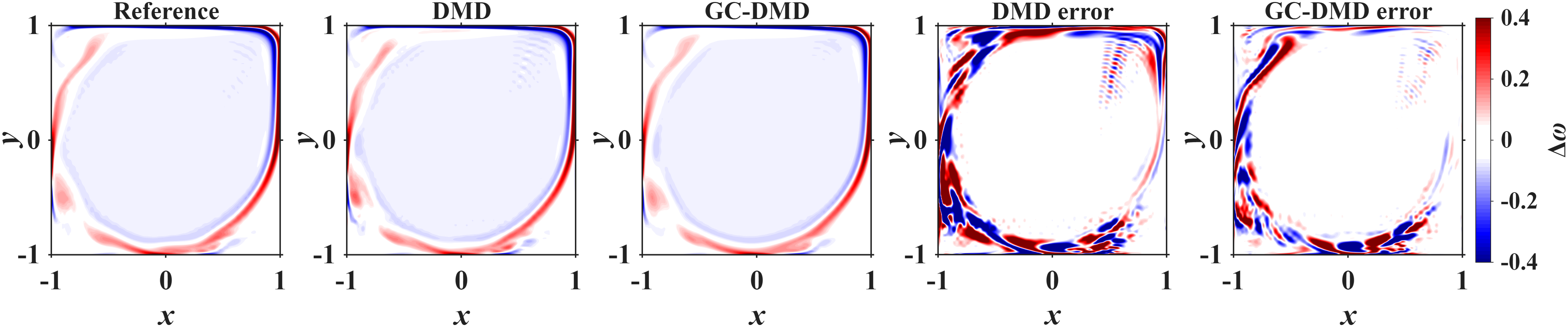}

        {\small (b)}
    \end{minipage}

    \par\vspace{2mm}

    % Row 2: t = 1000 and t = 1200
    \begin{minipage}[t]{0.8\textwidth}
        \centering
        \includegraphics[width=\linewidth]{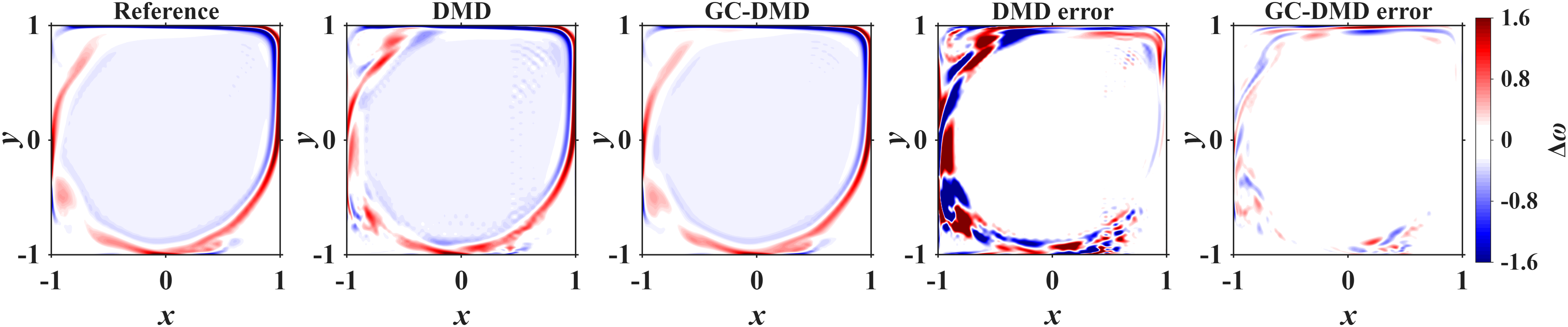}

        {\small (c)}
    \end{minipage}
    \hfill
    \begin{minipage}[t]{0.8\textwidth}
        \centering
        \includegraphics[width=\linewidth]{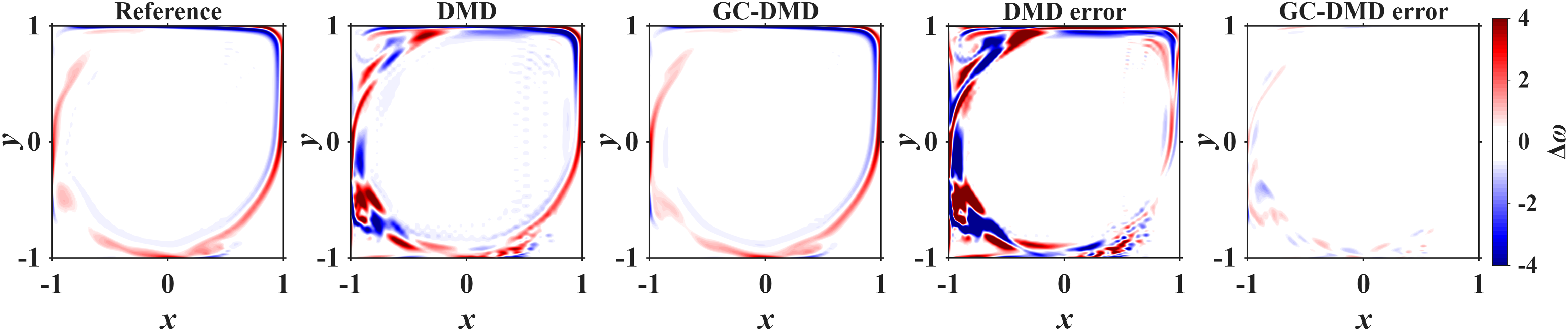}

        {\small (d)}
    \end{minipage}

    \caption{Reconstruction and recursive prediction performance of DMD and GC-DMD for the multifrequency cavity flow at $Re=19{,}000$. (a) Evolution of the relative full-field error at snapshot $k$ $E_{\psi}(k)$ over nondimensional time. The reduced-order trajectories are recursively propagated from the first snapshot, with the identification and independent prediction evaluation intervals separated at $k=800$ ($t=5080$). The vertical dashed lines indicate the selected snapshot indices. (b)-(d) Reference and predicted vorticity fields, together with the corresponding signed vorticity error distributions $\Delta\omega$, at $k=800$, $1000$, and $1200$, corresponding approximately to nondimensional times $t=5080$, $5100$, and $5120$, respectively.}
    \label{fig:cavity19_prediction}
\end{figure*}

The predictive performance of the identified operators is further evaluated through recursive flow-field reconstruction and prediction. Figure~\ref{fig:cavity19_prediction} compares the evolution of the relative full-field error at snapshot $k$ $E_{\psi}(k)$ and the corresponding vorticity fields and spatial error distributions at representative snapshots.

As shown in Fig.~\ref{fig:cavity19_prediction}(a),  the reduced-order trajectories are initialized from the first snapshot and propagated recursively throughout the observation interval, with the identification and independent prediction evaluation segments separated at $k=800$ ($t=5080$). Both methods exhibit relatively low errors over the identification interval, while during the prediction interval, the errors of the two methods evolve markedly differently. The relative full-field error of standard DMD increases rapidly and exhibits substantial fluctuations, whereas GC-DMD maintains considerably lower errors throughout the prediction interval. This difference becomes increasingly pronounced over longer prediction horizons, demonstrating that GC-informed regularization substantially reduces error accumulation during recursive evolution.

The spatial comparisons in Figs.~\ref{fig:cavity19_prediction}(b)-(d) further illustrate this improvement. At the end of the identification interval ($k=800$, $t=5080$), both methods reproduce the principal vortical structures, although differences remain in their signed vorticity error distributions. As the recursive evolution proceeds to $k=1000$ ($t=5100$) and $k=1200$ ($t=5120$), the DMD results exhibit increasingly pronounced deviations from the reference vorticity field, particularly near the cavity boundaries and regions of concentrated vorticity. In contrast, GC-DMD maintains a closer agreement with the reference field and substantially reduces the spatial prediction errors, even at the final snapshot. These results demonstrate that GC-informed regularization improves the long-horizon predictive accuracy of the reduced-order model for multifrequency cavity flows.

\subsubsection{Spectral stability and prediction robustness}

To further understand the improved long-horizon predictive performance of GC-DMD, the spectral characteristics of the identified dynamic modes are examined. At the fixed SVD truncation rank of $r=115$, the DMD and GC-DMD modes are independently ranked according to the criterion in Eq.~(6), and the evolution of the flow-field error is evaluated as progressively more dominant dynamic modes are retained.

\begin{figure}[width=\columnwidth,pos=!htbp]
    \centering

    \includegraphics[width=0.95\columnwidth]{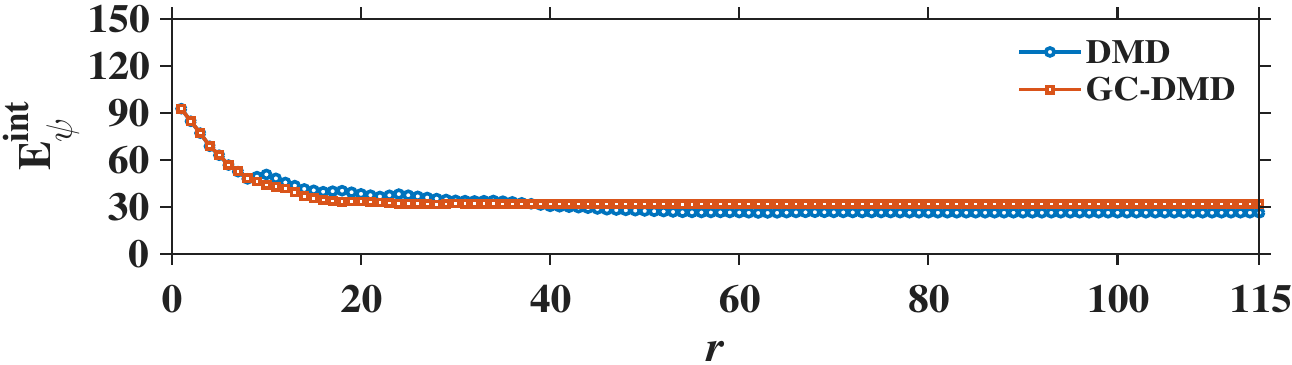}
    \par\vspace{-1mm}
    {\small (a)}

    \medskip

    \includegraphics[width=0.95\columnwidth]{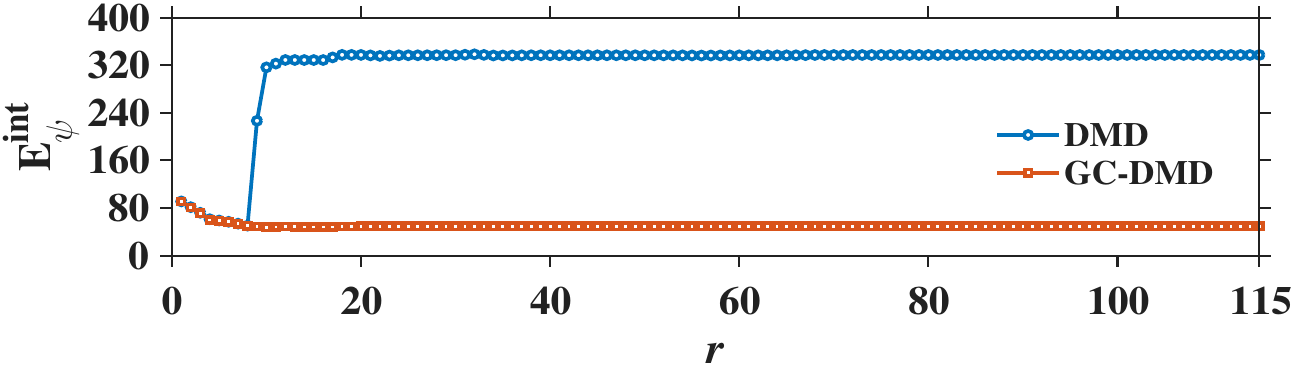}
    \par\vspace{-1mm}
    {\small (b)}

    \medskip

    \includegraphics[width=0.6\columnwidth]{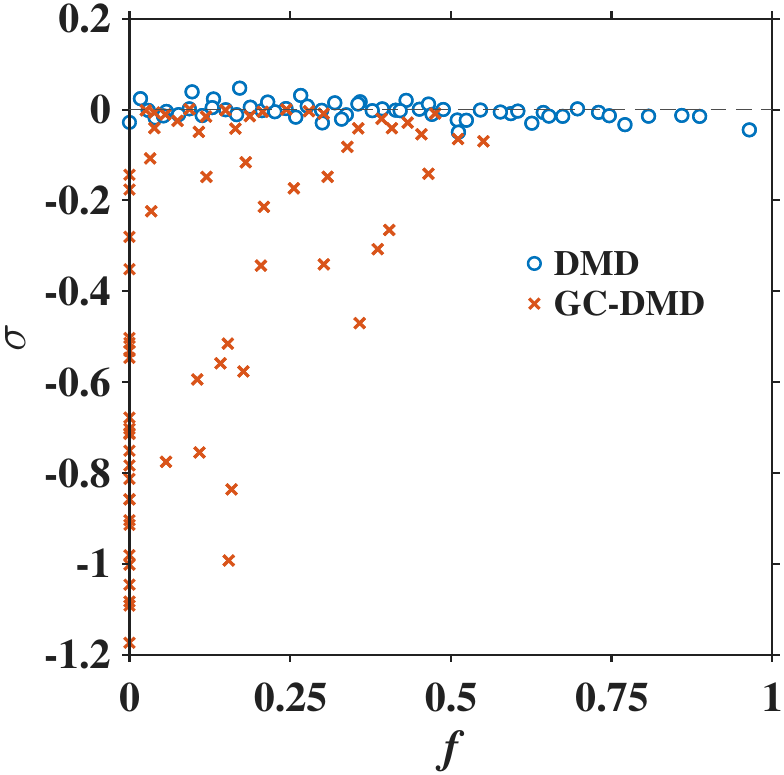}
    \par\vspace{-1mm}
    {\small (c)}

    \caption{Modal contributions and spectral characteristics of DMD and GC-DMD for the multifrequency cavity flow at $Re=19{,}000$ and $r=115$. (a) Relative full-field error over the identification interval versus the number of retained dominant dynamic modes. (b) Relative full-field error over the prediction interval versus the number of retained dominant dynamic modes. (c) Frequency-growth-rate distributions of the identified dynamic modes.}
    \label{fig:cavity19_modal_stability}
\end{figure}

Figures~\ref{fig:cavity19_modal_stability}(a) and (b) show the interval relative full-field error $E_{\psi}^{\mathrm{int}}$, over the identification and prediction intervals, respectively, as functions of the number of retained dominant dynamic modes. Over the identification interval, DMD achieves a lower reconstruction error than GC-DMD when sufficient modes are retained, indicating better reproduction of the sampled flow dynamics. However, over the prediction interval, the DMD error increases abruptly upon inclusion of the ninth dominant mode and remains substantially higher as additional modes are retained, whereas GC-DMD maintains comparatively low prediction errors without similar amplification. The frequency-growth-rate distributions in Fig.~\ref{fig:cavity19_modal_stability}(c) provide a spectral interpretation of this behavior. Although both methods exhibit modes with positive growth rates $\sigma$, GC-DMD produces fewer unstable modes, with generally smaller positive growth rates than those of DMD. This indicates that GC-informed regularization reduces both the prevalence and the growth intensity of unstable dynamic components, consistent with the eigenvalue distributions in Fig.~\ref{fig:cavity19_modal}(a). Together, these observations suggest that the improved long-horizon prediction of GC-DMD is closely associated with enhanced spectral stability, despite its somewhat higher reconstruction error over the identification interval.

\begin{figure}[width=\columnwidth,pos=!htbp]
    \centering

    \includegraphics[width=0.95\columnwidth]{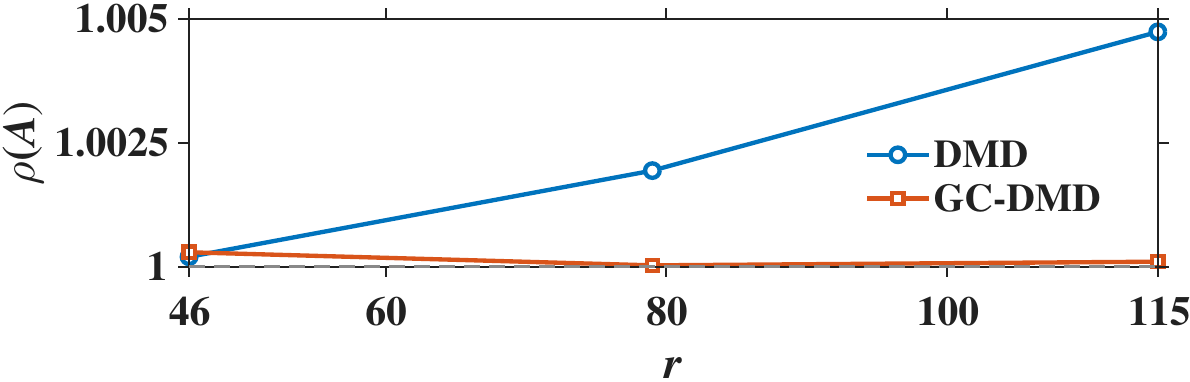}
    \par\vspace{-1mm}
    {\small (a)}

    \medskip

    \includegraphics[width=0.95\columnwidth]{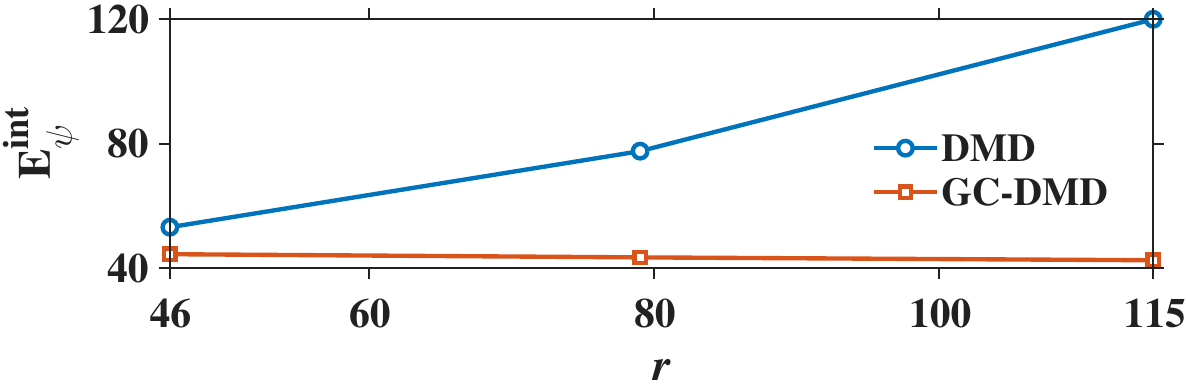}
    \par\vspace{-1mm}
    {\small (b)}

    \caption{Influence of SVD truncation rank on the spectral stability and long-horizon prediction of DMD and GC-DMD for the multifrequency cavity flow at $Re=19{,}000$. (a) Spectral radius versus retained SVD rank. (b) Mean relative full-field prediction error versus retained SVD rank. The three ranks correspond to cumulative SVD energy thresholds of 99.9\%, 99.99\%, and 99.999\%.}
    \label{fig:cavity19_rank_stability}
\end{figure}

To further assess the robustness of the identified operators, the influence of SVD truncation rank is examined in Fig.~\ref{fig:cavity19_rank_stability}, considering cumulative SVD energy thresholds of 99.9\%, 99.99\%, and 99.999\%. As the rank increases, GC-DMD maintains a spectral radius $\rho(A)$ close to unity while progressively reducing the mean long-horizon prediction error. In contrast, the spectral radius of DMD increases beyond unity, accompanied by substantial error amplification. This agrees with the observation of Xie and Tang~\cite{xie2024regularized} that retaining additional modes does not necessarily improve long-term DMD prediction when unstable components are introduced. The results indicate that GC-informed regularization improves the robustness of the identified operator to SVD truncation, allowing additional spatial information to be retained without the pronounced prediction deterioration observed in standard DMD.

Taken together, these results suggest that the improved long-horizon prediction of GC-DMD is associated with both operator sparsification and enhanced spectral stability. By suppressing unstable dynamic components, GC-informed regularization mitigates error amplification during recursive prediction and improves robustness to SVD truncation rank, providing a more reliable reduced-order representation of multifrequency flow dynamics.

\section{GC-informed operator sparsification in broadband transonic jet flow}

The preceding cavity-flow analysis demonstrates that GC-informed regularization improves the predictive performance of DMD in multifrequency flows while producing a substantially sparser evolution operator. To further evaluate the proposed framework under more complex flow dynamics, a broadband turbulent jet at Mach number $Ma=0.9$ is considered. Particular attention is given to the sparsification of the high-dimensional evolution operator, its relationship with predictive generalization, and the structural characteristics of the retained couplings.

\subsection{Flow configuration and reduced-order representation}

The present study employs the publicly available Mach 0.9 turbulent jet dataset from the SPOD benchmark of Schmidt et al.~\cite{schmidt2018spectral}, originally obtained from the large-eddy simulations of Br\`es et al.~\cite{bres2018importance}. The jet operates at a diameter-based Reynolds number of approximately $Re=1.0\times10^6$. The dataset comprises 5,000 snapshots of the azimuthally symmetric ($m=0$) pressure component of a round turbulent jet, sampled on a $175\times39$ spatial grid with a uniform nondimensional time interval of $\Delta t c_\infty/D=0.2$, where $c_\infty$ denotes the ambient speed of sound and $D$ is the jet nozzle diameter. The spatial coordinates are normalized by $D$.

The reduced-order models are constructed using the pressure fluctuation field, with the discrete state vector defined as $\boldsymbol{q}=\widetilde{\boldsymbol{p}}=\boldsymbol{p}^{\mathrm{raw}}-\overline{\boldsymbol{p}^{\mathrm{raw}}}$, where the overbar denotes the temporal mean calculated from the identification snapshots.

\begin{figure}[width=\columnwidth,pos=!htbp]
    \centering
    \includegraphics[width=\linewidth]{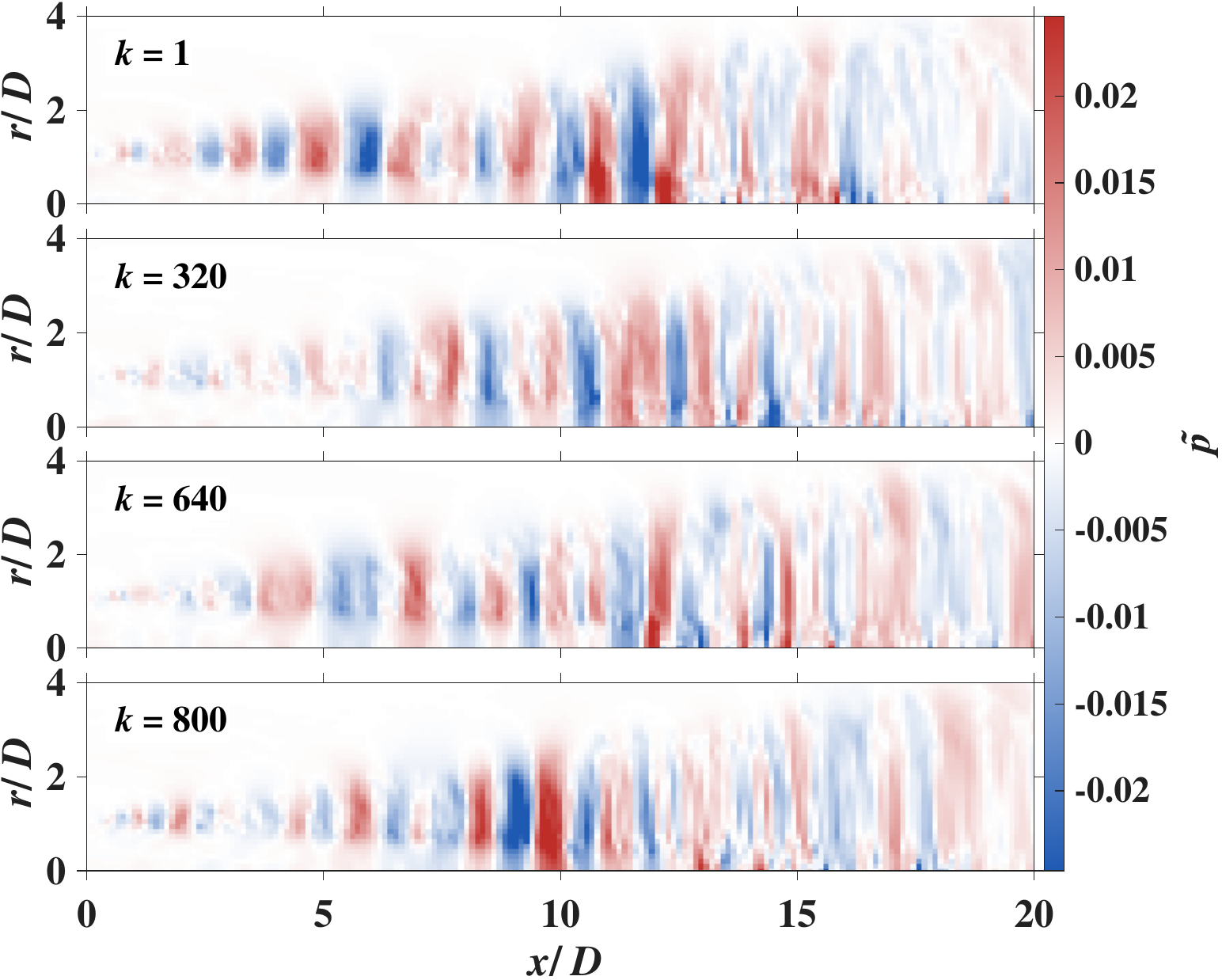}
    \caption{Representative instantaneous pressure-fluctuation fields of the Mach 0.9 turbulent jet at snapshots 1, 320, 640, and 800.}
    \label{fig:jet_configuration}
\end{figure}

Figure~\ref{fig:jet_configuration} presents representative instantaneous pressure-fluctuation fields at four selected snapshots. The flow exhibits complex spatiotemporal fluctuations, with coherent structures developing and evolving along the downstream direction. The spatial distribution contains fluctuations over multiple scales, reflecting the complex dynamics of the turbulent jet and posing additional challenges for reduced-order modeling.

The first 800 consecutive snapshots of the dataset are used for reduced-order modeling, with the first 640 constituting the identification segment and the remaining 160 reserved for independent testing. The temporal mean pressure field is subtracted from the snapshots, and the resulting fluctuation fields are projected onto a SVD basis constructed from the identification data. To balance the retention of spatial flow structures against computational efficiency, a truncation rank of $r=300$ is adopted, capturing approximately $99.70\%$ of the cumulative SVD energy. The identification segment is further divided chronologically into 512 training and 128 validation snapshots. The training subset is used to estimate the GC structure and identify candidate evolution operators, while the validation subset determines the optimal regularization parameter through one-step prediction errors. The final operators are then evaluated on the 160 test snapshots through one-step prediction, using the observed state at each preceding time step. Both DMD and GC-DMD employ the same reduced coordinates and data partitions to ensure a consistent comparison.

\subsection{GC-informed operator sparsification and predictive generalization}

Based on the reduced-order representation established above, the coupling characteristics and predictive generalization of the evolution operators identified by DMD and GC-DMD are further investigated.

\begin{figure}[width=\columnwidth,pos=!htbp]
    \centering
    \includegraphics[width=\columnwidth]{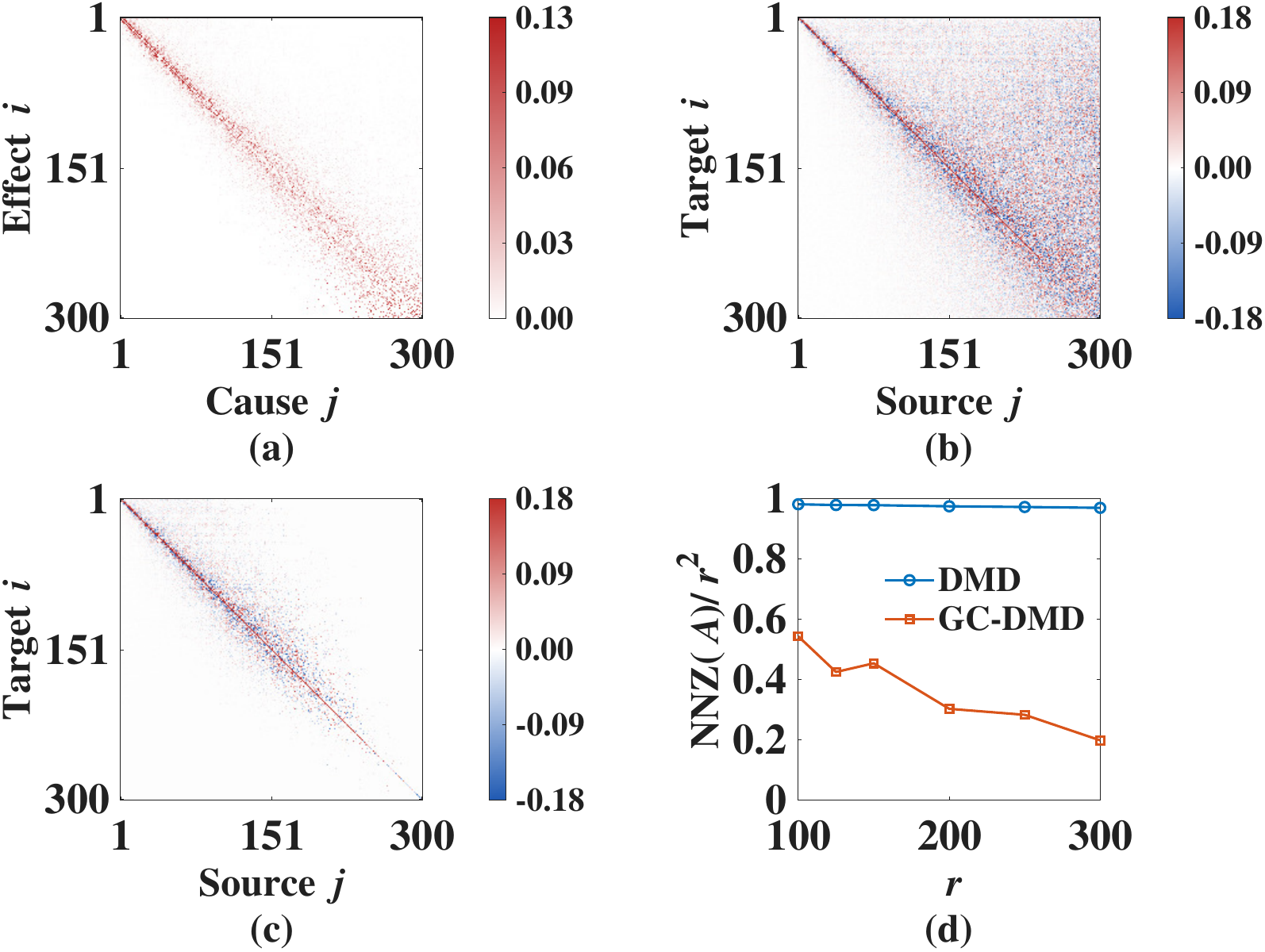}
    \caption{GC-informed operator sparsification for the Mach 0.9 turbulent jet. (a) Normalized conditional GC matrix at $r=300$. (b) Standard DMD operator at $r=300$. (c) GC-DMD operator at $r=300$. (d) Nonzero operator coefficient ratio as a function of the retained SVD rank.}
    \label{fig:jet_sparsity}
\end{figure}

Figure~\ref{fig:jet_sparsity}(a) presents the normalized conditional GC matrix at $r=300$, revealing a pronounced near-diagonal banded structure in which directional predictive dependencies are predominantly concentrated among SVD modes with nearby indices. The corresponding evolution operators are compared in Figs.~\ref{fig:jet_sparsity}(b) and (c). While the standard DMD operator contains numerous off-diagonal coefficients distributed throughout the matrix, GC-DMD substantially suppresses weakly supported couplings while retaining a predominantly near-diagonal structure consistent with the GC matrix. At $r=300$, the off-diagonal sparsity increases from 3.0\% for DMD to 82.1\% for GC-DMD, demonstrating a substantial reduction in the number of nonzero off-diagonal coefficients. The dependence of operator sparsification on SVD truncation rank is further examined in Fig.~\ref{fig:jet_sparsity}(d), which presents the nonzero operator coefficient ratio (NNZ$(A)/r^{2}$) as a function of the retained SVD rank. Here, the nonzero coefficient ratio is defined as the number of operator coefficients satisfying $|A_{ij}|>10^{-4}$ divided by the total number of coefficients, $r^2$, consistent with the threshold adopted in Section~3.3. The standard DMD operator remains nearly dense across the investigated ranks, whereas GC-DMD consistently exhibits substantially lower nonzero coefficient ratios. As the retained rank increases from $r=100$ to $r=300$, the nonzero coefficient ratio of GC-DMD decreases from approximately 0.54 to 0.20, indicating increasingly pronounced sparsification at higher reduced-order dimensions. Together with the cavity-flow results in Section~4.3, these observations suggest that GC-informed regularization can promote compact operator structures in both multifrequency and broadband flow regimes.

\begin{figure}[width=\columnwidth,pos=!htbp]
    \centering

    \includegraphics[width=0.87\columnwidth]{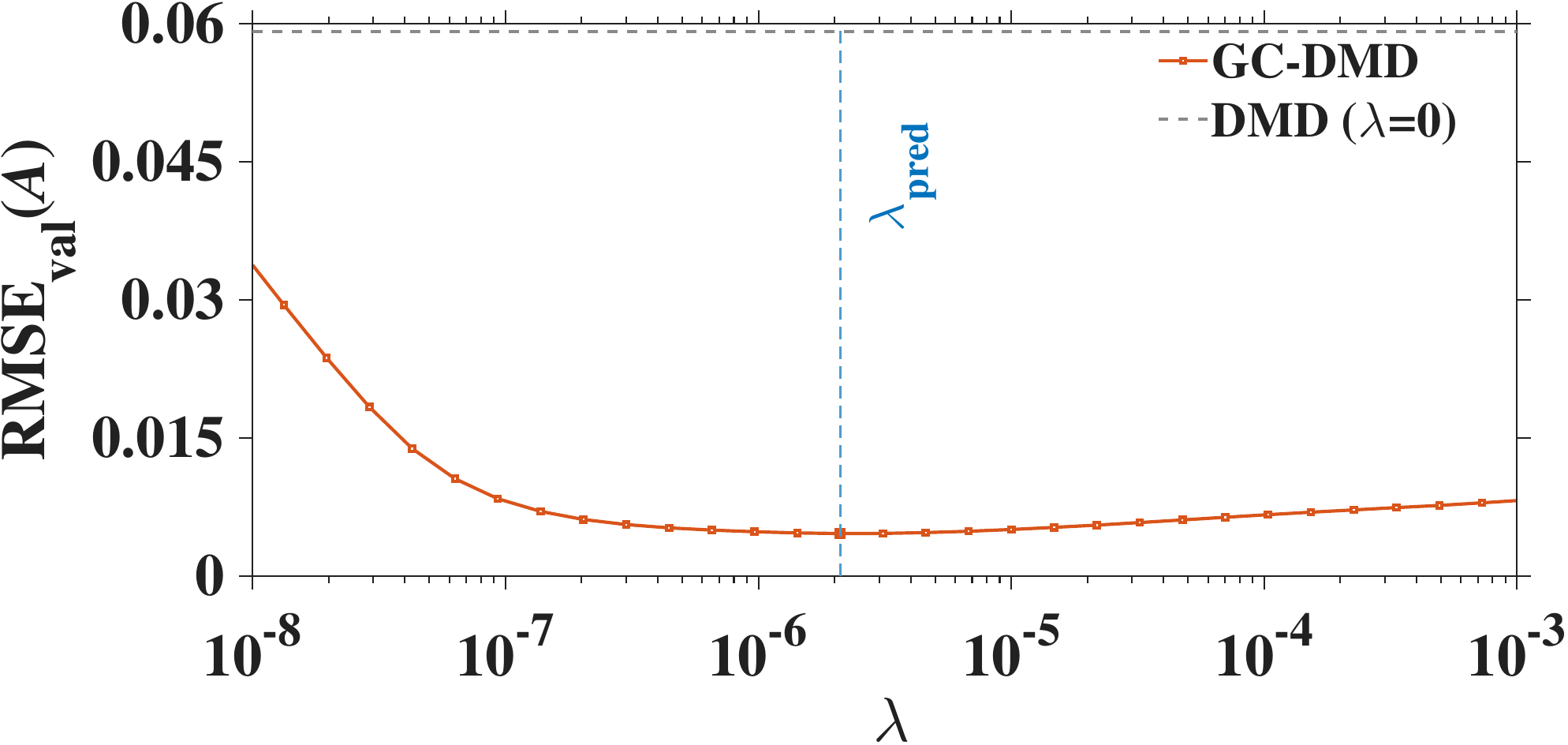}
    \par\smallskip
    {\small (a)}

    \medskip

    \includegraphics[width=0.87\columnwidth]{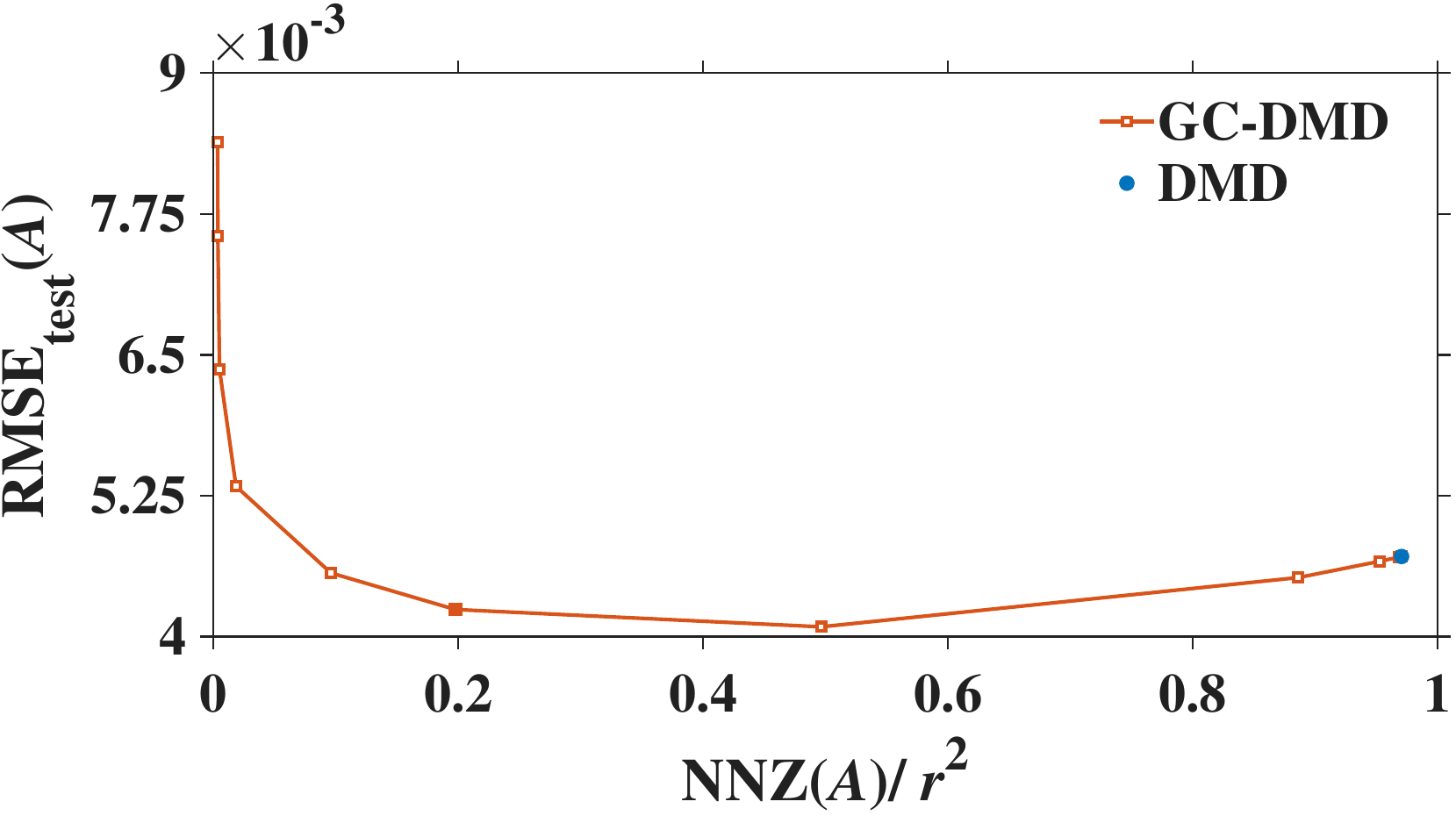}
    \par\smallskip
    {\small (b)}

    \caption{Predictive generalization of DMD and GC-DMD for the Mach 0.9 turbulent jet. (a) Validation one-step RMSE versus the regularization parameter $\lambda$. (b) Test one-step RMSE versus the nonzero operator coefficient ratio along the GC-DMD regularization path.}
    \label{fig:jet_generalization}
\end{figure}

The predictive generalization of the identified operators is subsequently evaluated using independent test data. Figure~\ref{fig:jet_generalization}(a) presents the validation one-step RMSE as a function of the regularization parameter $\lambda$. An appropriate level of GC-informed regularization substantially reduces the validation RMSE relative to standard DMD, with the optimal parameter selected at the minimum of the validation-error curve. This indicates that suppressing weakly supported operator couplings can improve predictive generalization rather than merely reducing model complexity. To further examine the trade-off between operator sparsity and predictive generalization, Fig.~\ref{fig:jet_generalization}(b) presents the test one-step RMSE as a function of the nonzero operator coefficient ratio along the GC-DMD regularization path. Each point corresponds to a different regularization parameter $\lambda$, and the nonzero coefficient ratio is evaluated using the same threshold as in Fig.~\ref{fig:jet_sparsity}(d). The standard DMD result is included as a reference. The curve exhibits a nonmonotonic relationship between operator complexity and prediction error, with the minimum test RMSE occurring at a nonzero coefficient ratio of approximately 50\%. In contrast, the validation-selected GC-DMD operator retains only 19.8\% of the coefficients while achieving a test RMSE close to this minimum. Even when the nonzero coefficient ratio decreases to approximately 8\%, the highly sparsified operator maintains a test RMSE comparable to that of standard DMD. These results indicate that a substantial fraction of the coefficients in the nearly dense DMD operator can be suppressed without compromising predictive accuracy. GC-informed regularization therefore enables a favorable balance between operator compactness and predictive generalization in broadband turbulent flow.

\subsection{Influence of coupling structure on predictive generalization}

The preceding results demonstrate that GC-informed regularization substantially reduces the number of nonzero operator coefficients while improving predictive generalization. However, operator regularization changes not only the number of retained coefficients but also their locations within the evolution operator. To further examine the influence of coupling organization on predictive performance independently of the degree of sparsification, an equal-sparsity comparison is conducted.

To ensure a consistent comparison, the nonzero operator coefficient ratio is fixed at approximately 19.8\%, corresponding to the GC-DMD operator obtained with the validation-selected regularization parameter $\lambda_{\mathrm{pred}}$, as discussed in Section~5.2. Four strategies are considered for selecting the retained coefficient locations: GC-selected, near-diagonal, energy-local, and random selection. For GC-selected, the retained locations are determined by the nonzero coefficients of the GC-DMD operator at $\lambda_{\mathrm{pred}}$. For near-diagonal selection, the off-diagonal locations are ranked in ascending order of the SVD modal-index separation $|i-j|$. For energy-local selection, they are ranked in ascending order of the modal-energy difference $|\log_{10}(E_i/E_j)|$, where $E_i$ denotes the energy of the $i$th SVD mode. For both ranking-based strategies, coefficient locations are selected sequentially from the corresponding ordered lists until the prescribed number is reached. For random selection, the off-diagonal locations are selected randomly. All four strategies preserve the diagonal entries and retain the same number of off-diagonal coefficient locations, ensuring an equal sparsity level.

For each strategy $s$, a binary structural matrix $\boldsymbol{C}^{(s)}\in\{0,1\}^{r\times r}$ is constructed according to the selected coefficient locations:
\begin{equation}
C_{ij}^{(s)}=
\begin{cases}
1, & (i,j)\in\mathcal{S}_s,\\
0, & (i,j)\notin\mathcal{S}_s,
\end{cases}
\label{eq:jet_structure_matrix}
\end{equation}
where $\mathcal{S}_s$ denotes the set of retained coefficient locations. Following Eq.~(14), the evolution operator for each strategy is identified by replacing the original GC matrix with the corresponding structural matrix $\boldsymbol{C}^{(s)}$:
\begin{equation}
\begin{aligned}
\boldsymbol{A}^{(s)}
=\arg\min_{\boldsymbol{A}}\Bigg[
&\frac{1}{2(m-1)}
\left\|\boldsymbol{Q}_2-\boldsymbol{A}\boldsymbol{Q}_1\right\|_F^2\\
&+\lambda_s\sum_{i=1}^{r}\sum_{j=1}^{r}
(1-C_{ij}^{(s)})|A_{ij}|
\Bigg].
\end{aligned}
\label{eq:jet_structure_regularization}
\end{equation}

For all four strategies, the operator coefficients are re-estimated using the training data defined in Section~5.1, and their predictive performance is evaluated on the same independent test set using the test one-step RMSE. The random selection is repeated 30 times, with the mean RMSE and standard deviation reported.

\begin{figure}[width=\columnwidth,pos=!htbp]
    \centering
    \includegraphics[width=\linewidth]{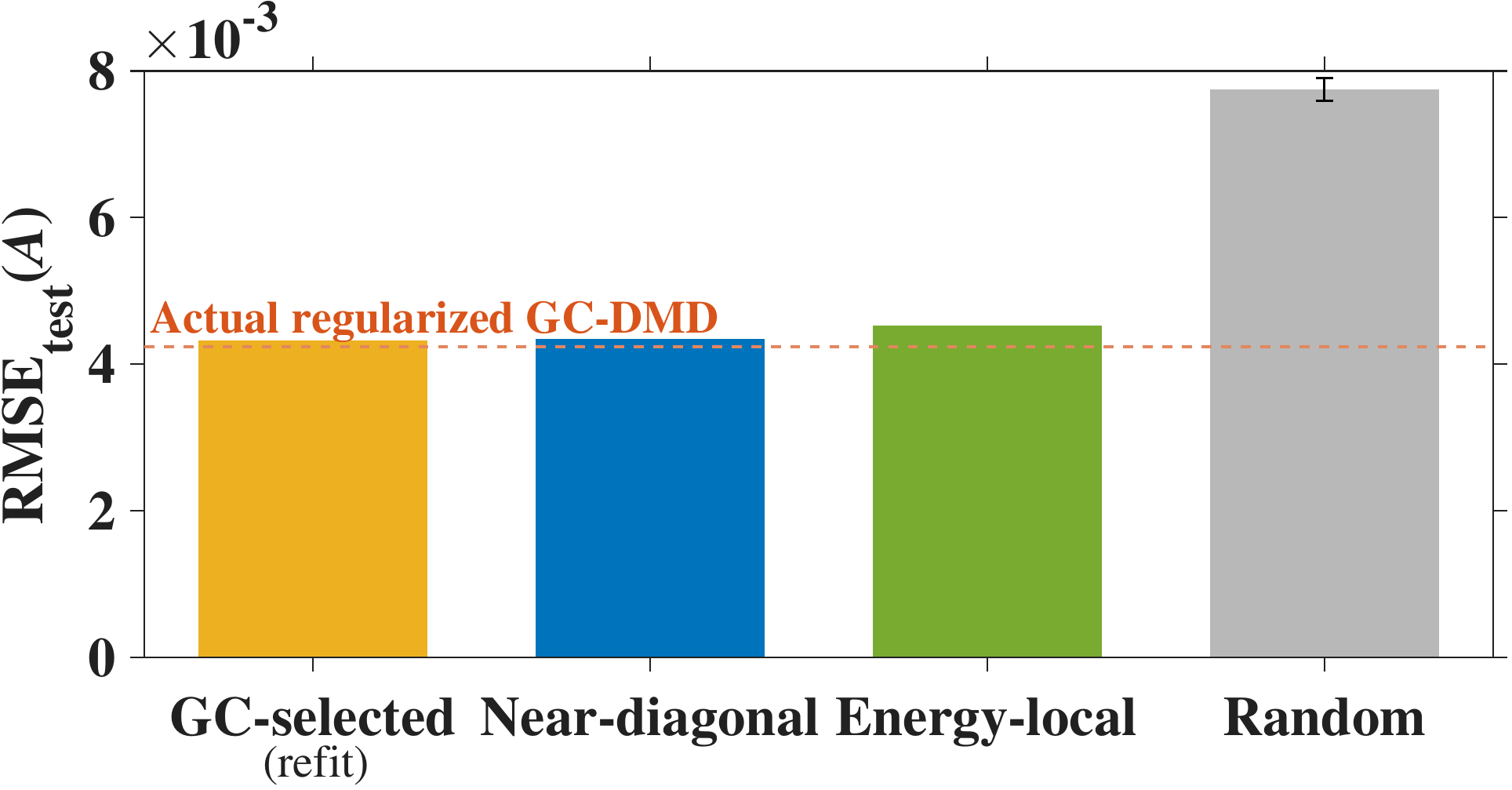}
    \caption{Comparison of test one-step RMSE for four coefficient-selection strategies at an equal sparsity level. The dashed line indicates the original GC-DMD operator, and the error bar represents the standard deviation over 30 random selections.}
    \label{fig:jet_structure}
\end{figure}

Figure~\ref{fig:jet_structure} compares the test one-step prediction errors obtained using the four strategies. The original GC-DMD operator, identified through GC-weighted regularization, is included as a reference. Among the four refitted operators, the GC-selected strategy achieves the lowest prediction error, followed closely by the near-diagonal and energy-local strategies, whereas random selection produces substantially larger errors. The favorable performance of the near-diagonal and energy-local strategies suggests that the predictive couplings retained by GC-DMD are closely associated with interactions among SVD modes with nearby indices and similar energy levels. Nevertheless, the slightly better performance of GC-selected indicates that the GC-derived coupling structure may capture additional predictive information beyond these two prescribed structural characteristics. These results suggest that the predictive performance of sparsified evolution operators depends not only on the number of retained coefficients but also on their structural arrangement.

Overall, the equal-sparsity comparison demonstrates that GC-informed operator sparsification is not merely a reduction in the number of coefficients but also involves the selection of a predictive coupling structure. Together with the preceding results, these findings suggest that both operator sparsity and the organization of retained interactions contribute to the predictive generalization of GC-DMD.

\section{Conclusions}

This study proposes a Granger Causality-informed Dynamic Mode Decomposition framework, named as GC-DMD, for modal decomposition, reduced-order modeling, and analysis of unsteady flows. GC-DMD incorporates directional predictive dependencies inferred from Granger causality into the identification of the standard DMD operator through regularization. It suppresses weakly supported couplings while preserving dynamically relevant interactions, providing a more accurate and reliable low-dimensional representation of the system dynamics. The method remains purely data-driven while incorporating information about the directional predictive structure of the dynamics into operator identification.

The proposed framework is evaluated using three complementary test cases. First, a linear system with a known coupling structure demonstrates that GC-derived directional information can improve operator identification by suppressing spurious interactions while retaining dominant couplings. Compared with standard DMD, GC-DMD reduces the relative operator-identification error by approximately 21\%, while a further comparison with conventional $L_1$ regularization supports the benefits of incorporating directional structural information. Second, lid-driven cavity flows with increasing dynamical complexity demonstrate improved long-horizon predictive performance. For predominantly periodic dynamics, GC-DMD preserves the principal spectral and modal characteristics while reducing prediction errors. In the multifrequency regime, its advantages become more pronounced, with fewer and weaker unstable modes, reduced error amplification during recursive prediction, and improved robustness to the choice of SVD truncation rank. Third, the Mach 0.9 turbulent jet demonstrates substantial operator sparsification together with improved prediction accuracy. At an SVD rank of $r=300$, the off-diagonal sparsity increases from 3.0\% for standard DMD to 82.1\% for GC-DMD. The retained couplings exhibit a pronounced near-diagonal organization in SVD coordinates, while comparisons at matched sparsity levels indicate that the GC-informed selection of retained couplings, rather than sparsity alone, contributes to predictive performance.

These findings demonstrate the potential of GC-derived directional information to guide the identification of sparse, predictive reduced-order models for complex unsteady flows. Nevertheless, the present framework relies on first-order linear evolution operators, which may limit its ability to represent nonlinear dynamics and delayed interactions. In addition, conditional GC is estimated in SVD coordinates, making the inferred dependencies specific to the chosen reduced representation. Future work may therefore extend GC-informed regularization to nonlinear dynamical models and higher-order models with explicit time delays, while exploring alternative reduced representations that better capture spatially localized interactions. Such developments could further broaden the applicability of causality-informed modal analysis and reduced-order modeling to more complex flow dynamics.

\appendix
\section{Comparison with standard $L_1$ regularization}
\label{app:L1}

To further evaluate the benefits of incorporating GC-derived directional structural information beyond conventional sparsity promotion, GC-DMD is compared with standard $L_1$-regularized DMD (L1-DMD) using 20 distinct five-dimensional linear systems. Following the VAR(1) formulation introduced in Section~3, the systems are constructed with different prescribed evolution operators determined by independent random seeds. These operators exhibit different directional coupling structures while maintaining comparable sparsity and stable dynamics. An additional forcing is added as independent Gaussian white noise with zero mean and standard deviation $\sigma=0.1$, and the data are sampled at a uniform interval of $\Delta t=1$. After initial simulation of 500 steps, 150, 400, and 450 snapshots are subsequently allocated to training, validation, and testing, respectively.

For a consistent comparison, all three methods use the same data within each system. Conventional L1-DMD identifies the evolution operator by minimizing the following uniformly regularized objective:

\begin{equation}
\begin{aligned}
\boldsymbol{A}^{\mathrm{L1-DMD}}
=\arg\min_{\boldsymbol{A}}\Bigg[
&\frac{1}{2(m-1)}
\left\|\boldsymbol{Q}_2-\boldsymbol{A}\boldsymbol{Q}_1\right\|_F^2\\
&+\lambda\sum_{i=1}^{r}\sum_{j=1}^{r}|A_{ij}|
\Bigg].
\end{aligned}
\label{eq:app_l1_objective}
\end{equation}

For both regularized methods, $\lambda$ is selected independently from the same set of candidate values by minimizing the one-step validation RMSE. Across the 20 systems, both regularized methods improve operator identification compared with standard DMD. GC-DMD achieves the lowest mean relative operator error, with reductions of approximately 27.0\% and 12.5\% compared with DMD and L1-DMD, respectively. The mean test RMSE is reduced by approximately 0.9\% and 0.4\%, respectively. These results demonstrate that GC-informed regularization improves operator identification beyond conventional sparsity promotion while providing modest additional gains in predictive generalization.

\begin{figure}[width=\columnwidth,pos=!htbp]
    \centering
    \begin{minipage}[t]{0.48\linewidth}
        \centering
        \includegraphics[width=\linewidth]{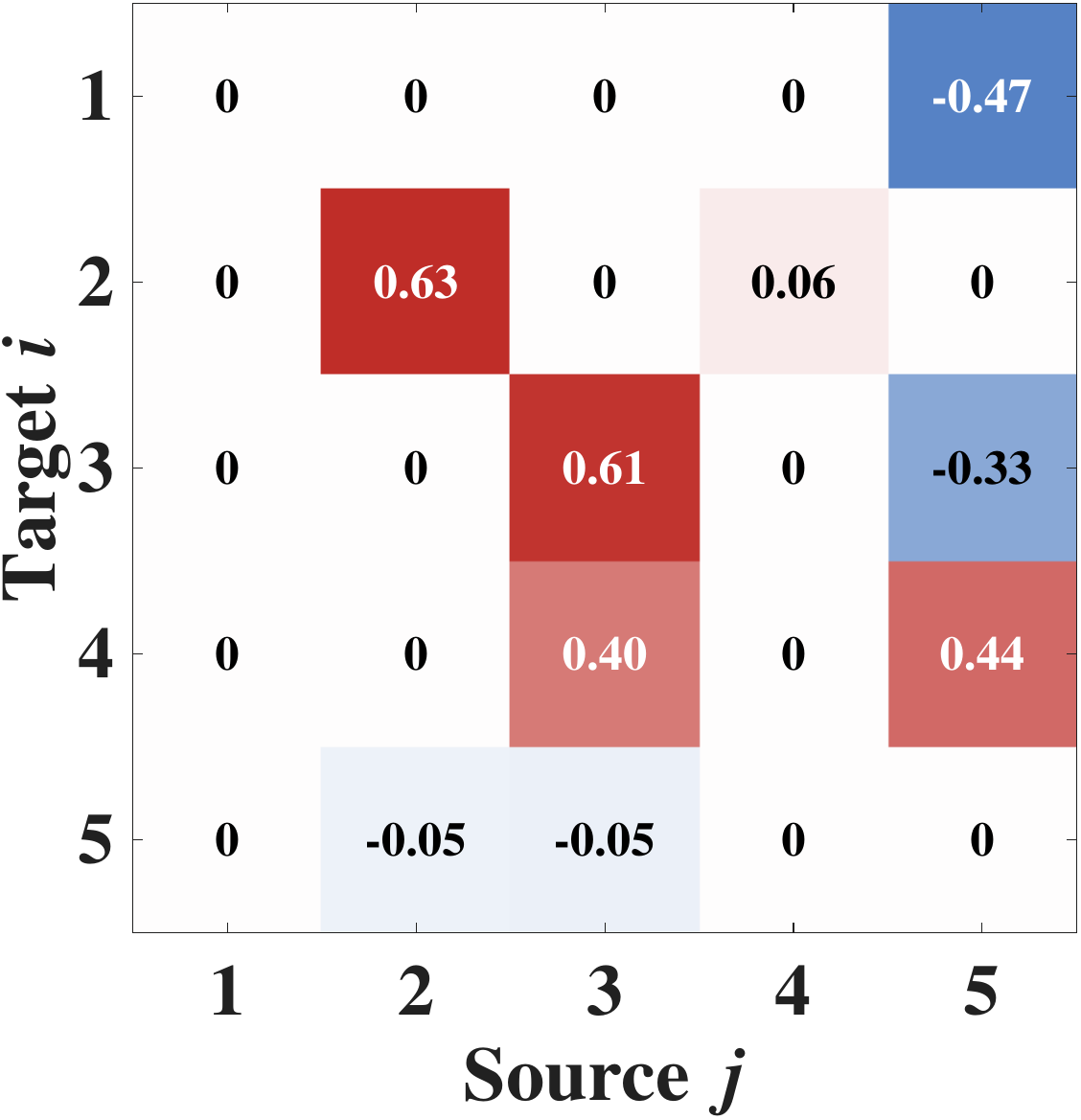}
        {\small (a)}
    \end{minipage}
    \hfill
    \begin{minipage}[t]{0.48\linewidth}
        \centering
        \includegraphics[width=\linewidth]{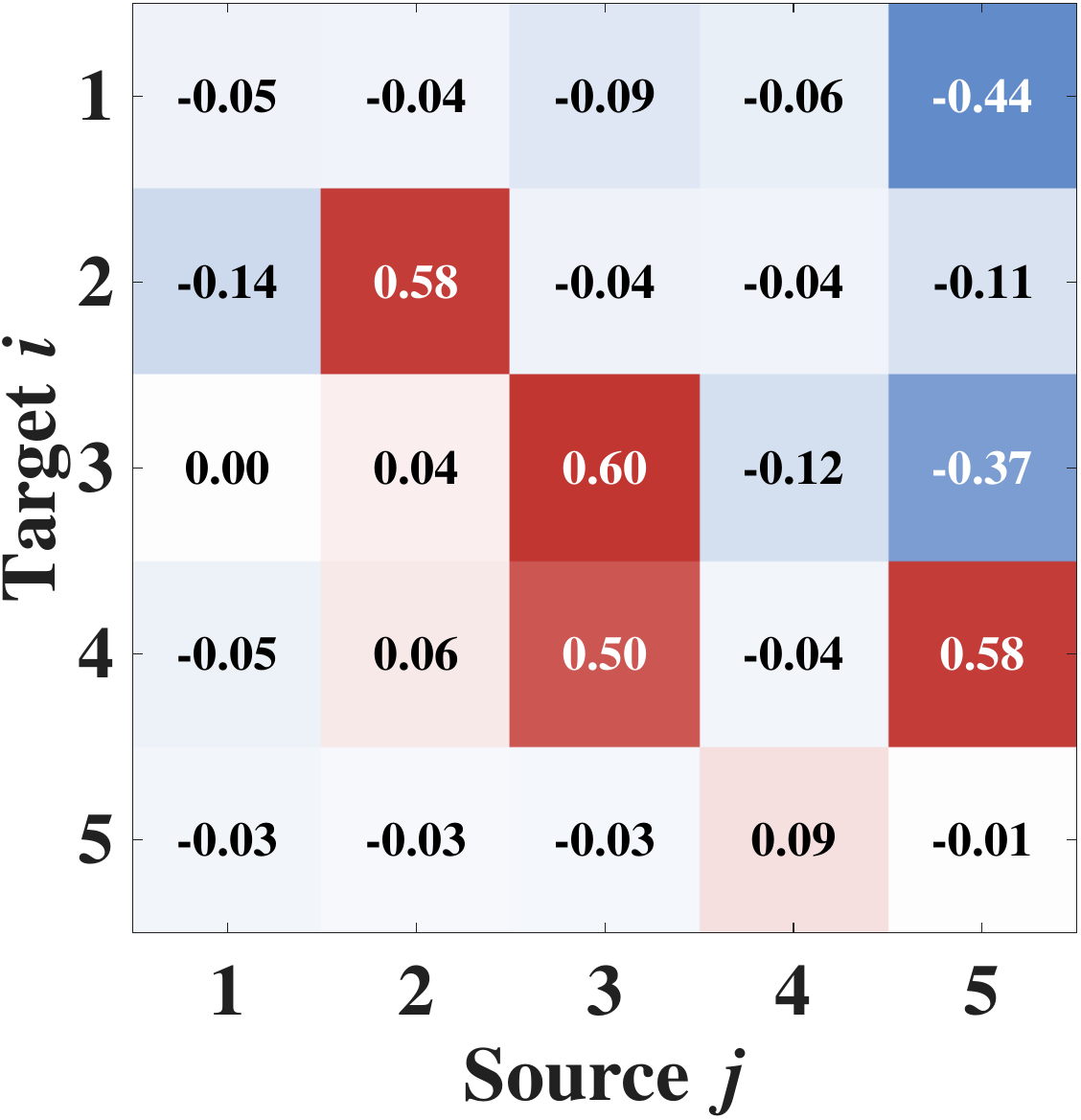}
        {\small (b)}
    \end{minipage}

    \par\vspace{1mm}

    \begin{minipage}[t]{0.48\linewidth}
        \centering
        \includegraphics[width=\linewidth]{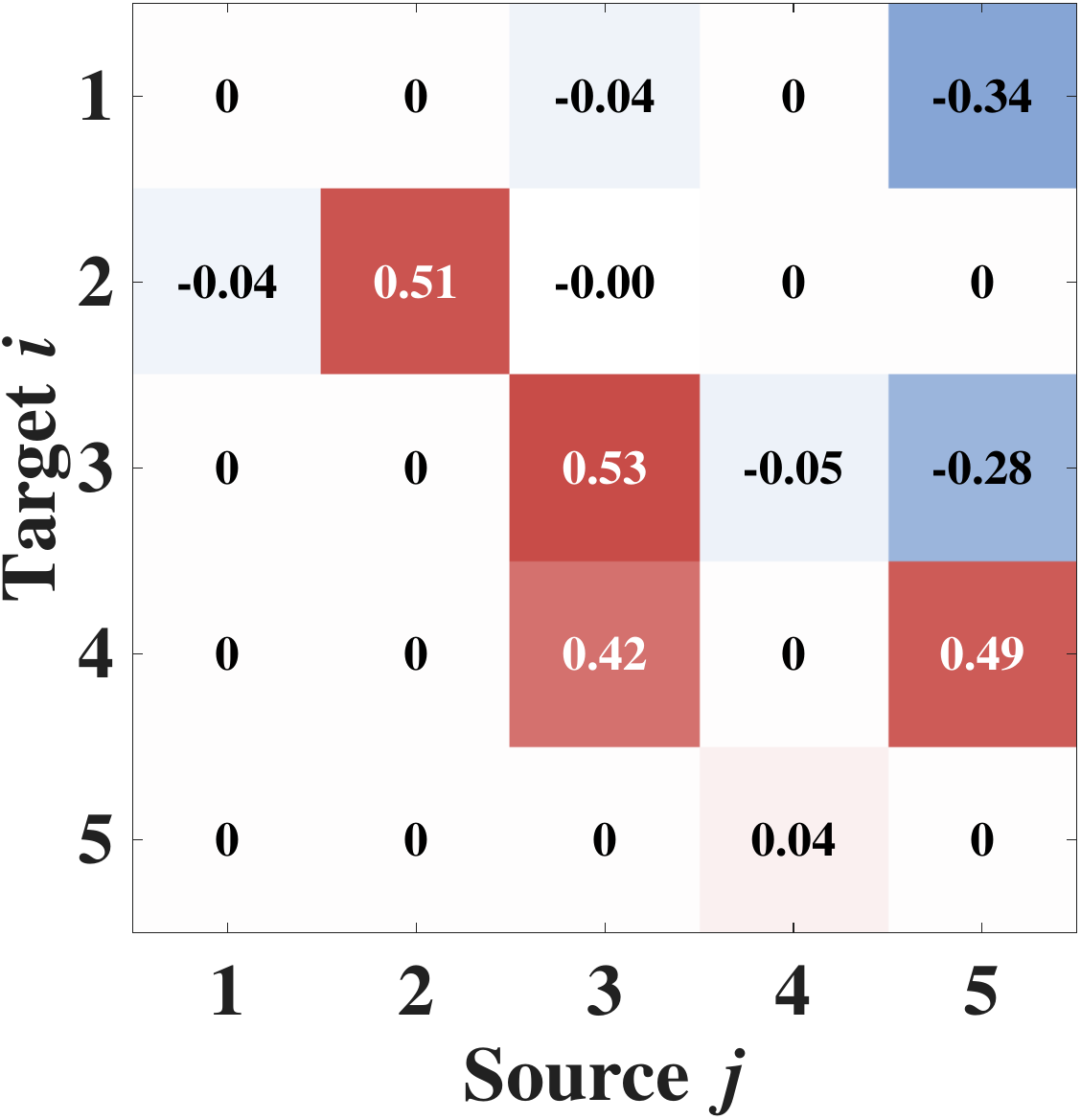}
        {\small (c)}
    \end{minipage}
    \hfill
    \begin{minipage}[t]{0.48\linewidth}
        \centering
        \includegraphics[width=\linewidth]{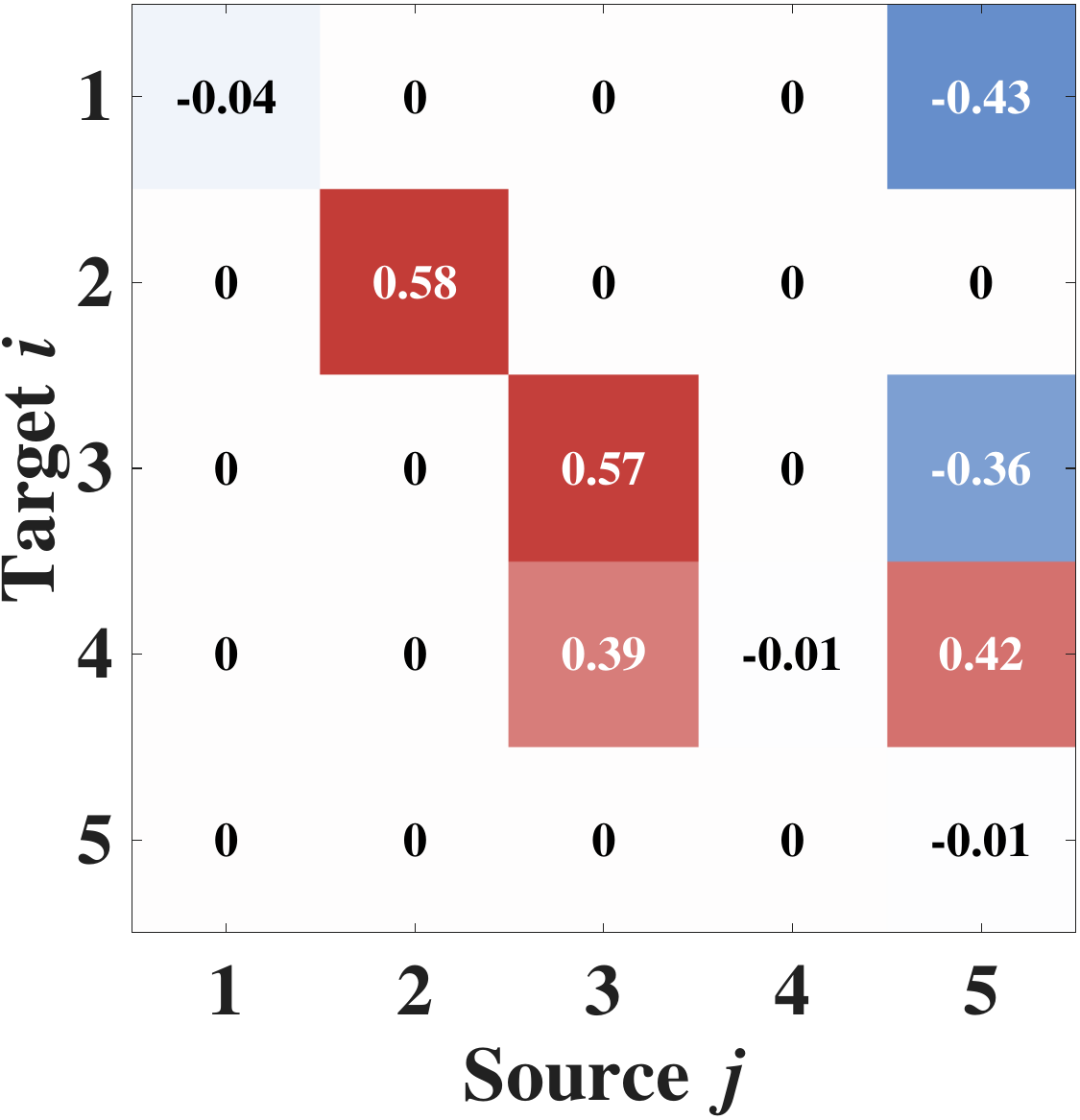}
        {\small (d)}
    \end{minipage}

    \caption{Comparison of the prescribed and identified evolution operators for a five-dimensional linear system. (a) Prescribed operator $\boldsymbol{A}^{\mathrm{true}}$. (b) Standard DMD. (c) L1-DMD. (d) GC-DMD. All panels use the same color scale, and the numerical values indicate the corresponding operator coefficients.}
    \label{fig:app_l1_operators}
\end{figure}

To illustrate the influence of directional regularization on individual operator coefficients, Figure~\ref{fig:app_l1_operators} compares the prescribed evolution operator with those identified by DMD, L1-DMD, and GC-DMD for one of the 20 systems. Standard DMD produces a relatively dense operator containing numerous spurious weak couplings that are absent from the true system. Both uniform $L_1$ and GC-informed regularization effectively suppress these redundant interactions, resulting in substantially sparser operators. Compared with L1-DMD, GC-DMD produces a sparser operator while preserving the dominant coupling structure, which can be attributed to the incorporation of physically interpretable directional dependencies into the regularization process. Moreover, the estimated coefficients of the principal interactions are generally closer to their true values, indicating improved recovery of both the coupling structure and its associated strengths. These observations highlight the advantage of using GC-derived structural information to guide operator identification rather than relying solely on uniform sparsity promotion.

\begin{figure}[width=\columnwidth,pos=!htbp]
    \centering
    \includegraphics[width=0.8\linewidth]{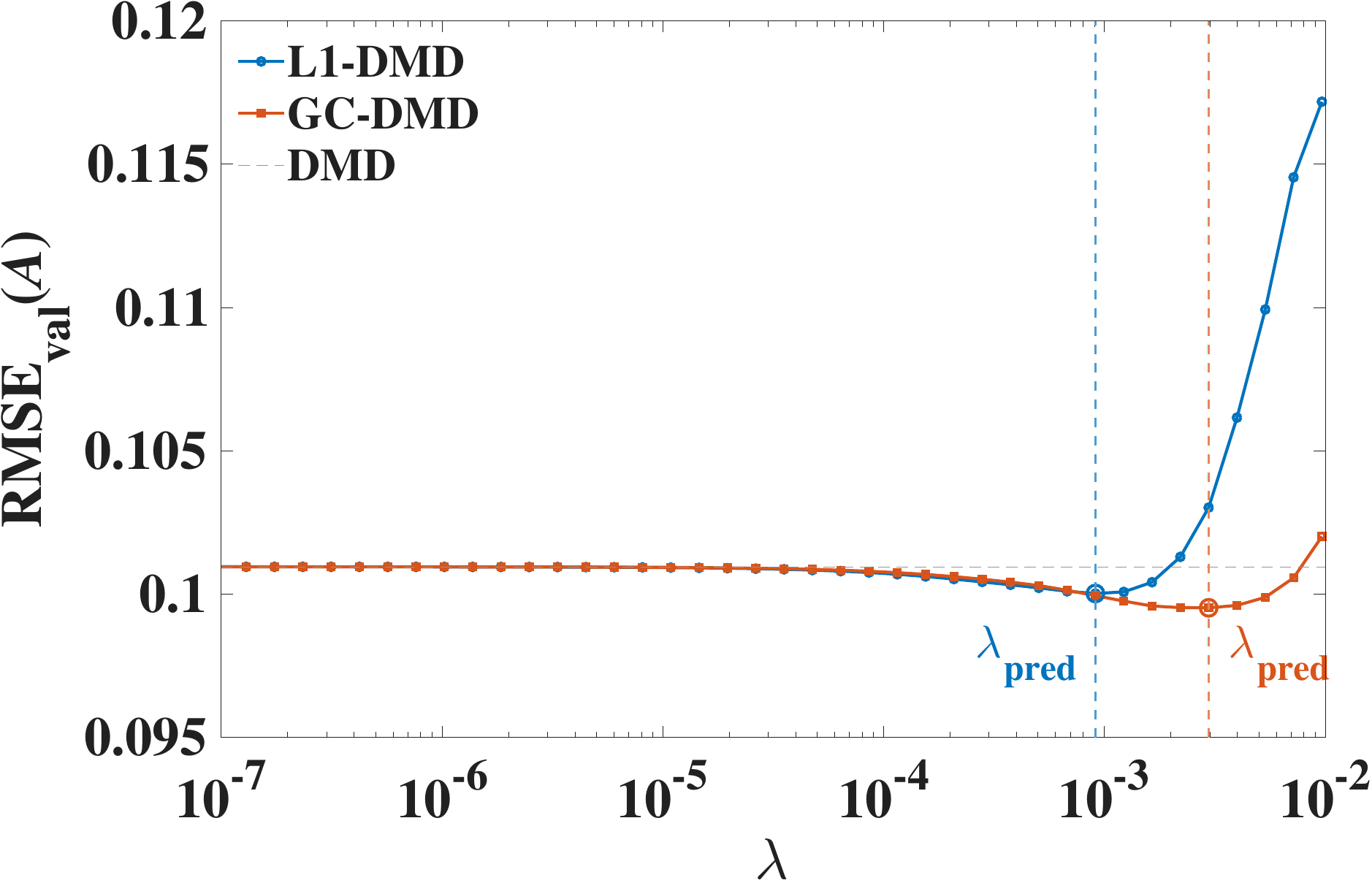}
    \caption{Comparison of validation RMSE as a function of the regularization parameter $\lambda$ for L1-DMD and GC-DMD. The dashed horizontal line denotes the unregularized DMD baseline, while the vertical dashed lines indicate the validation-selected regularization parameters for the two methods.}
    \label{fig:app_l1_lambda}
\end{figure}

For the same system examined in Fig.~\ref{fig:app_l1_operators}, Fig.~\ref{fig:app_l1_lambda} compares the validation RMSE of L1-DMD and GC-DMD over a range of regularization parameters. Both methods achieve lower validation errors than standard DMD with an appropriate amount of regularization, indicating that sparsity promotion can improve predictive generalization. GC-DMD achieves a slightly lower minimum validation error than L1-DMD, suggesting an additional benefit from incorporating GC-derived directional information. Moreover, as $\lambda$ increases beyond its optimal value, the validation error of GC-DMD grows more slowly than that of L1-DMD, indicating reduced sensitivity to the regularization strength. These results suggest that GC-informed regularization not only improves predictive generalization but also provides greater robustness to the choice of $\lambda$ in the illustrated case.

Overall, this comparison demonstrates that GC-informed regularization offers advantages over uniform $L_1$ regularization in identifying sparse evolution operators. By incorporating directional predictive dependencies into the regularization process, GC-DMD achieves more accurate recovery of the underlying coupling structure, together with modest improvements in predictive generalization. The reduced sensitivity to stronger regularization observed in the representative case further highlights the potential benefit of GC-derived structural priors for operator identification.

\section{Additional validation for the near-chaotic cavity flow at $Re=30{,}000$}
\label{app:cavity30k}

To further assess the applicability of GC-DMD to more complex flow dynamics, an additional cavity-flow case at $Re=30{,}000$ is considered. Compared with the multifrequency flow at $Re=19{,}000$ examined in Section~4.3, this higher-Reynolds-number case exhibits more irregular temporal fluctuations and requires a higher-dimensional SVD representation, posing greater challenges for operator identification and recursive prediction. A total of 1,200 consecutive snapshots are considered, covering the nondimensional time interval from approximately $t=4500$ to $t=4620$. Consistent with the procedure adopted in Section~4, the first 800 snapshots constitute the identification segment, further divided into 640 training snapshots and 160 validation snapshots, while the remaining 400 snapshots form the independent prediction evaluation segment. A SVD truncation rank of $r=162$, corresponding to a cumulative energy threshold of 99.999\%, is adopted.

\begin{figure}[width=\columnwidth,pos=!htbp]
    \centering
    \includegraphics[width=\columnwidth]{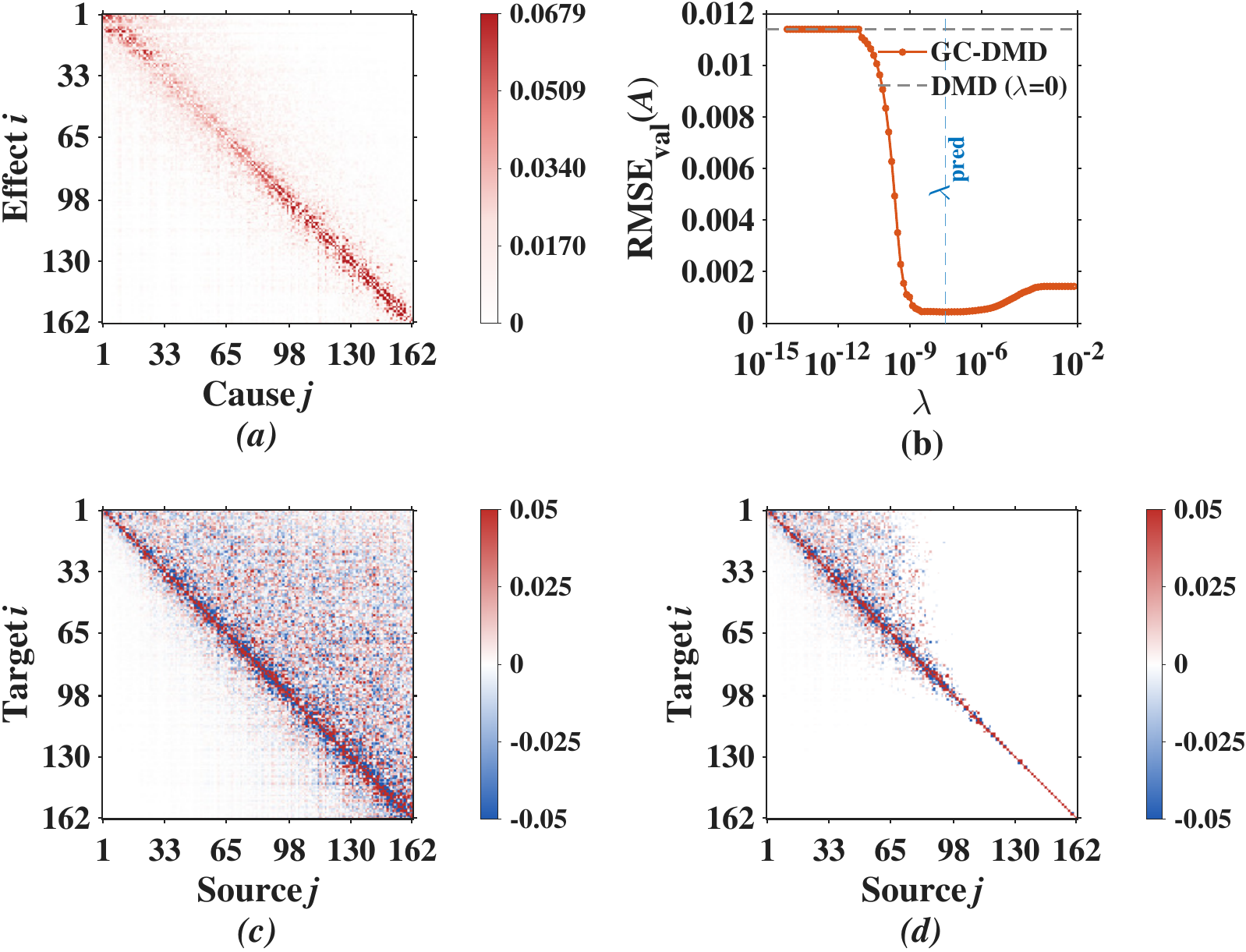}
    \caption{GC-informed operator identification for the near-chaotic cavity flow at $Re=30{,}000$. (a) Normalized conditional GC strength. (b) One-step validation RMSE as a function of the regularization parameter $\lambda$. (c) DMD evolution operator. (d) GC-DMD evolution operator.}
    \label{fig:app_cavity30k_operator}
\end{figure}

Figure~\ref{fig:app_cavity30k_operator} presents the GC-derived directional dependencies, validation performance, and evolution operators identified by DMD and GC-DMD. As shown in Fig.~\ref{fig:app_cavity30k_operator}(a), stronger GC interactions are predominantly concentrated near the main diagonal, indicating that the most pronounced directional predictive dependencies tend to occur between SVD coordinates with relatively close mode indices. Figure~\ref{fig:app_cavity30k_operator}(b) shows the one-step validation RMSE as a function of the regularization parameter $\lambda$. The validation error decreases substantially at an appropriate regularization strength, indicating improved predictive generalization compared with standard DMD, whereas excessive regularization leads to a subsequent increase in error. The corresponding evolution operators are compared in Figs.~\ref{fig:app_cavity30k_operator}(c) and \ref{fig:app_cavity30k_operator}(d). Standard DMD exhibits numerous weak off-diagonal coefficients distributed throughout the operator, whereas GC-DMD suppresses a substantial proportion of these interactions, producing a more concentrated, near-diagonal structure. Quantitatively, the off-diagonal sparsity increases from approximately 23.8\% for DMD to 81.4\% for GC-DMD, using the threshold-based definition $|A_{ij}|<10^{-3}$ for $i\neq j$. These results demonstrate that GC-informed regularization remains effective in identifying a more compact evolution operator for the higher-Reynolds-number flow.

\begin{figure*}[width=\textwidth,pos=!htbp]
    \centering

    % (a) Relative full-field error
    \begin{minipage}[t]{0.8\textwidth}
        \centering
        \includegraphics[width=\linewidth]{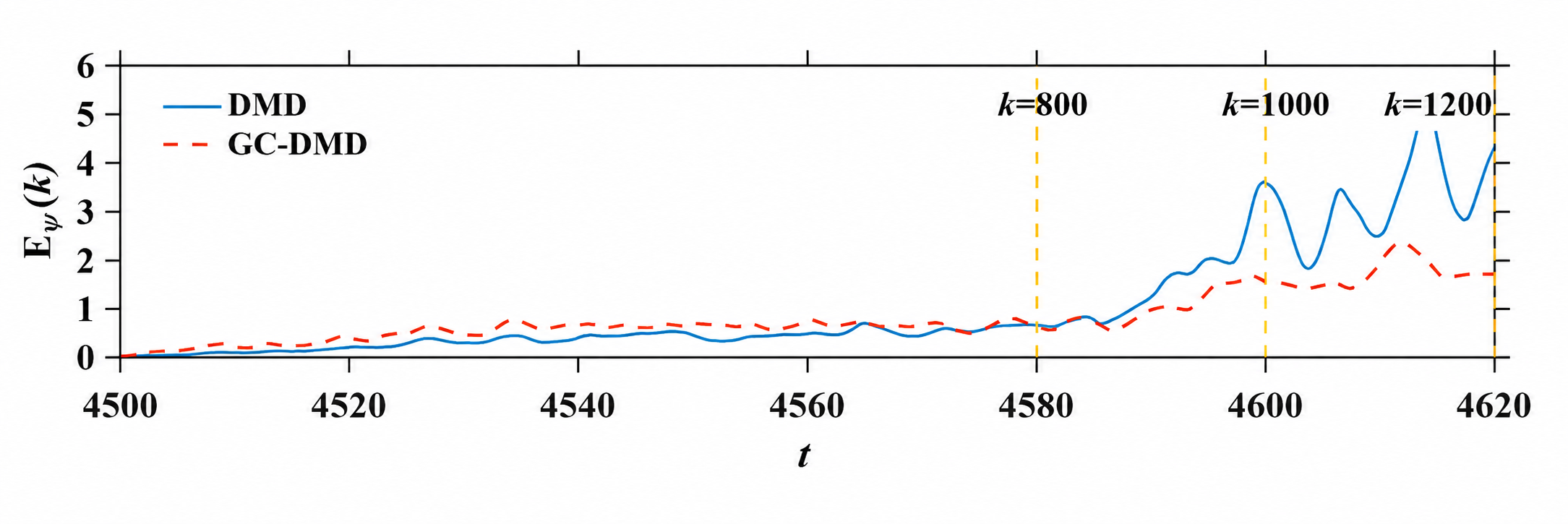}

        {\small (a)}
    \end{minipage}

    \par\vspace{2mm}

    % (b) Reconstruction at t = 800
    \begin{minipage}[t]{0.8\textwidth}
        \centering
        \includegraphics[width=\linewidth]{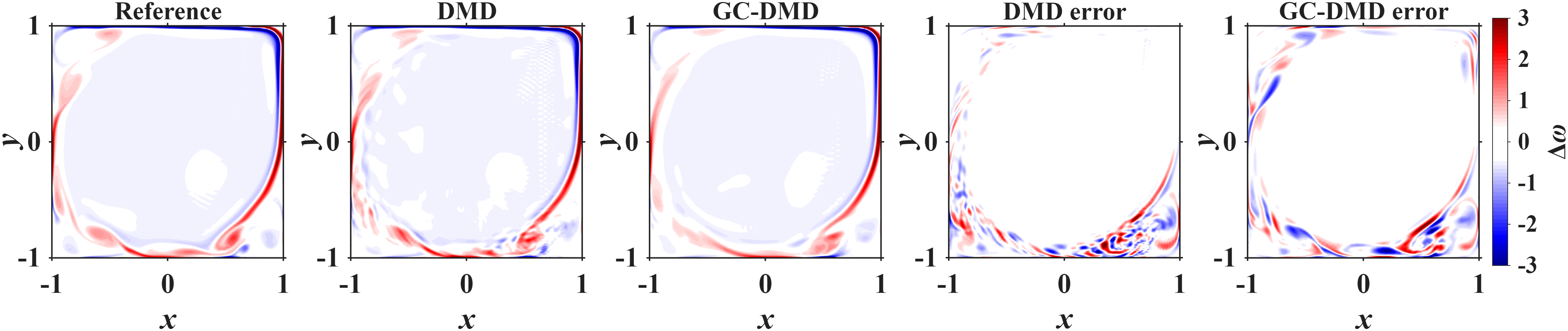}

        {\small (b)}
    \end{minipage}

    \par\vspace{2mm}

    % (c) Prediction at t = 1000
    \begin{minipage}[t]{0.8\textwidth}
        \centering
        \includegraphics[width=\linewidth]{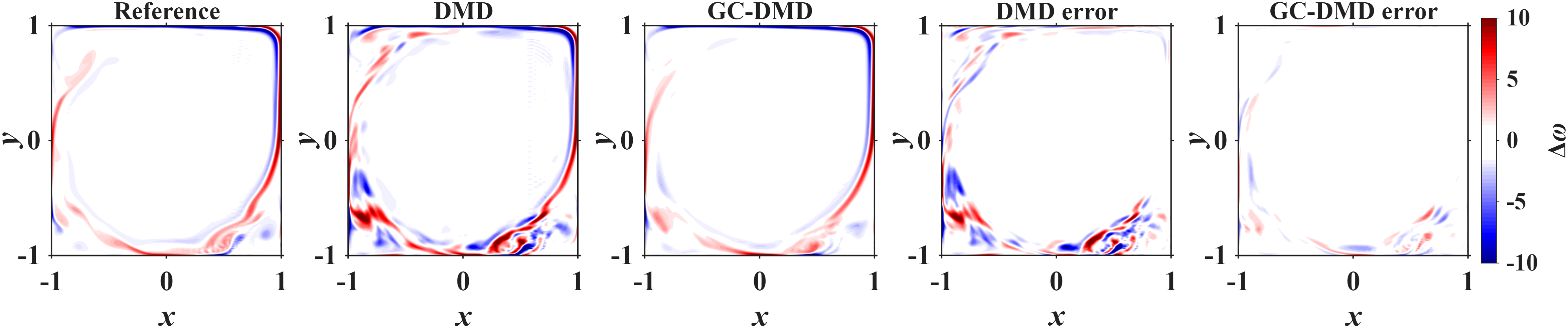}

        {\small (c)}
    \end{minipage}

    \par\vspace{2mm}

    % (d) Prediction at t = 1200
    \begin{minipage}[t]{0.8\textwidth}
        \centering
        \includegraphics[width=\linewidth]{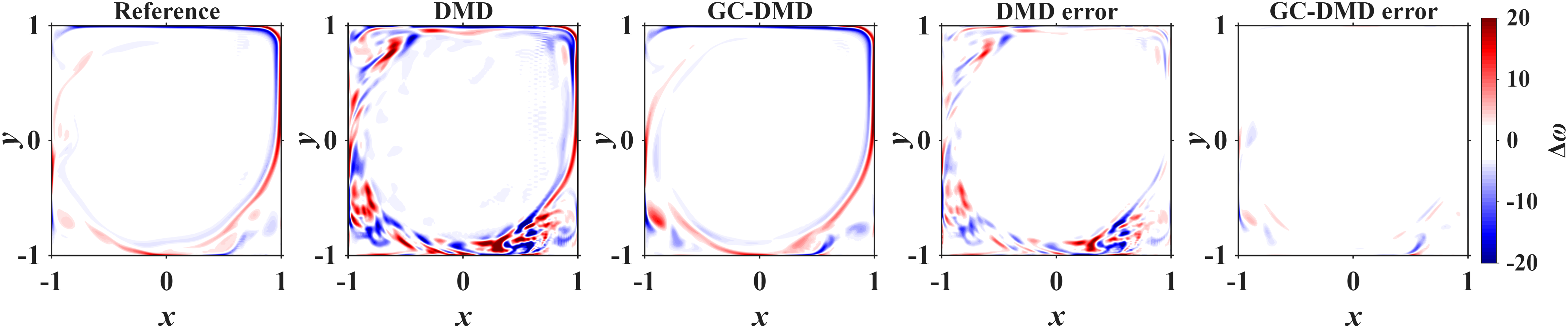}

        {\small (d)}
    \end{minipage}

    \caption{Reconstruction and recursive prediction performance of DMD and GC-DMD for the near-chaotic cavity flow at $Re=30{,}000$. (a) Evolution of the relative stream function error at snapshot $k$ $E_{\psi}(k)$ over nondimensional time. The reduced-order trajectories are recursively propagated from the first snapshot, with the identification and independent prediction evaluation intervals separated at $k=800$ ($t=4580$). The vertical dashed lines indicate the selected snapshot indices. (b)-(d) Reference and predicted vorticity fields, together with the corresponding signed vorticity error distributions $\Delta\omega$, at $k=800$, $1000$, and $1200$, corresponding approximately to nondimensional times $t=4580$, $4600$, and $4620$, respectively.}
    \label{fig:app_cavity30k_prediction}
\end{figure*}

The corresponding reconstruction and recursive prediction performance is presented in Fig.~\ref{fig:app_cavity30k_prediction}. The reduced-order trajectories are initialized from the first snapshot and propagated recursively throughout the observation interval, with the identification and independent prediction evaluation segments separated at $k=800$ ($t=4580$). Figure~\ref{fig:app_cavity30k_prediction}(a) shows the evolution of the relative full-field error at snapshot $k$ $E_{\psi}(k)$ defined in Section~4.1. Both methods exhibit comparable errors over the identification interval, but their performance diverges during independent prediction. The error of standard DMD increases rapidly and exhibits pronounced fluctuations, whereas GC-DMD maintains lower errors with slower overall accumulation, particularly over longer prediction horizons.

The spatial comparisons in Figs.~\ref{fig:app_cavity30k_prediction}(b)-(d) further illustrate this difference. At $k=800$ ($t=4580$), both methods reproduce the principal vortical structures, although differences remain in their signed vorticity error distributions. As recursive evolution proceeds to $k=1000$ ($t=4600$) and $k=1200$ ($t=4620$), the DMD results exhibit increasingly pronounced deviations from the reference vorticity field, particularly near the cavity boundaries and regions of concentrated vorticity. In contrast, GC-DMD maintains closer agreement with the reference field and generally reduces the spatial prediction errors, even at the final snapshot. These observations indicate that the improved operator identification translates into more accurate long-horizon prediction under increasingly irregular flow dynamics.

Overall, the $Re=30{,}000$ case provides complementary evidence that GC-informed regularization can promote a sparse, structured evolution operator while reducing long-horizon prediction errors. Together with the multifrequency results in Section~4.3, these findings provide further evidence of the effectiveness of GC-derived structural information in reducing redundant couplings and mitigating recursive prediction-error accumulation.

\printcredits

\section*{Acknowledgements}

\begingroup
\setlength{\emergencystretch}{3em}
\noindent
This work was supported by the National Natural Science Foundation of China (52602671, U2441211).
\par
\endgroup

\bibliographystyle{cas-model2-citation-order}
%\bibliographystyle{cas-model2-names}

\bibliography{cas-refs}

@article{taira2017modal,
  title={Modal analysis of fluid flows: An overview},
  author={Taira, Kunihiko and Brunton, Steven L and Dawson, Scott TM and Rowley, Clarence W and Colonius, Tim and McKeon, Beverley J and Schmidt, Oliver T and Gordeyev, Stanislav and Theofilis, Vassilios and Ukeiley, Lawrence S},
  journal={AIAA journal},
  volume={55},
  number={12},
  pages={4013--4041},
  year={2017},
  publisher={American Institute of Aeronautics and Astronautics}
}

@article{rowley2017model,
  title={Model reduction for flow analysis and control},
  author={Rowley, Clarence W and Dawson, Scott TM},
  journal={Annual Review of Fluid Mechanics},
  volume={49},
  pages={387--417},
  year={2017},
  publisher={Annual Reviews}
}

@article{schmid2022dynamic,
	title={Dynamic mode decomposition and its variants},
	author={Schmid, Peter J},
	journal={Annual Review of Fluid Mechanics},
	volume={54},
	number={1},
	pages={225--254},
	year={2022},
	publisher={Annual Reviews}
}

@article{rowley2004model,
	title={Model reduction for compressible flows using POD and Galerkin projection},
	author={Rowley, Clarence W and Colonius, Tim and Murray, Richard M},
	journal={Physica D: Nonlinear Phenomena},
	volume={189},
	number={1-2},
	pages={115--129},
	year={2004},
	publisher={Elsevier}
}

@article{braconnier2011towards,
	title={Towards an adaptive POD/SVD surrogate model for aeronautic design},
	author={Braconnier, Thierry and Ferrier, Marc and Jouhaud, J-C and Montagnac, Marc and Sagaut, Pierre},
	journal={Computers \& Fluids},
	volume={40},
	number={1},
	pages={195--209},
	year={2011},
	publisher={Elsevier}
}

@article{noack2016snapshots,
  title={From snapshots to modal expansions--bridging low residuals and pure frequencies},
  author={Noack, Bernd R},
  journal={Journal of Fluid Mechanics},
  volume={802},
  pages={1--4},
  year={2016},
  publisher={Cambridge University Press}
}

@article{schmid2010dynamic,
	title={Dynamic mode decomposition of numerical and experimental data},
	author={Schmid, Peter J},
	journal={Journal of fluid mechanics},
	volume={656},
	pages={5--28},
	year={2010},
	publisher={Cambridge University Press}
}

@book{tu2013dynamic,
	title={Dynamic mode decomposition: Theory and applications},
	author={Tu, Jonathan H},
	year={2013},
	publisher={Princeton University}
}

@book{kutz2016dynamic,
	title={Dynamic mode decomposition: data-driven modeling of complex systems},
	author={Kutz, J Nathan and Brunton, Steven L and Brunton, Bingni W and Proctor, Joshua L},
	year={2016},
	publisher={SIAM}
}

@article{hemati2014dynamic,
	title={Dynamic mode decomposition for large and streaming datasets},
	author={Hemati, Maziar S and Williams, Matthew O and Rowley, Clarence W},
	journal={Physics of Fluids},
	volume={26},
	number={11},
	year={2014},
	publisher={AIP Publishing}
}

@article{zhang2019online,
	title={Online dynamic mode decomposition for time-varying systems},
	author={Zhang, Hao and Rowley, Clarence W and Deem, Eric A and Cattafesta, Louis N},
	journal={SIAM Journal on Applied Dynamical Systems},
	volume={18},
	number={3},
	pages={1586--1609},
	year={2019},
	publisher={SIAM}
}

@article{bai2020dynamic,
	title={Dynamic mode decomposition for compressive system identification},
	author={Bai, Zhe and Kaiser, Eurika and Proctor, Joshua L and Kutz, J Nathan and Brunton, Steven L},
	journal={AIAA Journal},
	volume={58},
	number={2},
	pages={561--574},
	year={2020},
	publisher={American Institute of Aeronautics and Astronautics}
}

@article{jovanovic2014sparsity,
	title={Sparsity-promoting dynamic mode decomposition},
	author={Jovanovi{\'c}, Mihailo R and Schmid, Peter J and Nichols, Joseph W},
	journal={Physics of Fluids},
	volume={26},
	number={2},
	year={2014},
	publisher={AIP Publishing}
}

@article{wang2016nuclear,
	title={Nuclear norm regularised dynamic mode decomposition},
	author={Wang, Shaobo and Qing, Xiangyun},
	journal={IET Signal Processing},
	volume={10},
	number={6},
	pages={626--632},
	year={2016},
	publisher={Wiley Online Library}
}

@article{wynn2013optimal,
	title={Optimal mode decomposition for unsteady flows},
	author={Wynn, Andrew and Pearson, DS and Ganapathisubramani, Bharathram and Goulart, Paul J},
	journal={Journal of Fluid Mechanics},
	volume={733},
	pages={473--503},
	year={2013},
	publisher={Cambridge University Press}
}

@article{azencot2019consistent,
	title={Consistent dynamic mode decomposition},
	author={Azencot, Omri and Yin, Wotao and Bertozzi, Andrea},
	journal={SIAM Journal on Applied Dynamical Systems},
	volume={18},
	number={3},
	pages={1565--1585},
	year={2019},
	publisher={SIAM}
}

@article{rains2024constrained,
	title={Constrained optimized dynamic mode decomposition with control for physically stable systems with exogeneous inputs},
	author={Rains, Jacob and Wang, Yi and House, Alec and Kaminsky, Andrew L and Tison, Nathan A and Korivi, Vamshi M},
	journal={Journal of Computational Physics},
	volume={496},
	pages={112604},
	year={2024},
	publisher={Elsevier}
}

@article{tu2014spectral,
	title={Spectral analysis of fluid flows using sub-Nyquist-rate PIV data},
	author={Tu, Jonathan H and Rowley, Clarence W and Kutz, J Nathan and Shang, Jessica K},
	journal={Experiments in Fluids},
	volume={55},
	number={9},
	pages={1805},
	year={2014},
	publisher={Springer}
}

@article{xie2024regularized,
	title={Regularized dynamic mode decomposition algorithm for time sequence predictions},
	author={Xie, Xiaoyang and Tang, Shaoqiang},
	journal={Theoretical and Applied Mechanics Letters},
	volume={14},
	number={5},
	pages={100555},
	year={2024},
	publisher={Elsevier}
}

@article{baddoo2021physics,
	title={Physics-informed dynamic mode decomposition (piDMD)},
	author={Baddoo, Peter J and Herrmann, Benjamin and McKeon, Beverley J and Kutz, J Nathan and Brunton, Steven L},
	journal={arXiv preprint arXiv:2112.04307},
	year={2021}
}

@article{ding2026understanding,
	title={Understanding the global interdependencies of SDGs via causal network analysis},
	author={Ding, Yu and Chen, Xi and Shuai, Chenyang and Wang, Chenglong and Xu, Zhenci and Zhao, Bu},
	journal={Sustainable Production and Consumption},
	year={2026},
	publisher={Elsevier}
}

@incollection{guo2010granger,
	title={Granger causality: theory and applications},
	author={Guo, Shuixia and Ladroue, Christophe and Feng, Jianfeng},
	booktitle={Frontiers in computational and systems biology},
	pages={83--111},
	year={2010},
	publisher={Springer}
}

@article{shojaie2022granger,
	title={Granger causality: A review and recent advances},
	author={Shojaie, Ali and Fox, Emily B},
	journal={Annual review of statistics and its application},
	volume={9},
	pages={289--319},
	year={2022},
	publisher={Annual Reviews}
}

@incollection{bossomaier2016transfer,
	title={Transfer entropy},
	author={Bossomaier, Terry and Barnett, Lionel and Harr{\'e}, Michael and Lizier, Joseph T},
	booktitle={An introduction to transfer entropy: Information flow in complex systems},
	pages={65--95},
	year={2016},
	publisher={Springer}
}

@article{miersch2025evaluating,
	title={Evaluating the robustness of pcmci+ for causal discovery of flood drivers},
	author={Miersch, Peter and G{\"u}nther, Wiebke and Runge, Jakob and Zscheischler, Jakob},
	journal={Artificial Intelligence for the Earth Systems},
	volume={4},
	number={4},
	pages={240114},
	year={2025},
	publisher={American Meteorological Society}
}

@article{martinez2024decomposing,
	title={Decomposing causality into its synergistic, unique, and redundant components},
	author={Mart{\'\i}nez-S{\'a}nchez, {\'A}lvaro and Arranz, Gonzalo and Lozano-Dur{\'a}n, Adri{\'a}n},
	journal={Nature Communications},
	volume={15},
	number={1},
	pages={9296},
	year={2024},
	publisher={Nature Publishing Group UK London}
}

@article{martinez2023causality,
	title={Causality analysis of large-scale structures in the flow around a wall-mounted square cylinder},
	author={Mart{\'\i}nez-S{\'a}nchez, {\'A}lvaro and L{\'o}pez, Esteban and Le Clainche, Soledad and Lozano-Dur{\'a}n, Adri{\'a}n and Srivastava, Ankit and Vinuesa, Ricardo},
	journal={Journal of Fluid Mechanics},
	volume={967},
	pages={A1},
	year={2023},
	publisher={Cambridge University Press}
}

@article{massaro2026interpreting,
	title={Interpreting coherence in wall turbulence},
	author={Massaro, Daniele and Hussain, Fazle},
	journal={Nature Reviews Physics},
	pages={1--15},
	year={2026},
	publisher={Nature Publishing Group UK London}
}

@article{wang2026causal,
	title={Causal analysis of a turbulent shear flow model},
	author={Wang, Yongyi and Kou, Jiaqing and Noack, Bernd R and Zhang, Weiwei},
	journal={Journal of Fluid Mechanics},
	volume={1035},
	pages={A18},
	year={2026},
	publisher={Cambridge University Press}
}

@inproceedings{tissot2014granger,
	title={Granger causality in wall-bounded turbulence},
	author={Tissot, Gilles and Lozano-Dur{\'a}n, Adrian and Cordier, Laurent and Jim{\'e}nez, Javier and Noack, Bernd R},
	booktitle={Journal of Physics: Conference Series},
	volume={506},
	number={1},
	pages={012006},
	year={2014}
}

@article{towne2023database,
	title={A database for reduced-complexity modeling of fluid flows},
	author={Towne, Aaron and Dawson, Scott TM and Br{\`e}s, Guillaume A and Lozano-Dur{\'a}n, Adri{\'a}n and Saxton-Fox, Theresa and Parthasarathy, Aadhy and Jones, Anya R and Biler, Hulya and Yeh, Chi-An and Patel, Het D and others},
	journal={AIAA journal},
	volume={61},
	number={7},
	pages={2867--2892},
	year={2023},
	publisher={American Institute of Aeronautics and Astronautics}
}

@article{gunjal2023granger,
	title={Granger Causality for prediction in Dynamic Mode Decomposition: Application to power systems},
	author={Gunjal, Revati and Nayyer, Syed Shadab and Wagh, SR and Stankovic, AM and Singh, NM},
	journal={Electric Power Systems Research},
	volume={225},
	pages={109865},
	year={2023},
	publisher={Elsevier}
}

@article{nathaniel2025deep,
	title={Deep Koopman operators for causal discovery},
	author={Nathaniel, Juan and Roesch, Carla and Buch, Jatan and DeSantis, Derek and Rupe, Adam and Lamb, Kara D and Gentine, Pierre},
	journal={Communications Physics},
	volume={8},
	number={1},
	pages={513},
	year={2025},
	publisher={Nature Publishing Group UK London}
}

@book{partington2004linear,
  title={Linear operators and linear systems: an analytical approach to control theory},
  author={Partington, Jonathan R},
  number={60},
  year={2004},
  publisher={Cambridge University Press}
}

@article{srivastava2026operator,
  title={Operator theoretic causality analysis of fluid flows using linearized dynamics: A. Srivastava et al.},
  author={Srivastava, Ankit and Jim{\'e}nez-Fern{\'a}ndez, V{\'\i}ctor and Cattafesta III, Louis N and Dawson, Scott TM},
  journal={Theoretical and Computational Fluid Dynamics},
  volume={40},
  number={4},
  pages={22},
  year={2026},
  publisher={Springer}
}

@article{hirsh2020centering,
  title={Centering data improves the dynamic mode decomposition},
  author={Hirsh, Seth M and Harris, Kameron Decker and Kutz, J Nathan and Brunton, Bingni W},
  journal={SIAM Journal on Applied Dynamical Systems},
  volume={19},
  number={3},
  pages={1920--1955},
  year={2020},
  publisher={SIAM}
}

@article{kou2017improved,
  title={An improved criterion to select dominant modes from dynamic mode decomposition},
  author={Kou, Jiaqing and Zhang, Weiwei},
  journal={European Journal of Mechanics-B/Fluids},
  volume={62},
  pages={109--129},
  year={2017},
  publisher={Elsevier}
}

@article{liu2021data,
  title={Data-driven modal decomposition of transient cavitating flow},
  author={Liu, Yunqing and Long, Jincheng and Wu, Qin and Huang, Biao and Wang, Guoyu},
  journal={Physics of Fluids},
  volume={33},
  number={11},
  year={2021},
  publisher={AIP Publishing}
}

@article{granger1969investigating,
  title={Investigating causal relations by econometric models and cross-spectral methods},
  author={Granger, Clive WJ},
  journal={Econometrica: journal of the Econometric Society},
  pages={424--438},
  year={1969},
  publisher={JSTOR}
}

@article{barnett2014mvgc,
  title={The MVGC multivariate Granger causality toolbox: a new approach to Granger-causal inference},
  author={Barnett, Lionel and Seth, Anil K},
  journal={Journal of neuroscience methods},
  volume={223},
  pages={50--68},
  year={2014},
  publisher={Elsevier}
}

@article{benjamini1995controlling,
  title={Controlling the false discovery rate: a practical and powerful approach to multiple testing},
  author={Benjamini, Yoav and Hochberg, Yosef},
  journal={Journal of the Royal statistical society: series B (Methodological)},
  volume={57},
  number={1},
  pages={289--300},
  year={1995},
  publisher={Wiley Online Library}
}

@article{beck2009fast,
  title={A fast iterative shrinkage-thresholding algorithm for linear inverse problems},
  author={Beck, Amir and Teboulle, Marc},
  journal={SIAM journal on imaging sciences},
  volume={2},
  number={1},
  pages={183--202},
  year={2009},
  publisher={SIAM}
}

@article{chang2026post,
  title={Post-selection inference for causal effects after causal discovery},
  author={Chang, Ting-Hsuan and Guo, Zijian and Malinsky, Daniel},
  journal={Biometrika},
  volume={113},
  number={1},
  pages={asaf073},
  year={2026},
  publisher={Oxford University Press}
}

@article{lozano2009grouped,
  title={Grouped graphical Granger modeling for gene expression regulatory networks discovery},
  author={Lozano, Aur{\'e}lie C and Abe, Naoki and Liu, Yan and Rosset, Saharon},
  journal={Bioinformatics},
  volume={25},
  number={12},
  pages={i110--i118},
  year={2009},
  publisher={Oxford University Press}
}

@article{arbabi2017study,
  title={Study of dynamics in post-transient flows using Koopman mode decomposition},
  author={Arbabi, Hassan and Mezi{\'c}, Igor},
  journal={Physical Review Fluids},
  volume={2},
  number={12},
  pages={124402},
  year={2017},
  publisher={APS}
}

@article{schmidt2018spectral,
  title={Spectral analysis of jet turbulence},
  author={Schmidt, Oliver T and Towne, Aaron and Rigas, Georgios and Colonius, Tim and Bres, Guillaume A},
  journal={Journal of Fluid Mechanics},
  volume={855},
  pages={953--982},
  year={2018},
  publisher={Cambridge University Press}
}

@article{bres2018importance,
  title={Importance of the nozzle-exit boundary-layer state in subsonic turbulent jets},
  author={Br{\`e}s, Guillaume A and Jordan, Peter and Jaunet, Vincent and Le Rallic, Maxime and Cavalieri, Andr{\'e} VG and Towne, Aaron and Lele, Sanjiva K and Colonius, Tim and Schmidt, Oliver T},
  journal={Journal of Fluid Mechanics},
  volume={851},
  pages={83--124},
  year={2018},
  publisher={Cambridge University Press}
}

\end{document}